# Inferring Hidden Motives: A Utility Bayesian Model of learning the values of others

Gregory Stanley, Jun Zhang, and Richard L. Lewis

***Abstract***

Cooperation depends on judging whom to trust: identifying the strengths of other people's morally relevant social motives, called social preferences, and refining those beliefs after observing their behavior over time. We model this process with a Utility Bayesian Model (UBM): a continuous probability distribution over the parameters of another person's utility function, reweighted after each observed choice. In repeated binary dictator games with randomized payoffs, the UBM captured participants' predictions of preprogrammed agents and of one another better than non-Bayesian alternatives and over $15,000$ models that represent others as discrete types: people track motives as graded positions in a continuous space, not as categories. Because participants alternated between choosing and predicting, the framework separately estimates preferences and beliefs about preferences. Choosers' own utilities, fit with a seven-parameter function selected from $526$ candidate forms, showed that most people place reliably positive weight on a stranger's payoff but weight their own roughly nine times more; that aversion to being ahead exceeds aversion to being behind, reversing the canonical ordering; that sensitivity to others' payoffs and to inequalities grows more than proportionally with their size; and that antisocial preferences are more common than prior estimates suggest. Predictors' initial beliefs captured the dominance of self-interest but expected aversion to being behind to outweigh aversion to being ahead. Together, these parameter distributions locate any individual's social preferences, and beliefs about them, in a common seven-dimensional space, enabling commensurable cross-study comparisons, meta-analyses, and empirically informed priors for future Bayesian models of social cognition.

# 1 Introduction

Understanding other people's social preferences is central to cooperation. To decide whom to trust, reciprocate, avoid, forgive, or help, people must infer how others value their own outcomes, the outcomes of others, and inequalities between them. This cognitive problem is difficult because social preferences are hidden and must be inferred from behavior that is often compatible with multiple underlying motives.

This paper focuses on outcome-based social preferences: preferences over the final distribution of payoffs between people. An outcome-based preference cares only about who ends up with what, not about how the distribution came about. It sets aside the many other motives that shape social behavior, including intentions, reciprocity, reputation, self-image, social norms, and strategic reasoning. Those phenomena matter, but they attach to the process and context of an interaction, whereas outcome-based preferences attach to its result. Isolating them is the natural starting point, because preferences over outcomes are the part of social motivation that every other motive ultimately operates on top of.

We study these preferences with the simplest social choice that can express them: the binary dictator game. A chooser selects between two fixed allocations, option A and option B, each specifying a payoff for the chooser and a payoff for the other player. Because the other player is passive and the choice is one-shot, there is nothing to retaliate against, negotiate over, or strategically anticipate. Reciprocity, reputation, and within-round strategy have no structural foothold, so what remains in the choice is evidence about how the chooser values the outcomes themselves: their own payoff, the other's payoff, and the relationship between them. The game is engineered to strip away everything except the outcome-based core.

What it cannot strip away is the multiplicity inside that core, and this is what makes even so simple a game worth studying. The moment a second person's payoff enters, four quantities arise together and cannot be separated: what the chooser gains, what the other gains, the gap between the two, and the absolute size of the stakes. No social choice can be made simpler than this without removing the second person altogether, because adding any payoff for another introduces the self-other gap and the absolute stakes in the same stroke. Nor is this multiplicity peculiar to dictator games. Deeper game trees, more options, more players, and simultaneous moves all add structure on top of these quantities, but none subtracts them: extra levels collapse into expected payoffs at the point of choice, and added players or simultaneity introduce strategic and higher-order reasoning without ever eliminating the underlying payoff dimensions. These four quantities are not a feature of one experimental paradigm. They are present in any game, however complex, and in any real social situation, however far from a laboratory. Splitting an inheritance, deciding how much to tip, choosing whether to undercut a colleague, voting for a tax policy that benefits you at others' expense: each is at bottom a judgment about how much one's own payoff is worth against someone else's, and whether the gap between them matters in itself.

Because so many of these quantities are present at once, a single choice rarely reveals a single motive. The same allocation can be chosen out of generosity, out of guilt, out of a wish to come out ahead, or out of indifference, and these motives leave overlapping traces in behavior. The mathematics of the payoffs is precise, but the motive behind the choice is not pinned down by it. Other designs can appear to sidestep this ambiguity but do not actually resolve it. For instance, a standard continuous dictator game with a fixed sum to divide holds the total constant, which makes giving to the other identical to reducing the gap and so conflates altruism with a distaste for inequality rather than controlling for either. The ambiguity is not an artifact of any one task. It is intrinsic to outcome-based choice. This is the problem social creatures must face. We must judge one another's motives from behavior that is consistent with several of them, and we must do it well enough to decide whom to trust, because the cost of misreading is being exploited. That inference problem, reading hidden preferences from ambiguous choices, is precisely what a Bayesian observer is built to solve, and it motivates the model we develop here.

Research in behavioral economics and psychology has produced influential models of such preferences, including inequity aversion; Equity, Reciprocity, and Competition; social welfare; Constant Elasticity of Substitution; and maximin-efficiency models (Fehr & Schmidt, 1999; Bolton & Ockenfels, 2000; Andreoni & Miller, 2002; Charness & Rabin, 2002; Engelmann & Strobel, 2004). These models have clarified stable individual differences in altruism, fairness, self-interest, and inequality aversion. However, two interrelated methodological restrictions limit how well existing models generalize.

The first restriction is that many studies fit models to a narrow portion of the possible payoff landscape. Allocation tasks often use small, hand-selected payoff sets, which can make logically distinct motives behaviorally indistinguishable. For example, standard dictator games are constant-sum and cannot decouple altruism from social comparison. Broader payoff exploration is necessary for distinguishing psychological dimensions that collapse under narrower designs (Andreoni & Miller, 2002; Murphy & Ackermann, 2014).

The second restriction is that researchers typically compare only a small set of inherited utility functions. A utility function is just an equation that takes the situation a person is in, such as how many apples they have, the weather outside, or a particular split of money between themselves and someone else, and returns a number standing for how much they value that situation. In social settings, its parameters carry psychological meaning: how much weight a person puts on their own payoff, on another's, on the gap between them. But a utility function can only recover distinctions it is capable of representing. If the candidate function class omits relevant psychological terms, fitted parameters may absorb multiple motives at once, producing estimates that are stable but conceptually misleading. If a utility function lacks a separate altruism term, concern for others loads onto inequality aversion. If it lacks separate terms for advantageous and disadvantageous inequality, envy-like and guilt-like responses collapse. If self-interest is fixed rather than estimated, unusual but theoretically meaningful patterns such as negative self-weighting cannot be detected. Limited payoff exploration and limited utility-function exploration can each distort inference.

These limitations matter not only for measuring social preferences but also for modeling how people learn them. Cooperation depends on updating beliefs about counterparts: who is generous, who is selfish, who is competitive. Bayesian models are well-suited to this problem because they formalize how uncertain beliefs are updated as evidence accumulates (Camerer & Ho, 1999; Griffiths, Kemp, & Tenenbaum, 2008; Hampton, Bossaerts, & O'Doherty, 2008; Yoshida, Dolan, & Friston, 2008; Fudenberg & Levine, 2009; Devaine, Hollard, & Daunizeau, 2014; Kleiman-Weiner, Saxe, & Tenenbaum, 2017). But a model of social-preference learning depends on the utility function embedded in its likelihood term. If that function conflates motives, the learning model inherits the confusion.

A further problem is that researchers often infer beliefs about social preferences from choice data alone. Choices are not pure measures of beliefs: they reflect the chooser's own preferences, strategic incentives, and expectations about others simultaneously. Explicit predictions of a counterpart's choices provide a more direct measure of beliefs, reducing the confound between what a participant wants and what they expect another person to do.

To address these problems, we used repeated binary dictator games in which participants encountered identifiable counterparts over multiple rounds, alternating between chooser and predictor roles. Payoffs for both players were randomly drawn from $\{1, 2, 3, 4, 5\}$, spanning all $625$ possible binary dictator-game structures defined by that grid. The task isolates outcome-based social preferences from within-round strategic reasoning while covering a broad range of self-regarding, other-regarding, and inequality-sensitive tradeoffs. A "dyad" is a series of games between the same two players. Each participant was involved in multiple dyads, and in some cases would alternate between being in the role of the chooser or the predictor, who would predict choices rather than making choices themselves.

The present approach integrates four mutually reinforcing methodological elements.

First, we use randomized binary dictator-game payoffs to sample a broad multidimensional outcome space, allowing utility functions to be evaluated across many different tradeoffs rather than over a small set of hand-picked games.

Second, we compare $526$ candidate utility functions varying along psychologically meaningful dimensions: self-interest, altruism, social comparison, asymmetric envy and guilt terms, nonlinear payoff sensitivity, reference dependence, payoff ratios, min-max forms, etc. The goal is not to ask which canonical model performs best, but to search a wider space of possible utility structures and identify which distinctions are worth preserving after penalizing model complexity.

Third, we elicit explicit predictions of counterpart choices. These predictions provide a more direct behavioral window into participants' beliefs about others' preferences than choices alone, while choice data continue to supply estimates of the chooser's own preferences.

Fourth, we formalize social-preference learning with a Utility Bayesian Model. Observers maintain probabilistic beliefs over continuous social-preference parameters, updated after each observed choice via a likelihood term supplied by the utility function and a softmax choice rule. Parameter values that make the observed choice more likely gain posterior weight, while values that make it unlikely lose weight. Across repeated meetings, initially uncertain beliefs become more concentrated as evidence accumulates.

This integrated design lets us ask several linked questions. Do people update beliefs about others' social preferences in a way better captured by a continuous Bayesian model than by non-Bayesian or discrete typological alternatives? Which utility function best supports such inference across a broad payoff space? What distribution of self-interest, altruism, envy, guilt, and nonlinear payoff sensitivity is revealed when these dimensions are estimated jointly? And what do those distributions imply about the structure of human social motivation?

The paper proceeds in the order these questions require. We first specify the Utility Bayesian Model and establish through simulation that its machinery recovers known preferences, that posteriors converge toward a chooser's true preferences over repeated observations, and that prior variance and choice stochasticity affect update speed as Bayesian theory predicts. We then validate it against human behavior in a setting with known ground truth, an experiment in which people predict the choices of artificial agents whose preferences we fixed in advance, and show that a continuous Bayesian model outperforms both non-Bayesian and discrete-typological alternatives. Only then do we turn to human-on-human data: first to select, from 526 candidate utility functions, the form that best captures how people choose and predict, and then to use that form to estimate the full distribution of social preferences in the population. The final step asks what that distribution reveals about social motivation, and what the broader method implies beyond the case studied here.

# 2 Utility Bayesian Model: Formulation and Implementation

## 2.1 Level of Description

The Utility Bayesian Model (UBM) formalizes the problem of inferring hidden motives from ambiguous observations, as described in the introduction. It does so at a functional (a.k.a. computational) level (Marr, 1982) in the sense that it specifies the information-processing problem an observer must solve without claiming that the brain literally performs the computation this way. Although the model uses explicit probabilistic representations and grid-based approximations for clarity, real observers may approximate the same inference through other mechanisms such as sampling-based heuristics.

Bayesian models of social inference have consistently outperformed reinforcement-based and heuristic accounts in contexts where latent preferences must be inferred from ambiguous behavior (Camerer & Ho, 1999; Erev & Roth, 1998; Hampton, Bossaerts, & O'Doherty, 2008), and they offer greater interpretability than more opaque approaches such as neural networks, in the sense that each component can be assigned a psychological meaning even if the underlying implementation is unknown. The architecture below makes this concrete: every component has a psychological meaning.

Next, we detail how our Utility Bayesian Model integrates prior beliefs with observed chooser decisions to produce updated posterior beliefs about social preference parameters.

## 2.2 Utility Bayesian Model Mechanics: Prior, Likelihood, and Posterior

Observers' initial beliefs about a chooser's social preferences are represented by a population prior over a multidimensional parameter space (Gelman, et al., 2003). This prior assumes that social preference parameters are normally distributed in the population. Before observing any individual's choices, one predicts their behavior as if randomly sampled from this population (Lee & Wagenmakers, 2014). We model each preference dimension with a Gaussian (normal) prior distribution[1] characterized by population-level parameters (means and standard deviations). Once observations of the chooser's actual decisions become available, the model updates from these population priors to individual-level posterior beliefs, although after the first round, we do not impose any specific parametric form on subsequent posteriors.

[1] The gaussian prior is truncated within parameter bounds for the mean parameters, which is between -1 and 1 for self-interest, altruism, envy, and guilt and is 0.02 – 2.00 for the exponent and 0.02 – 4.00 for temperature.

Specifically, we parameterize social preferences using:

- $V_{ii}$: self-interest, indicating how strongly the chooser values their own payoffs.
- $V_{ij}$: altruism, indicating how strongly the chooser values the recipient's payoffs.
- $\alpha_{ij}$: envy, capturing negative reactions to disadvantageous inequality.
- $\beta_{ij}$: guilt, capturing negative reactions to advantageous inequality.
- $\gamma$: exponent, capturing nonlinear sensitivity to payoff magnitudes and differences.
- $\tau$: temperature in the softmax function, determining stochasticity in choice behaviors.

Here the chooser is player $i$ and the predictor is player $j$. $V$ indicates value (i.e., $V_{ij}$ reflects how much player $i$ values player $j$. Loss aversion parameters $\lambda_{ii}$ and $\lambda_{ij}$ are introduced later in the paper.

We use $\mu$ for mean parameters and $\sigma$ for standard deviation parameters. For instance, $\mu(V_{ij})$ and $\sigma(V_{ij})$ are the mean and standard deviation parameters for altruism. Note that chooser parameters are single values, unlike predictor parameters, which are distributions that can be characterized by a mean and standard deviation for each preference dimension.

This is an example utility function:

$$U_i(A) = V_{ii}\left(\pi_i^A\right)^\gamma + V_{ij}\left(\pi_j^A\right)^\gamma - \alpha_{ij} \times max\left(\pi_j^A - \pi_i^A, 0\right)^\gamma - \beta_{ij} \times max\left(\pi_i^A - \pi_j^A, 0\right)^\gamma$$

For the utility for player $i$ in option A, where $\pi_i^A$ is the payoff for player $i$ in option A and $\pi_j^A$ is the payoff for player $j$ in option A.

## 2.3 Likelihood Function and Softmax Choice Rule

After observing a chooser's decision, we compute the likelihood of this choice given a parameter vector $\theta = \{V_{ii}, V_{ij}, \alpha_{ij}, \beta_{ij}, \gamma, \tau\}$ through a logit-type or softmax function (Luce, 1959; McFadden, 1974):

$$p(choose\ option\ A|\theta) = \frac{e^{U(A,\theta)/\tau}}{e^{U(A,\theta)/\tau} + e^{U(B,\theta)/\tau}}$$

Here $\tau$ (temperature parameter) controls how deterministic or stochastic choices are: higher $\tau$ yields more random, less deterministic choices, whereas lower $\tau$ increases the probability of choosing the option with higher subjective utility. Subjective utility, $U(A, \theta)$, is computed from the parameters $\theta$ by weighting each payoff dimension accordingly. The likelihood then follows from translating these utilities into choice probabilities through the softmax rule[2].

## 2.4 Grid-Based and Particle Filter Posterior Updating

By default, we used a discrete grid-based approximation method for posterior updating because it is simple and transparent. The multidimensional parameter space is discretized into a finite grid (with resolution defined by $n$ bins per dimension). At each grid point, we compute likelihood probabilities, forming a likelihood probability mass function (PMF) over the parameter space. This likelihood PMF is directly comparable to the prior PMF because both are defined over identical parameter grids.

Multiplying the prior and likelihood PMFs point-by-point yields the posterior PMF, representing updated beliefs after observing each choice:

$$p(\theta|choice) \propto p(choice|\theta) \times p(\theta)$$

The posterior PMF subsequently serves as the prior for future belief updates, creating a self-perpetuating Bayesian cycle. This iterative self-updating mechanism ensures that a single set of initial parameters is sufficient to simulate participant responses across multiple games[3].

However, because the number of points in the PMF grids is the number of bins per dimension raised to the number of dimensions ($n^k$), full grid-based updating becomes computationally infeasible for higher-dimensional models. Instead, we also switch to using particle filter-based posterior updating, which replaces the full PMF grids with much smaller samples of weighted probability particles.

[2] Throughout this paper, the Utility Bayesian Model generates simulated choices and predictions using a fixed softmax temperature of $\tau$ = 1.0. This response temperature standardizes the model's own outputs and sets a theoretical minimum achievable loss. It does not characterize participant behavior. By contrast, the likelihood function uses a separate softmax temperature fitted individually to each participant. This likelihood temperature reflects participants' beliefs about how stochastically their chooser-counterpart behaves, and governs Bayesian belief updating.

For predictor participants specifically, this likelihood temperature conflates two conceptually distinct forms of stochasticity: (1) the predictor's belief about the randomness of their chooser-counterpart's choices, and (2) the predictor's own stochasticity when submitting predictions. These could in principle be modeled with separate parameters, since a predictor might be highly confident that their counterpart is random while themselves being a very consistent predictor, or vice versa. We collapse them into a single parameter because the short game series does not provide sufficient data to identify both independently. Notably, unlike the social-preference parameters ($V_{ii}$, $V_{ij}$, $\alpha_{ij}$, $\beta_{ij}$, etc.), $\tau$ has no corresponding variance parameter because it is not a dimension of the belief distribution being updated. It is a fixed assumption about the shape of the likelihood function used in those updates.

[3] Updates were skipped whenever participants failed to respond within the allotted time, mitigating potential contamination from inattentive responses.

We verified that the particle filter (PF) reproduces the full grid posterior on synthetic data while varying the number of parameters $k \in [1, 2, 3, \ldots 9]$ and the ratio of probability particles sampled from the full grid, called the sampling ratio within $[0.05, 0.10, 0.15, 0.20, \ldots 0.95]$. We simulated $50$ artificial choosers paired with $50$ artificial predictors across 20 random payoff dictator games, where the predictors used grids of nine bins per dimension. We provided identical prior parameters to grid-based and PF-based predictors and compared the posteriors recovered from their final round posteriors. In *all* cases, the correlation exceeded $0.993$, demonstrating that the PF faithfully reproduces grid-based updating, typically multiple orders of magnitude faster. We used PF-based updating for the section, "Results: Joint Parameter Distributions in Human-on-human Experiment," which relied on high-dimensional models and we used grid-based updating in the followings sections: "Simulation Methods and Results", "Illustrating Belief Updates", and "Model Validation: Comparing Bayesian and Alternative Cognitive Models".

## 2.5 Optimization and Model Evaluation

Our Utility Bayesian Model follows the sequence of steps participants experience:

1. First, predictor-participants predict their chooser-counterparts' choice.
2. Second, they observe the actual choice.
3. Finally, they revise their beliefs based on this observation. They use these updated beliefs to make predictions in the next round, or more specifically the next round that they are matched with the same counterpart and assigned the role of predictor.

Crucially, participants' predictions reflect posterior beliefs revised based on choices observed in the previous round. Thus, our model follows the psychologically plausible causal sequence: predict choice → observe choice → update belief, as shown in Figure 1.

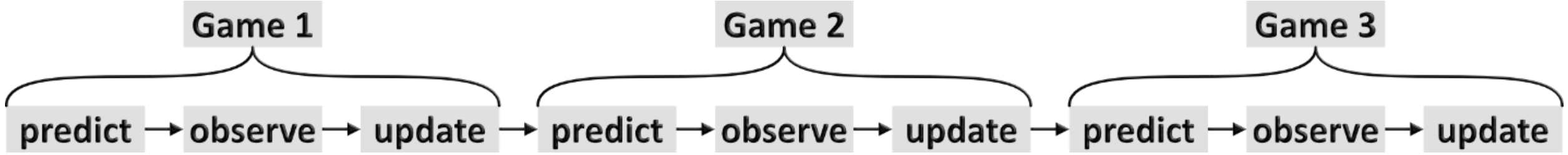

Figure 1: Predict, observe, update:
Within each game, players in the predictor role, first predict their counterpart's choice, second observe their counterpart's choice, and third update their beliefs about their counterpart's social preferences. These revised beliefs (posteriors) become the preconceptions (priors) used to make predictions in the next game. Choice observations are an input to the model and predictions are an output of the model.

Our agent function separately maintains stable chooser parameters (used exclusively to generate choices) and dynamically updates predictor parameters (used exclusively for generating predictions). Rather than using predictions or prediction errors, predictor beliefs evolve solely based on observed chooser actions, separating Bayesian updating evidence (chooser actions) from model evaluation criteria (participant predictions).

Model evaluation utilizes the negative log-likelihood (NLL), comparing posterior-generated model predictions against actual participant predictions:

$$NLL(\theta) = -\sum_{t=1}^{T} ln[p(observed\ choice_t|\theta)],$$

Here, participant parameter vectors ($\theta$) directly map to total negative log-likelihood losses, allowing the optimization process to adjust only initial parameters to minimize this cumulative loss across all observations.

For parameter estimation, we first apply simulated annealing (SA), a global optimization method robust to non-differentiable and complex loss functions, initially exploring the parameter space broadly and gradually shifting focus toward regions of lower loss as it "cools" (Kirkpatrick, Gelatt Jr, & Vecchi, 1983). We subsequently refine these parameter estimates using the local optimization algorithm L-BFGS-B (Byrd, Lu, Nocedal, & Zhu, 1995), which efficiently searches for more precise minima using gradient-based methods. Parameters identified by SA serve as starting points for L-BFGS-B, allowing a smooth transition from global exploration to local exploitation. Ultimately, we select the parameter set that yields the lowest NLL across both optimization steps[4].

Figure 2 illustrates the dynamic evolution of beliefs through the Bayesian updating process across sequential games. The likelihood functions (top row), generated from observed choices (middle row), inform the posterior beliefs (bottom row), demonstrating how initial uncertainty progressively sharpens into precise beliefs over successive observations.

---

[4] We implemented a nonlinear penalty on parameter values during the estimation process, calculated by squaring each mean parameter, multiplying by a penalty weight of $0.05$, and adding these values to the loss function. This penalty was essential because most utility functions exhibit scale invariance in their parameters, meaning infinitely many parameter sets differing by a constant multiplier could produce identical predicted behavior (e.g., parameters $\{.1, .2, .4\}$ versus $\{.2, .4, .8\}$). Without this penalty, the L-BFGS-B optimization algorithm frequently failed to converge to stable solutions, terminating prematurely.

The penalty itself was subtracted from the final reported loss values, thereby minimally impacting the primary analyses or conclusions. Nonetheless, this method slightly biases parameter estimates toward zero, marginally compressing extreme parameter values and subtly affecting the shape of parameter distributions.

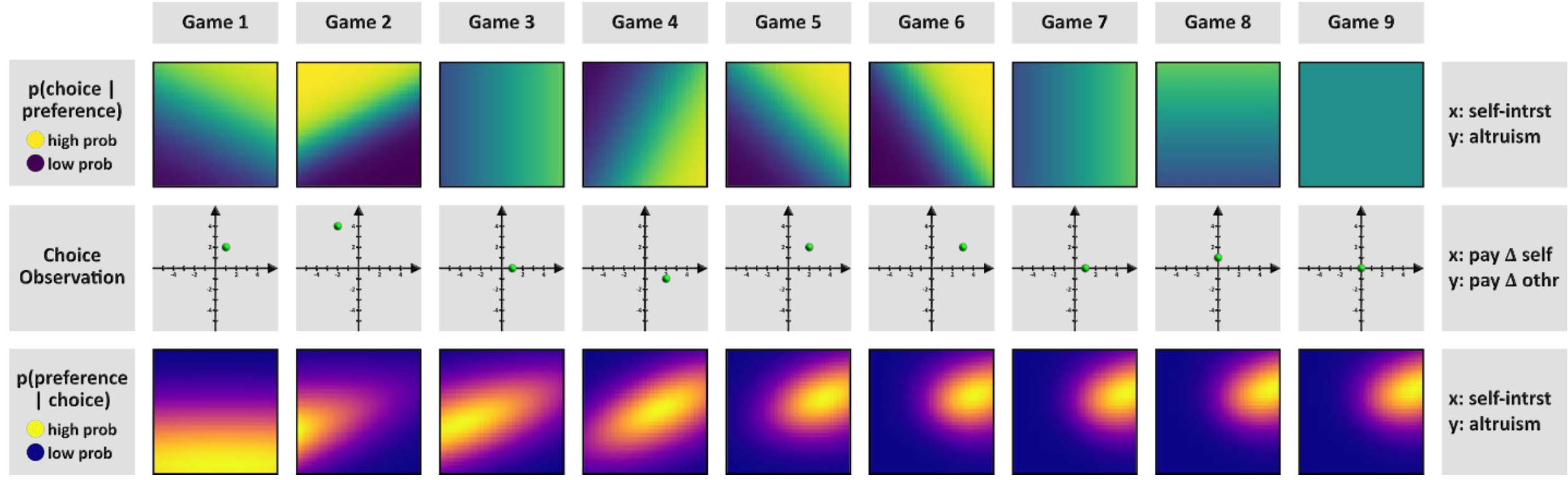

Figure 2: Bayesian belief updating over nine sequential decisions:
Each column illustrates one "game" (choice observation) and how that single observation updates the agent's preference belief state. Top row: Likelihood of the observed choice given different preference parameters, shown as heatmaps over the 2D preference space ($x$-axis = self-interest weight, $y$-axis = altruism weight). Yellow regions indicate high probability of the observed choice under those preference values, and purple regions indicate low probability. Middle row: The choice observation in each game, represented by the green dot in a payoff-difference space ($x$-axis = Δ payoff to self, $y$-axis = Δ payoff to other). Bottom row: Posterior distribution of the preference parameters after observing that game's choice, shown as a probability distribution heatmap (yellow = high posterior density, purple = low). As the games progress from 1 to 9, the posterior distributions become more concentrated, reflecting increasing certainty about the agent's true preference parameters. This figure demonstrates how Bayesian updating uses each new choice to incrementally refine the belief over the preference model. Interestingly, the likelihood distributions (top row) become sharper in proportion to the distance between the chosen option and the origin of the payoff space (middle row), reflecting the fact that the model treats extreme choices to be more diagnostic of preferences than moderate choices. Indeed, the game nine choice of $(3, 3)$ vs. $(3, 3)$ is uninformative, resulting in a flat likelihood and thus there is no belief update.

Having articulated our model's mechanics, notation, and evaluation strategies, we next validate the model with a simulation.

# 3 Testing our Model and Methods with Simulations

Having explained *how* the UBM works, we turn to demonstrating *that* the UBM works. For the UBM to work, beliefs about social preferences need to become more accurate as a result of repeated observations, provided that the player being observed makes consistent choices. Indeed, this learning should be fastest when the softmax temperature is low, creating a consistent signal, and when preconceptions are easiest to overwrite—the priors have high variance. Additionally, we needed to establish that the machinery to fit parameters to the UBM works in the sense that it converges towards accurate parameters. To earn trust in our model and methods, we replicated the conditions of our experiments with synthetic participants in a simulation to test belief update accuracy, belief update speeds, and our parameter optimizer.

## 3.1 Simulation Setup

We generated 945 simulated dyads by varying parameter values for predictors and choosers:

- **Predictor parameters**
  - *Self-interest mean* $\mu(V_{ii})$: fixed at 1.0
  - *Self-interest standard deviation* $\sigma(V_{ii})$: fixed at 1.0
  - *Altruism mean* $\mu(V_{ij})$: seven levels [-1.0, -0.667, -0.333, 0.0, 0.333, 0.667, 1.0]
  - *Altruism standard deviation* $\sigma(V_{ij})$: three levels [0.5, 1.0, 1.5]
  - *Softmax temperature* $\tau$: three levels [0.5, 1.75, 3.0]
- **Chooser parameters**
  - *Self-interest* $V_{ii}$: fixed at 1.0
  - *Altruism mean* $V_{ij}$: five levels [-1.0, -0.5, 0.0, 0.5, 1.0]
  - *Softmax temperature* $\tau$: three levels [0.5, 1.75, 3.0]

Every possible combination of these levels yielded $7 \times 3 \times 3 \times 5 \times 3 = 63$ predictors $\times$ 15 choosers $=$ 945 unique dyads. Each dyad matched one "predictor" bot to one "chooser" bot. The simulation proceeded for 24 rounds per dyad. On each round:

1. A randomized payoff structure from $\{1, 2, 3, 4, 5\}$ was assigned to the binary choice, simulating the variety of potential payoffs observed in a real experiment.
2. The "chooser" bot used its utility function $U_i(A) = V_{ii}\left(\pi_i^A - \pi_i^B\right) + V_{ij}\left(\pi_j^A - \pi_j^B\right)$ with its softmax temperature τ to probabilistically select one of the two payoff options.
3. Simultaneously, the "predictor" bot used an analogous function to produce a predicted choice (and so introduced "semi-stochastic" noise when $\tau \neq 0$).

At the end of 24 rounds, we fitted our Utility Bayesian Model (UBM) at the dyad level—that is, each predictor's parameter estimates were inferred from those 24 rounds against a single chooser. We focused on $\mu(V_{ij})$, the altruism mean, for most analyses. Although other parameters could also be examined, altruism is arguably the most morally salient to human observers.

## 3.2 Simulation Results

### *1. Predictor Estimates Converge to the Chooser's True Altruism*

We paired synthetic choosers with synthetic predictors to see if the predictors could figure out how altruistic their chooser-counterparts actually were by the end of 24 rounds. Figure 3 tracks the round-by-round correlation between the predictors belief about their chooser-counterpart's altruism and the altruism parameter we actually assigned to those choosers. The predictors beliefs converge towards the truth, reaching a correlation of 0.886 by the final round.

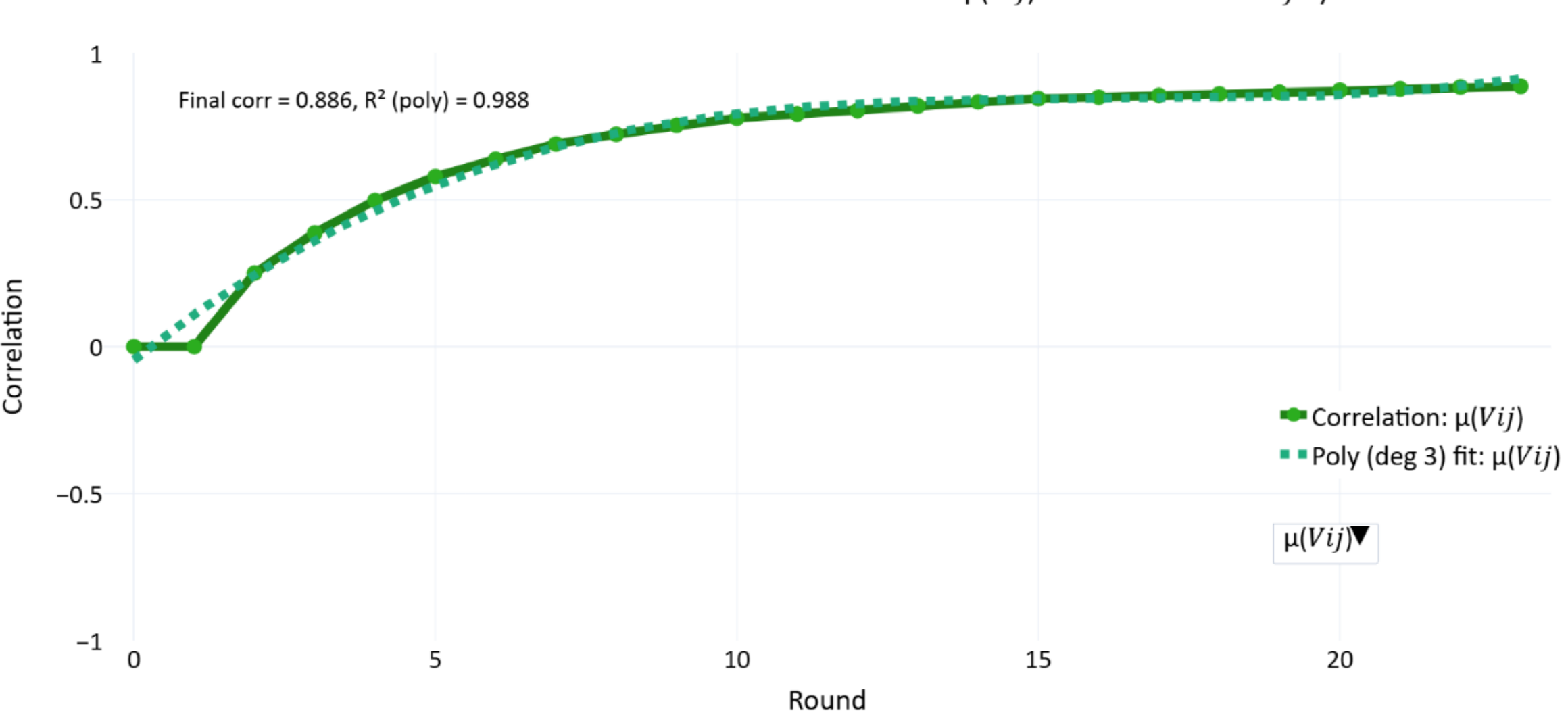

Figure 3: How well synthetic predictors figure out how altruistic their counterparts are across 24 rounds: The solid green line plots the Pearson correlation, by round, between the altruism means fitted for each *predictor* bot ($\mu(V_{ij})$) and the "ground-truth'' altruism programmed into its *chooser* counterpart. The dashed green line gives the least-squares fit. The trajectory verifies that the Utility Bayesian Model progressively revises beliefs toward the chooser's true parameter with continued evidence, demonstrating reliable convergence in iterated interactions.

### *2. Beliefs update faster when the variance in priors is high and temperature is high*

In principle, two things should determine how accurately a predictor homes in on its counterpart: how readable the counterpart's choices are, and how open-minded the predictor starts out. A chooser with a high softmax temperature behaves erratically, making it hard to infer their motives. A predictor that begins with a tight, confident prior is harder to move even when the signal is consistent. A working Bayesian model should show worse recovery under both conditions.

Table 1 tests this. We stratified dyads into terciles on two variables: the chooser's softmax temperature $\tau$ (how noisy their choices were) and the predictor's prior width $\sigma(V_{ij})$ (how uncertain they started), then computed mean correlations between true and inferred chooser altruism over the final three rounds.

The results are as expected. Moving right across the table, from low to high chooser $\tau$, recovery drops in every row, falling from $r \approx 0.93–0.95$ at low $\tau$ to $r \approx 0.80–0.83$ at high $\tau$. When a chooser behaves randomly, there is simply less information to learn from. Moving down the table, from narrow to wide predictor priors, recovery improves, especially at low $\tau$, where it rises from $r = 0.927$ (narrow prior) to $r = 0.953$ (medium prior) with a slight drop at the widest prior ($r = 0.942$), suggesting diminishing returns once the prior is already fairly diffuse. Even the least favorable condition, a narrow prior facing a high-τ chooser, the predictor still learns the chooser's motives, $r = 0.795$.

Table 1: Mean Pearson correlations between true chooser altruism $V_{ij}$ and predictor mean altruism estimates $\mu(V_{ij})$ over the final three rounds, stratified by chooser randomness τ and predictor prior width $\sigma(V_{ij})$. $n$=105 per cell.

| | **τ ∈ [0.50, 0.78]** | **τ ∈ [0.78, 1.26]** | **τ ∈ [1.26, 3.00]** |
|---|---|---|---|
| **$\sigma(V_{ij})$ ∊ [0.50, 0.83]** | 0.927 | 0.858 | 0.795 |
| **$\sigma(V_{ij})$ ∊ [0.76, 1.17]** | 0.953 | 0.913 | 0.813 |
| **$\sigma(V_{ij})$ ∊ [1.56, 4.00]** | 0.942 | 0.925 | 0.843 |

### *3. Beliefs also update faster when conditions are favorable*

We tested speed as well as accuracy. We defined belief update speed as the first round in which a predictor's altruism estimate crossed halfway from its starting position to the chooser's true value, normalized by total rounds. A score of 0 means the predictor arrived immediately and a score of 1 means it never crossed the halfway point within the allotted rounds. Higher values indicate slower updating.

We regressed update speed onto the same two variables for all 945 dyads. Higher chooser τ significantly slowed updating ($\beta = +0.025$, $p = .013$): when a chooser's behavior is noisy, the predictor's beliefs take longer to zero in on the truth. Wider predictor priors sped up updating even more strongly ($\beta = -0.138$, $p < .001$): a predictor who starts out less certain is easier to move when evidence arrives. Overall model fit was modest but highly significant ($R^2 = 0.037$, $F(2, 942) = 18.29$, $p < .001$). The low $R^2$ reflects that crossing-time captures a single binary event across 24 rounds and is inherently noisy; the 3×3 table gives the cleaner view of the same pattern.

These findings confirm that the UBM converges towards the truth and does so faster when the prior is open and the signal is clear.

### *4. The Optimizer Accurately Recovers Predictor Parameters*

We correlated the true predictor altruism with the fitted predictor altruism across all 945 dyads, obtaining $r = 0.891$, shown in Figure 4. This strong correlation indicates that, even with random payoffs and semi-stochastic agent choices, our optimization procedure reliably estimates the predictor's underlying parameters. Hence, subsequent references to "fitted priors" can be considered accurate approximations of the predictor's initial beliefs. Section 7.10 reports the results of a parameter recovery simulation for $k = 1$ through $k = 9$ to test how well parameters are covered in larger models.

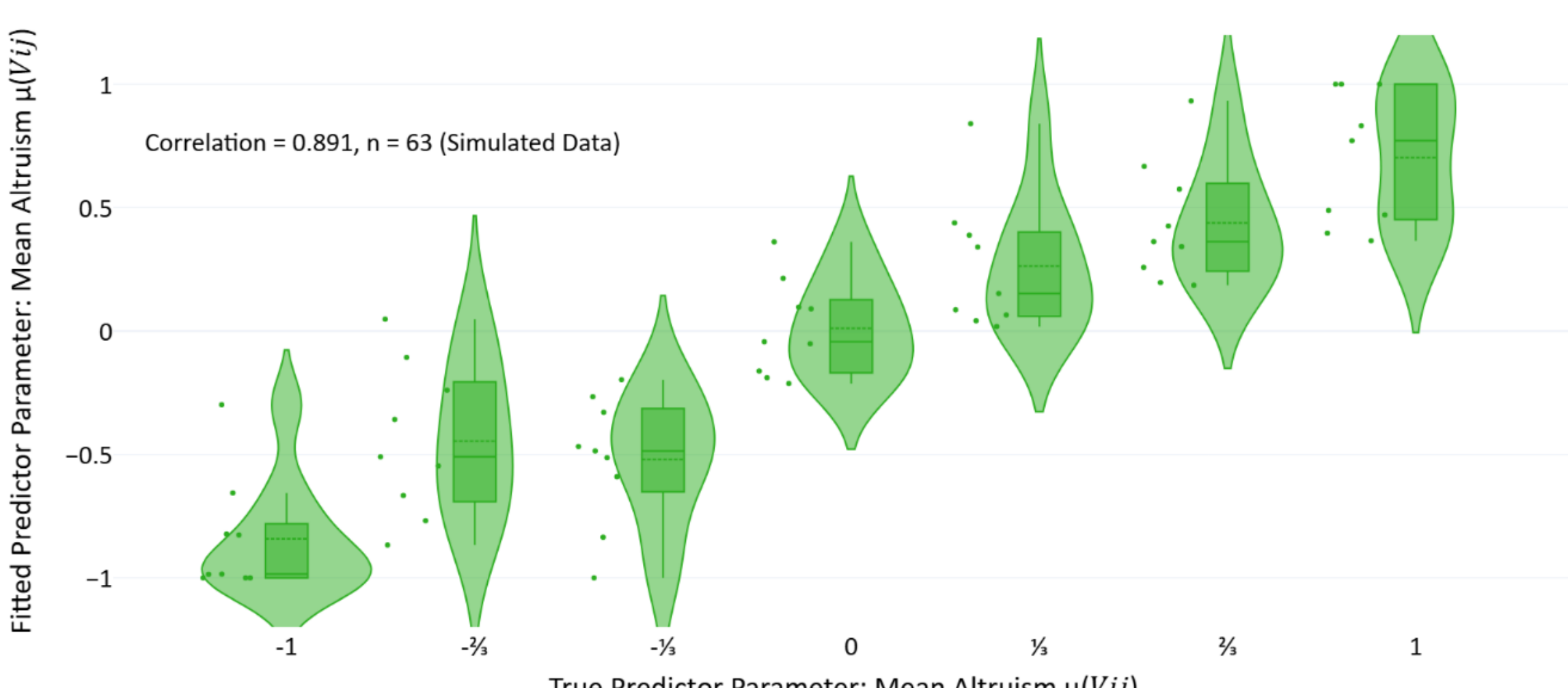

Figure 4: Validation of parameter recovery accuracy for altruism in predictor bots:
Violin plots display distributions of fitted predictor altruism parameter estimates $\mu(V_{ij})$ obtained after the first round of simulated interactions, grouped by the true altruism values programmed into the predictor bots (seven evenly spaced values from -1 to $+1$). Individual data points (green dots, $n = 63$) represent each simulated predictor who was paired with 15 simulated choosers. The strong positive correlation ($r = 0.891$) confirms that the optimization pipeline reliably recovers the programmed parameters, even with stochastic choices and random payoff structures.

These simulations validated that the UBM works as a Bayesian model should and that the optimizer recovers the true parameters. Yet, showing that the model is internally sound does not show that the model describes people. To evaluate the model against actual human behavior, we first describe the human-on-bot experiment that generated the data used in the following analyses.

# 4 Methods – Experiment 1: *Human-on-bot Social Preference Prediction Game Experiment*

## 4.1 Experimental Design Overview

Participants across all experiments engaged in repeated binary dictator games (bDG), designed to investigate how individuals infer and update beliefs about others' social preferences from observed choices. Each trial involved a chooser ("dictator") selecting between two potential payoff allocations (Option A vs. Option B). Retaliation, negotiation, or strategic reciprocity *within* each game was structurally impossible, ensuring that choices reflected pure, outcome-based social preferences.

Participants observed these decisions and predicted choosers' subsequent choices under new payoff scenarios. Predictions were hidden from choosers. Predictions provided a measure of participants' evolving beliefs about the choosers' underlying social preferences.

We conducted two complementary experiments, a human-on-bot experiment designed primarily for initial model validation, and a subsequent human-on-human experiment offering greater ecological validity and enabling detailed parameter estimation.

We refer to Experiment 1, the Social Preference Prediction Game, as the human-on-bot experiment and Experiment 2, the Iterated Binary Dictator Game experiment, as the human-on-human experiment to highlight one of the most salient differences between them: in the human-on-bot experiment human participants play preprogrammed computerized agents and in the human-on-human experiment, human participants play each other. While there are other differences, both experiments involve repeated interactions in binary dictator games where players predict their counterparts' choices over varied payoff structures.

## 4.2 The Social Preference Prediction Game

The **human-on-bot experiment** served primarily to validate our Bayesian cognitive model's ability to accurately describe how participants updated beliefs about chooser preferences. Participants ($n = 83$) interacted repeatedly with four identifiable, computer-programmed chooser-agents ("avatars"), each governed by distinct, fixed preference parameters. These preprogrammed preferences provided an objective reference point by which to assess participants' belief accuracy.

**Key Features**:

- **Controlled Payoff Structures:** Each avatar encountered eight distinct payoff-difference scenarios repeatedly, summarized in Table 2.
- **Prediction Elicitation:** Participants predicted avatar choices after observing past avatar decisions. Predictions were binary (Option A or Option B).
- **Abstract Payoff Presentation:** Payoffs appeared as abstract dots rather than numbers[5].
- **Constant Matching:** Participants repeatedly encountered the same four avatars, identified by consistent icons throughout the experiment.

[5] Using dots instead of numbers allows the method to be more easily adapted to studies involving nonhuman animals and non-literate populations. Payoffs were switched back to numbers on the Morality Game (human-human experiment) to avoid crowding the user-interface.

- **No Time Pressure:** Participants had unlimited time to respond.
- **Two Games Per Trial:** Each trial included an *observation phase*, where participants watch an avatar choose, and a *response phase* where participants predict the same avatar's choice in a subsequent binary dictator game, typically with a different payoff structure. Participants submit predictions only during the response phase, without seeing the avatar's actual choice.

Table 2 summarizes how often each avatar selected each option defined by payoff differences and Figure 5 illustrates the four avatar's preference profiles based on the regions of the payoff spaces containing their preferred options:

Table 2: Frequency of avatar choices across eight controlled payoff-difference scenarios in Experiment 2.
In the human-on-bot experiment, single choice observations were rarely diagnostic of an avatar's preference profile because different avatars often select the same option. This overlap in choices necessitates gradual belief updating.

| Payoff Difference ($\pi_i^A - \pi_i^B$, $\pi_j^A - \pi_j^B$) | Utilitarian | Selfish | Competitive | Masochistic |
|---|---|---|---|---|
| (-2, 4) | 16 | 0 | 0 | 4 |
| (2, 4) | 16 | 16 | 0 | 0 |
| (4, 2) | 16 | 16 | 16 | 0 |
| (4, -2) | 16 | 16 | 16 | 0 |
| (2, -4) | 0 | 16 | 16 | 0 |
| (-2, -4) | 0 | 0 | 16 | 4 |
| (-4, -2) | 0 | 0 | 0 | 4 |
| (-4, 2) | 0 | 0 | 0 | 4 |

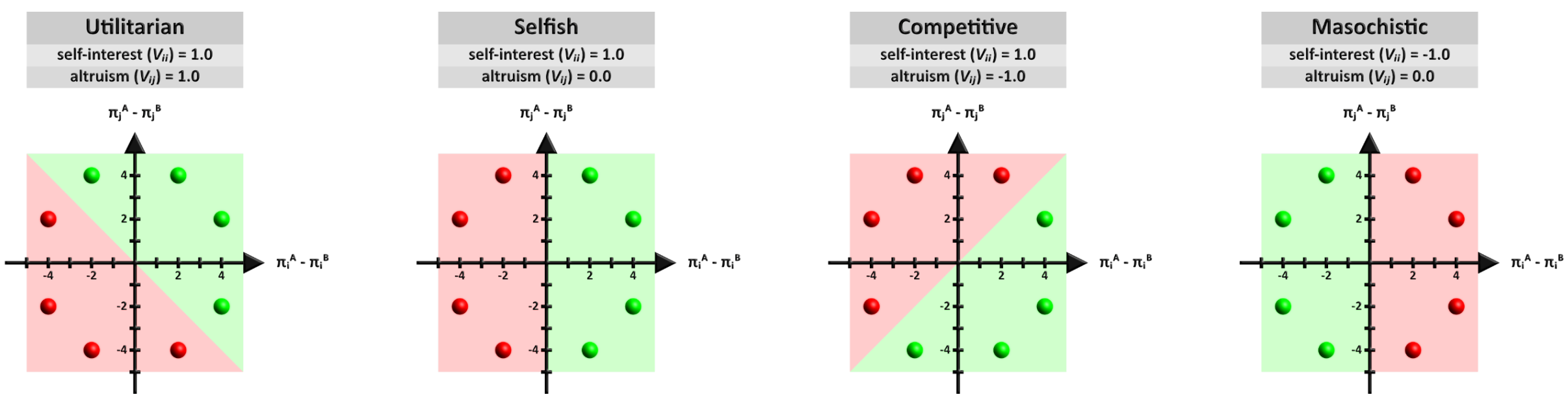

Figure 5: Avatar Preference Profiles:
Participants interacted with four distinct avatars, each characterized by a unique social preference profile: Utilitarian (96 games), Selfish (96 games), Competitive (96 games), and Masochistic (32 games). Avatars' choices were governed by a two-parameter utility function: $U_i(A) = V_{ii}(\pi_i^A - \pi_i^B) + V_{ij}\left(\pi_j^A - \pi_j^B\right)$ where parameters $V_{ii}$ (self-interest) and $V_{ij}$ (altruism) systematically varied across avatars, resulting in distinct decision patterns. Green regions in each payoff matrix indicate chosen outcomes, whereas red regions represent non-chosen outcomes. Visually distinct avatars represent each preference profile across the experiment although the avatar appearance to preference profile mapping was randomized across participants.

## 4.3 Predicted Belief Trajectories Across Avatar Types

With the experimental design established, we begin by illustrating how the model represents belief trajectories across the four avatar types. Figure 6 illustrates the Utility Bayesian Model's belief updating behavior. This visualization traces predictor belief trajectories for each of the four avatar types in the human-on-bot experiment—Utilitarian, Selfish, Competitive, and Masochistic—across repeated observations. Each panel shows a separate instantiation of the model's updating process for different participants, overlaid on a color-coded backdrop indicating regions of the parameter space that would generate the observed choices for each avatar. The green cross in every panel marks the model's initial prior. All belief trajectories radiate from this single prior because we fitted priors to participants across all their counterpart, rather than dyad by dyad. As decisions unfold, the lines (turquoise, blue, pink, and red) depict the evolving mean beliefs along the self-interest $\mu(V_{ii})$ and altruism $\mu(V_{ij})$ dimensions.

Close inspection of these paths highlights several core features of Bayesian inference. In panels where the priors are set with relatively large variance, the trajectories undergo rapid shifts. By contrast, minimal updating occurs when the priors are narrowly concentrated, as low variance expresses high confidence in beliefs. In such cases, the final beliefs end up near the original prior (green cross), indicating that new observations carry less weight for a model that is strongly peaked from the start.

Additionally, the zig-zag shape of the belief paths reflects the random distribution of payoff structures across trials. Had the bots made identical choices over identical payoffs, the belief trajectories would have been linear. Instead, each new observation nudged the mean estimates in slightly different directions because these variable choice observations can express variable willingness to sacrifice self-interest for the interests of others and vice versa. Despite this variability in payoffs, the avatar-bots choose according to stable social preferences, which is why these choice observations consistently nudge beliefs into one broad direction for each avatar. Unlike humans, these preprogrammed bots never make contradictory or uncharacteristic choices and that is why the beliefs never backtrack. For instance, the masochistic avatars always make choices that reduce their payoffs and so the red lines may zig up or down along the altruism axis, but never rightward along the self-interest axis.

Because the lines show only the model's mean parameters for each avatar type, they do not capture the standard deviations or covariances that also evolve across trials. Nonetheless, they demonstrate how repeated Bayesian updates gravitate toward true parameters, in this case $(1, 1)$ for Utilitarian, $(1, 0)$ for Selfish, $(1, -1)$ for Competitive, and $(-1, 0)$ for Masochistic avatars.

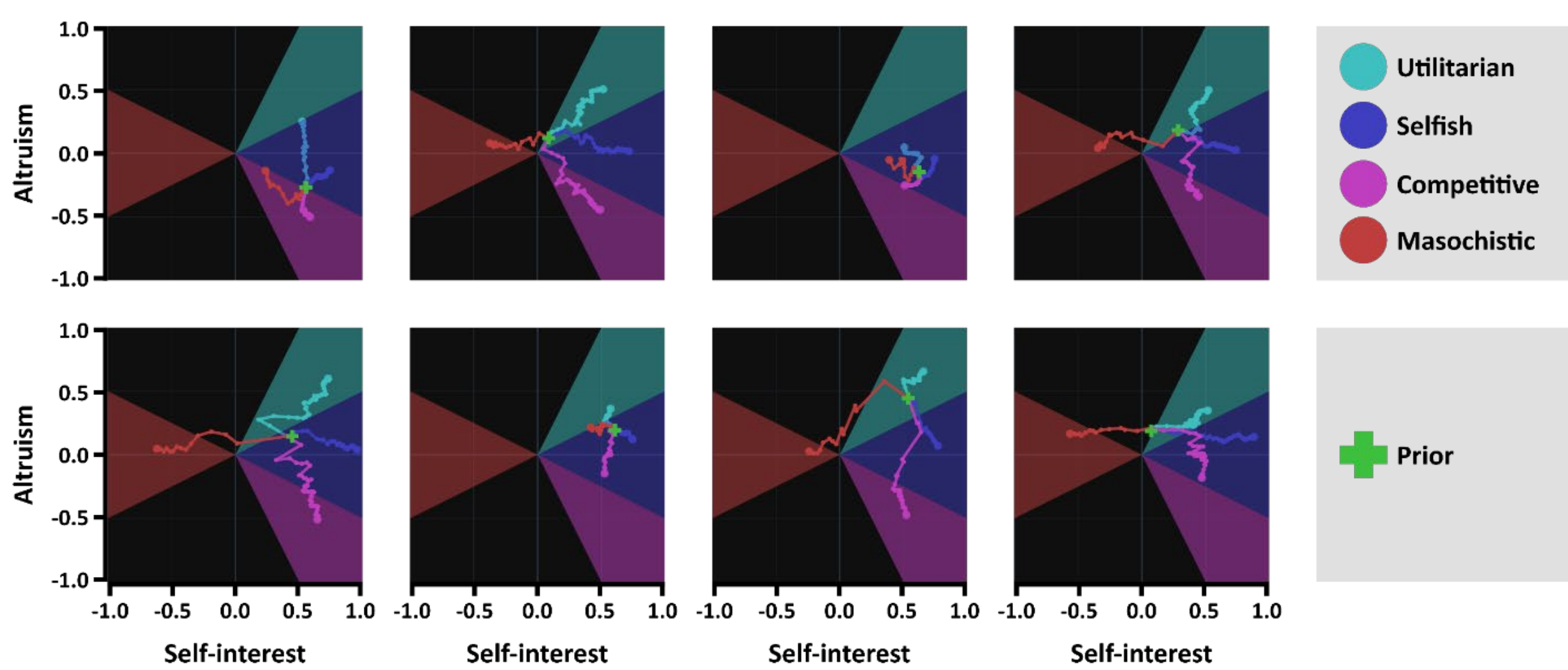

Figure 6: Belief Updates and Convergence Trajectories in 2D Parameter Space:

Each panel shows the evolution of mean beliefs for four avatar types—Utilitarian $(1, 1)$, Selfish $(1, 0)$, Competitive $(1, -1)$, and Masochistic $(-1, 0)$—in the $(V_{ii}, V_{ij})$ plane over repeated observations. The green cross indicates the shared initial prior for all avatars; colored lines trace how the model's mean parameters shift as it observes each avatar's choices. Shaded sectors represent the regions of parameter space compatible with each avatar's behavior. Panels where the prior variance is low (confident priors) exhibit minimal movement, while those with higher variance show large, rapid shifts in response to new data. The zig-zag shapes arise from variability in the payoff structures from trial to trial, illustrating how each observation exerts a distinct directional pull. These convergent trajectories demonstrate the model's capacity for iterative Bayesian updating in a continuous, two-dimensional representation of social preferences.

# 5 Model Validation: Comparing the UBM to Alternative Models

The simulations validated the UBM's internal properties: the optimizer recovers known parameters, posteriors update at rates Bayesian theory predicts, and updates respond appropriately to prior variance. None of that shows the UBM predicts human behavior better than alternative cognitive accounts of the same task. This section uses the human-on-bot data from Experiment 1 to address two nested questions. First, do observers need both memory of prior expectations and observation of new evidence, or does one alone suffice? Second, given that both are needed, do observers represent counterparts as instances of a few discrete preference types or as positions in a continuous parameter space?

The human-on-bot setup is the cleaner test ground for both questions. The bots' preferences were preprogrammed and participants never alternated roles, so each model's score isolates how well it accounts for predictor behavior without confounding it with the predictor's own preferences. We scored every model by total negative log-likelihood (NLL) on participants' predictions across all trials, with lower NLL indicating better fit. Table 3 reports the full comparison.

## 5.1 Memory, Observation, and Their Integration

Bayesian inference is fundamentally about combining stored beliefs with incoming evidence. A direct test of whether observers need this combination is to compare the UBM against models stripped to one component or the other. The Stochastic model assigns equal probability to all choices, using neither memory nor observation, and sets a chance baseline (reported NLL is the mean of $1000$ runs). The No Learning model is identical to the UBM except its priors are effectively frozen by constraining minimum and maximum standard deviation parameters to near zero, preventing meaningful updates from new observations. It uses memory without observation. The No Memory model disregards priors entirely, predicting solely from the immediately preceding choice. It uses observation without memory.

The UBM substantially outperformed all three ($\mathrm{NLL} = 5811.11$ with a $21 \times 21$ grid, versus values exceeding $10{,}000$ for each alternative). Neither memory alone nor observation alone suffices. The result does not uniquely identify Bayesian inference, since other architectures also integrate memory and evidence, but it rules out the simpler alternatives and is consistent with what Bayesian theory predicts about how the two ought to combine.

## 5.2 Typological versus Continuous Representations

That observers integrate memory with evidence leaves a separate question open: at what resolution do they represent the preferences they are updating? Discrete typologies of social preferences have a long history and clear practical appeal. Social Value Orientation classifies individuals as prosocial, individualistic, or competitive (Van Lange, Joireman, Parks, & Dijk, 2013; Balliet, Parks, & Joireman, 2009); public goods experiments identify conditional cooperators and free riders (Fischbacher & Gächter, 2010); folk psychology distinguishes good people from bad ones (Cosmides, Barrett, & Tooby, 2010; Krupp, Debruine, & Barclay, 2008). Each scheme trades resolution for parsimony and each has earned its place in the literature.

But choice data routinely show variation that those categories cannot accommodate. Altruism varies along a continuous spectrum rather than clustering into discrete types (Andreoni & Miller, 2002);

generosity and social concern are heterogeneous across a broad payoff space (Fisman, Kariv, & Markovits, 2007); the SVO Slider captures continuous variability that the traditional categorical SVO fails to register (Murphy, Ackermann, & Handgraaf, 2011). Existing Bayesian work points the same way: observers invert partners' actions to update continuous posteriors over preference weights (Jern, Lucas, & Kemp, 2017; Kleiman-Weiner, Saxe, & Tenenbaum, 2017), revise beliefs about counterparts in graded ways across repeated dictator tasks (Iriberri & Rey-Biel, 2013), and are well-described by hierarchical models that treat both preferences and beliefs as continuous latent variables (Aksoy & Weesie, 2014). None of this prior work, however, compared continuous and discrete representations across enough alternatives to settle the question.

We tested it directly. Because the UBM uses a discrete grid to approximate a continuous parameter space, the comparison is more accurately one of resolution than of continuous-versus-discrete in any strict sense. At low resolution sit typological models that recognize only a handful of preference types. At higher resolution sits the UBM, whose grid covers the same parameter space at much finer granularity. Figure 7 shows how each typological model can be read as a special case of the high-resolution grid in which most cells receive zero prior probability. To clarify, the probabilities assigned to the cells can shift continuously over time. These typological Bayesian models are discrete in the sense of carving up the hypothesis space into segments but the probabilities over those segments can update gradually.

We started with three illustrative typological Bayesian models. The Good versus Evil model categorizes altruism parameters as positive or negative. The Canonical Social Value Orientation model uses the familiar altruistic, selfish, and competitive types. The Perfect Oracle model starts with a prior perfectly calibrated to the experiment's actual distribution of avatar types, giving it the best possible starting guess in principle. Rather than fitting parameters, these models use the fixed priors shown in Figure 7. Despite its theoretically optimal prior, the Perfect Oracle yielded poor predictive performance ($\mathrm{NLL} = 17381.06$), indicating that participants do not rely on perfect population knowledge from the outset. The Good versus Evil and Canonical SVO models also underperformed substantially, each producing NLL values exceeding $14{,}000$. None of these typologies captures the dynamic belief updating participants actually exhibited.

Table 3: Model Comparison by Total Negative Log-Likelihood.

Loss values represent the sum of negative log-likelihoods (NLL) across all participants in the human-on-bot experiment. The "High-Resolution Bayesian" model (leftmost bar) shows the lowest NLL of 5811, outperforming all discrete and non-Bayesian alternatives.

| Model Name | Bayesian | Parameterization | Loss (Total NNL) |
|---|---|---|---|
| **Perfect Oracle** | True | Discrete | 17381.055 |
| **Canonical SVO** | True | Discrete | 16904.924 |
| **Good versus Evil** | True | Discrete | 14639.593 |
| **Stochastic** | False | Neither | 13272.419 |
| **No Memory** | False | Continuous | 12057.898 |
| **No Learning** | False | Continuous | 10683.986 |
| **Utility Bayesian** | True | Continuous | 5811.111 |

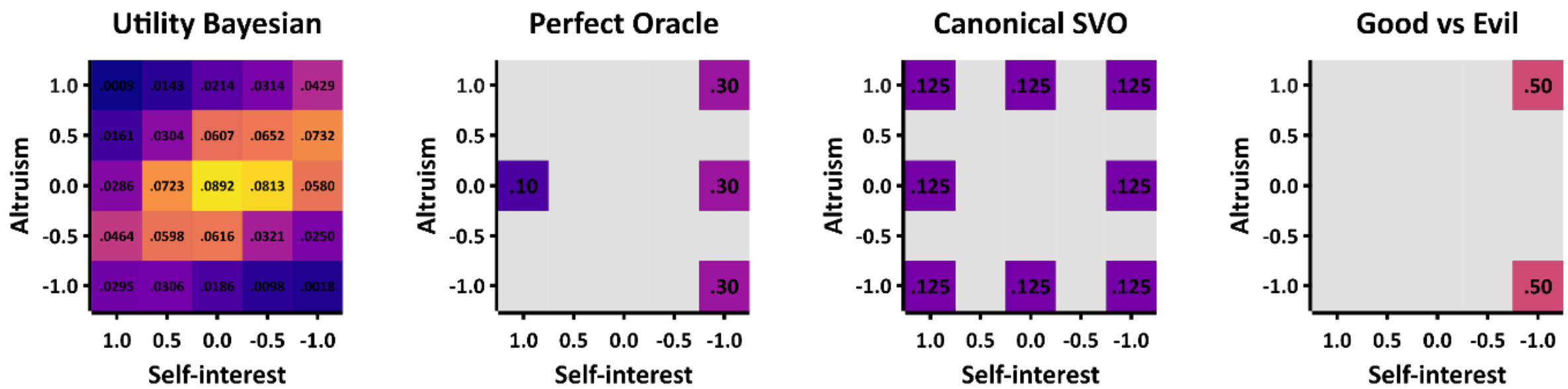

Figure 7: Illustrating Different Resolutions of Social Preference Hypothesis Spaces:

Each square shows a 5×5 parameter grid for $V_{ii}$ and $V_{ij}$ ranging from $-1$ to $+1$. The color in each cell corresponds to the prior probability assigned to that cell. The leftmost square depicts our high-resolution Utility Bayesian Model (here, 5 bins per dimension) with probabilities spread across all 25 cells. Panels B–D display lower-resolution "discrete" models—Perfect Oracle, Canonical SVO, and Good vs. Evil—where only a handful of cells carry nonzero probability (colored), effectively treating the rest as impossible (dark/empty). This demonstrates how simple discrete models can be seen as special cases of the higher-resolution approach, with fewer "bins" (or types) in their hypothesis space. In practice, our "high-resolution" approach uses finer grids (like 21×21) to approximate "continuous" preferences but we also used a 5×5 grid when comparing the UBM to typological models that represent types in somewhere in the 5×5 parameter space so that they UBM would be more comparable to the typological models. Note: The distinction between continuous and discrete Bayesian models pertains to their hypothesis spaces, not how probabilities are updated. Both continuous (UBM) and discrete (typological) models update probabilities continuously (gradually rather than stepwise), but only typological models explicitly forbid certain regions of the parameter space.

## 5.3 Searching the Space of Typological Bayesian Models

Three typological models is itself a minute fraction of what's possible. There are infinitely many ways to partition a continuous hypothesis space into discrete types, and a more demanding test asks whether *any* discrete partitioning could rival the UBM.

The logic behind these additional comparisons was that each typological hypothesis space can be thought of as a lower-resolution special case of our grid-based, high-resolution approach, where only a handful of parameter “bins” receive nonzero prior probability. We discretized the two-dimensional parameter space $(V_{ii}, V_{ij})$ into a 5×5 grid with equally spaced points between −1 and +1, then constructed discrete models by selecting different subsets of grid points as preference types. We evaluated every combination of three types ($\binom{25}{3} = 2{,}300$ models) and every combination of four types ($\binom{25}{4} = 12{,}650$ models), plus a random sample of five-type and six-type models given computational constraints. For each, we optimized the prior distribution over types and computed total NLL on participants' predictions. Then the best-performing model at the population level was refit participant by participant. Individual-level fitting for *every* model would have been computationally intractable. To rule out a resolution confound, we also reran the UBM at 5×5 rather than 21×21 resolution; results were substantively similar, confirming that the UBM's advantage is not only due to higher grid density.

Despite this extensive search, no typological model approached the UBM's predictive accuracy. The best discrete model reached an NLL slightly above $8{,}500$ at the population level, improving marginally to approximately $8{,}470$ with individual-level priors. The UBM's $5811$ remained well out of reach. Across more than $15{,}000$ typological alternatives, representing social preferences as a handful of discrete types failed to capture the inference participants performed.

This combinatorial approach inherently included many models with low face validity or minimal differentiation. But that breadth is itself what makes the result informative: even with the discrete model space sampled this exhaustively, no partition of it competes. To our knowledge, this is the largest discrete-versus-continuous Bayesian model comparison in the social preference literature.[6]

Several scope conditions are worth flagging. First, this is a functional-level description in Marr's (1982) sense: the result characterizes input-output relationships without specifying the algorithm by which observers perform the inference. Humans might approximate continuous Bayesian updating through sampling-based or particle-filter heuristics rather than literal grid computations, and adjudicating that question requires methods this paper doesn't deploy. Second, the analysis is restricted to two parameters (self-interest and altruism). Whether the same pattern holds for envy, guilt, and the other parameters estimated jointly in the parameter distribution results is an open empirical question, though we expect it to. Third, the experimental context provided clear, repeated feedback over many trials. In information-poor or cognitively demanding environments, simpler typological representations might prove more adaptive. The conclusion is not that humans always represent others continuously, but that in this kind of setting, they do.

[6] We acknowledge that we could have tested discrete models with types at irregular intervals within the parameter space, but we expect that any improvements to model fit would be marginal and not worth the computation expense.

Taken together, the analyses to this point provide grounds for trusting the model. Simulations confirmed that the optimizer recovers known parameters and posteriors converge appropriately. Applied to human-on-bot data, the UBM outperformed non-Bayesian alternatives that rely on memory or observation alone, and it outperformed typological Bayesian models that partition preferences into discrete types. What remains is to apply this validated framework to what it was built for: real interactions between real people, whose preferences are unknown, variable, and worth measuring. We begin by describing the human-on-human experiment that generated the data for everything that follows.

# 6 Methods – Experiment 2: Human-on-human Iterated Binary Dictator Game Experiment

The **human-on-human experiment** ($n = 73$) shifts from the control of preprogrammed agents to a relatively ecologically valid context—due to actual human interactions, role switching, and more varied payoff structures. Participants alternated between chooser and predictor roles, repeatedly encountering one another across multiple rounds. Participants played simultaneously in groups of four to eight players.

**Key Features**:

- **Randomized Payoff Structures:** Payoffs for both chooser and recipient positions were fully randomized within $\{1, 2, 3, 4, 5\}$, yielding a comprehensive space of $625$ unique payoff combinations.
- **Numeric Payoff Presentation:** Payoffs were shown as integers rather than abstract dots.
- **Probabilistic Matching:** Participants received explicit information about the probability of re-encountering each counterpart. We acknowledge this manipulation to be transparent, although future research will be needed to test the effect of matching probabilities on cooperation rates.
- **Private Predictions:** As with the human-on-bot experiment, predictions were private, never exposed to counterparts, to eliminate the confound of strategic signaling.
- **Audiovisual Feedback:** Participants received immediate audiovisual confirmation after each submission, represented by colored arrows clearly indicating chosen and predicted options, enhancing participant engagement and data reliability.

- **Time Pressure:** Predictions and choices were required within eight seconds per response, simulating realistic decision-making constraints. Time pressure is necessary in synchronous human-on-human multiplayer experiments to prevent slower participants from slowing down everyone else, which is why the human-on-human experiment involved time pressure but the human-on-bot experiment did not.

Table 4 contrasts procedural details between experiments:

Table 4: Contrasts between human-on-bot and human-on-human experiments highlighting complementary purposes and methodological variations.

| Dimension | Human-on-bot Experiment | Human-on-human Experiment |
|---|---|---|
| **Purpose** | Model validation | Parameter estimation |
| **Participants** | Human predictors; artificial choosers | Humans as predictors and choosers |
| **Payoff Structure** | Controlled (8 unique payoff structures) | Randomized (625 payoff structures) |
| **Matching Procedure** | Constant, identifiable avatars | Probabilistic human matching |
| **Payoff Presentation** | Abstract dots | Numeric integers |
| **Response Types** | Predictions only | Predictions and choices |
| **Data per Participant** | 160 responses | ~120 responses |
| **Time Pressure** | None | 8 seconds per response |
| **Platform URLs** | m3labexperiment.com | moralitygame.com |

## 6.1 Participants, Recruitment, and Procedure

Participants were undergraduate students recruited via standard university participant pools and compensated through course credits. Comprehension checks and attention-validation trials ensured data quality. Experiments were conducted remotely via standardized web platforms. Figure 8 shows the user-interface.[7]

### *Procedural Controls Across Experiments:*

- **Payoff Order Randomization:** All payoff presentations and orderings were randomized and counterbalanced, mitigating order effects.
- **Neutral Instructions:** Instructions were carefully neutral, explicitly avoiding normative or evaluative language.

[7] Although another related experiment explored 234 payoff structures in a separate sub-study within the human-bot paradigm (456 participants), these data are available in the GitHub repo and are reserved for potential supplementary analyses (e.g., Bayesian prior estimation). We clarify this briefly here to avoid confusion and clearly delineate it from our core human-bot and human-human datasets.

- **Performance Feedback:** Participants in the human-on-bot experiment received no accuracy feedback regarding their predictions or counterparts' subsequent choices[8], while predictor-participants in the human-on-human feedback saw their counterparts' choices after submitting predictions.

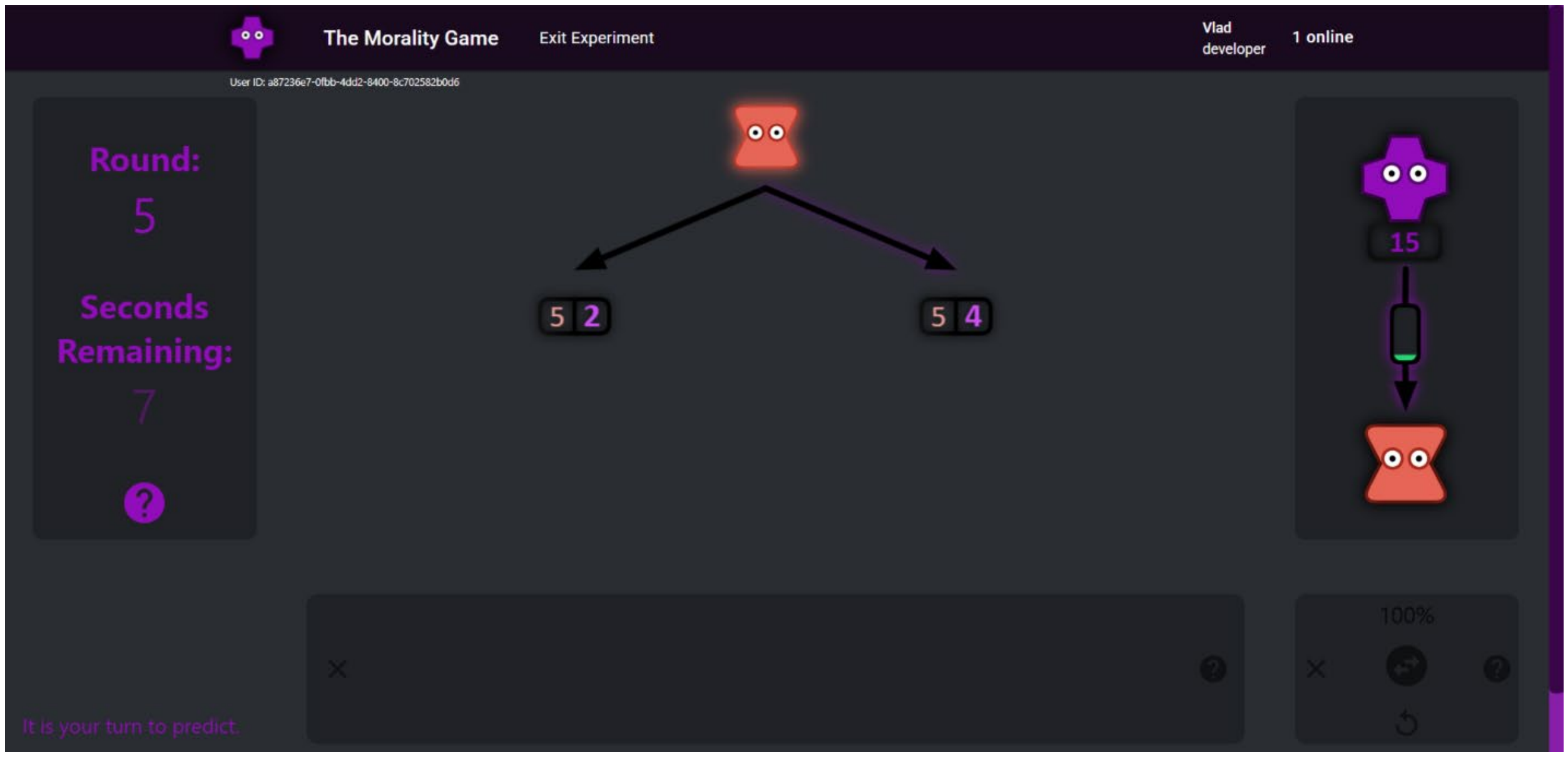

Figure 8: Human-on-human Experiment Stimuli:
This screenshot recreates the user interface from the perspective of a participant in the predictor role during round 5, with 7 seconds remaining to submit a response. The predictor (purple avatar) must predict whether their counterpart (chooser; orange avatar) will select the left or right option. The chooser's payoff is fixed at 5 points for either choice, whereas the predictor's payoff differs between options: 2 payoffs left and 5 payoffs right. Participants select options and submit predictions via mouse, touchscreen, or keyboard. The faint purple shadow around the right arrow indicates a pre-selected prediction. Upon submission (via click, tap, or spacebar), this shadow intensifies, providing visual feedback. At this point, the arrow's interior turns orange if the chooser already chose or will turn orange when the chooser submits their choice. The right-side panel displays the predictor's cumulative payoff (15 points) and includes a thermometer icon, where the green mercury level indicates the relatively low probability of these participants being rematched in subsequent rounds. Participants understand the matching procedure and can hover the cursor over the thermometer icon to discover the exact probability, which is not revealed by default to avoid visual clutter on the user-interface.

[8] This feedback was withheld in the human-bot experiment to avoid imposing the researcher's model upon the participants. Instead, we were interested in investigating the moral schemas that participants would spontaneously apply to the games.

# 7 Information Criterion Analysis: Determining Optimal Utility Structures

Thus far, we have been validating the UBM as a whole using simulations and the human-on-bot data from Experiment 1. Yet, the UBM inverts a forward choice model within its likelihood term and this component—a utility function within a choice probability function—can be swapped out without altering the fundamental way that Bayesian updates occur. The two-dimensional utility function tested already is only one option within a much larger menu of utility functions. This section is about selecting the best utility function on that menu and figuring out what makes it the best, what that means for human moral psychology, if it is even appropriate to select only one utility function to represent human social preferences. Once we find this essential ingredient to the UBM, we will use it in the following section to report the distribution of social preference parameters in our sample of participants.

## 7.1 Motivation and Theoretical Justification of Utility Form Variations

The menu of utility functions has $526$ items. Those items were generated by flipping on or off $17$ binary *utility settings* that determined each equation's functional form. Each setting refers to an algebraic property of these equations that captures a psychological mechanism hypothesized to influence how individuals represent and evaluate social outcomes, such as whether or not exponents should be included to represent nonlinear marginal utility or whether or not distinct parameters should apply to positive versus negative payoffs to represent psychological asymmetries between losses and gains. $526$ is far less than all possible combinations because dependencies between settings rendered some combinations redundant or meaningless (i.e. the setting that when turned on applies distinct exponents to each social preference term rather than one exponent parameter applied uniformly only matters if the setting that includes exponent parameters is active).

***(1) Conditional Welfare Mode***

Conditional welfare mode captures the tendency to weigh payoffs differently depending on relative social standing—whether one's payoff is ahead (advantageous) or behind (disadvantageous) of others, a concept called conditional welfare (Charness & Rabin, 2002).

**Example:**

$$U_i(A) = \begin{cases} V_{ii}\left(\pi_i^A - \pi_i^B\right) + (1 - V_{ii})\left(\pi_j^A - \pi_j^B\right), & if\ \pi_i^A \geq \pi_j^A \\ \lambda_{ii}\left(\pi_i^A - \pi_i^B\right) + (1 - \lambda_{ii})\left(\pi_j^A - \pi_j^B\right), & if\ \pi_i^A < \pi_j^A \end{cases}$$

### *(2) Reference-Dependent Altruism*

Reference-dependent altruism introduces the idea that altruistic preferences differ depending on whether the recipient's payoff is above or below a neutral reference point ($R = 3$), chosen as the middle payoff in $\{1, 2, 3, 4, 5\}$. This aligns with insights from reference-dependent models (Cox, Friedman, & Gjerstad, 2007). This utility setting is a variant of conditional welfare models.

**Example:**

$$U_i(A) = \begin{cases} V_{ii}\left(\pi_i^A - \pi_i^B\right) + (1 - V_{ii})\left(\pi_j^A - \pi_j^B\right), & if\ \pi_i^A \geq R \\ \lambda_{ii}\left(\pi_i^A - \pi_i^B\right) + (1 - \lambda_{ii})\left(\pi_j^A - \pi_j^B\right), & if\ \pi_i^A < R \end{cases}$$

### *(3) Min-Max Preferences (Rawlsian and Leontief Forms)*

Min-max functional forms model fairness considerations that focus on maximizing minimum payoffs, consistent with a desire for equality or fairness as famously advocated by Rawls (1971) and formalized in production economics by Leontief (1941). These forms differ in the placement of preference weights (e.g., self-interest and altruism parameters) relative to the min operator. The Rawlsian form applies a single altruism parameter externally, capturing the principle of maximizing the payoff of the worst-off individual without modulating individual contributions separately. However, the Leontief form incorporates separate parameters internally within the min operator, reflecting an agent-specific weighted consideration of everyone's outcome. Although neither Rawls nor Leontief originally proposed using the max operator, for completeness and symmetry, we tested both min and max variants.

**Rawlsian (external weighting) example:**

$$U_i(A) = V_{ij} \times min\left(\pi_i^A, \pi_j^A\right)^\gamma$$

**Leontief (internal weighting) example:**

$$U_i(A) = min\left(V_{ii}\left(\pi_i^A\right)^\gamma, V_{ij}\left(\pi_j^A\right)^\gamma\right)$$

### *(4) Welfare Efficiency Preferences*

Inspired by Engelmann and Strobel (2004), we included terms expressing preferences over the sum-total payoffs and the minimum payoff for both players. While this Rawlsian drive overlaps with the min-max forms described above, these forms specifically capture the *balance* between a utilitarian drive to maximize aggregate welfare with a Rawlsian drive to maximize the welfare of the least well-off.

**Example – last two terms:**

$$U_i(A) = \left(1 - V_{ii} - V_{ij}\right)\pi_i^A + V_{ii}\left(\frac{\pi_i^A + \pi_j^A}{2}\right) + V_{ij} \times min\left(\pi_i^A, \pi_j^A\right)$$

***(5) Relative Income Penalty***

To ensure that Bolton and Ockenfels's (2000) Equity, Reciprocity, and Competition (ERC) model could be part of our model space, we added their quadratic term that penalizes deviations from equality. This functional form is not redundant with payoff ratio social comparisons terms, described below, only because the term is squared by default.

**Example – second term:**

$$U_i(A) = V_{ii}\left(\pi_i^A\right) - \alpha_{ij}\left(\frac{\pi_i^A}{\pi_i^A + \pi_j^A} - \frac{1}{2}\right)^2$$

## *(6 & 7) Use of Exponential Parameters and Single versus Individual Exponents per Term*

Exponential parameters model nonlinear subjective utilities, aligning with diminishing marginal utility (Gossen, 1854; Marshall, 1920). We explored whether a single exponent across all terms suffices or distinct exponents per term are necessary, reflecting varied nonlinear evaluations for self-interest, altruism, and social comparison.

**Single exponent example:**

$$U_i(A) = V_{ii}\left(\pi_i^A - \pi_i^B\right)^{\gamma} + V_{ij}\left(\pi_j^A - \pi_j^B\right)^{\gamma} - \alpha_{ij}\left(\pi_i^A - \pi_j^A\right)^{\gamma}$$

**Individual exponents example:**

$$U_i(A) = V_{ii}\left(\pi_i^A - \pi_i^B\right)^{\gamma_1} + V_{ij}\left(\pi_j^A - \pi_j^B\right)^{\gamma_2} - \alpha_{ij}\left(\pi_i^A - \pi_j^A\right)^{\gamma_3}$$

***(8) Apply Exponents Directly to Payoffs***

When utility functions have exponents and include payoff differences or payoff ratios, the exponents can be applied to the payoffs before or after these transformations. The two placements encode different kinds of nonlinear sensitivity. Assuming $\gamma < 1$, curving each payoff before comparison makes a fixed payoff difference feel smaller as both payoffs get larger (Ottone & Ponzano, 2005), while curving the difference itself makes sensitivity diminish with the size of the payoff difference (or ratio) per se, even at the same absolute payoff sizes, which in a social comparison term is akin to inequality aversion with diminishing sensitivity (Hill & Neilson, 2007).

**Example of exponents applied inside the payoff difference:**

$$U_i(A) = \cdots \alpha_{ij}\left(\left(\pi_i^A\right)^{\gamma} - \left(\pi_j^A\right)^{\gamma}\right)$$

**Example of exponents applied outside the payoff difference:**

$$U_i(A) = \cdots \alpha_{ij}\left(\pi_i^A - \pi_j^A\right)^\gamma$$

### *(9 & 10) Single Payoffs and Payoff Ratios instead of Differences*

Computing utilities from single payoffs may be cognitively simpler than computing utilities from differences (Camerer C. F., 2003) and models the payoff-utility relation as absolute, not relative. Additionally, computing utilities from payoff ratios, not payoff differences, means that a payoff difference becomes less salient as the payoffs grow larger, aligning with nonlinear marginal utility and proportional fairness perspectives (Bolton & Ockenfels, 2000). Note that the social comparison term always uses payoff differences.

**Single payoff example:**

$$U_i(A) = V_{ii}\left(\pi_i^A\right)^\gamma + V_{ij}\left(\pi_j^A\right)^\gamma$$

**Payoff ratio example:**

$$U_i(A) = V_{ii}\left(\frac{\pi_i^A}{\pi_i^A + \pi_i^B} - \frac{1}{2}\right)^\gamma + V_{ij}\left(\frac{\pi_j^A}{\pi_j^A + \pi_j^B} - \frac{1}{2}\right)^\gamma$$

### *(11) Reference-Dependent Utility*

Inspired by Prospect Theory, reference-dependent utility evaluates outcomes against a neutral reference point, allowing for modeling of asymmetric perceptions of outcomes as gains or losses (Kahneman & Tversky, 1979).

Ideally, identifying reference-dependent preferences requires outcomes that span true gains and losses relative to a meaningful neutral (often zero or a status-quo allocation). However, because our experimental payoffs were restricted to positive values $\{1, 2, 3, 4, 5\}$, we set an artificial reference point at the midpoint of this range ($R$=3). Conceptually, this is akin to rescaling outcomes around an average midpoint ({-2, -1, 0, 1, 2}), but it is fundamentally different from genuine loss framing around zero or another psychologically meaningful reference.

**Example:**

$$U_i(A) = V_{ii}\left(\pi_i^A - R\right)^\gamma + V_{ij}\left(\pi_j^A - R\right)^\gamma$$

### *(12) Negativity Parameters (Loss Aversion)*

Negativity parameters model distinct sensitivity to negative deviations (losses) relative to positive deviations (gains) from the reference point, capturing the well-documented psychological phenomenon of loss aversion (Kahneman & Tversky, 1979).

As noted in the previous section, our lack of negative payoffs prevent participants from experiencing absolute losses. Instead, in this study, negativity parameters only weight relative losses—losses compared to the alternative payoff or the reference point payoff of 3.

**Example:**

$$U_i(A) = V_{ii} \times max\left(\pi_i^A - \pi_i^B, 0\right)^\gamma + \lambda_{ii} \times max\left(\pi_i^B - \pi_i^A, 0\right)^\gamma$$

### *(13) Negativity in Social Comparison (Envy/Guilt)*

Separate negativity parameters within social comparison terms reflect differentiated psychological responses to advantageous versus disadvantageous inequalities, consistent with inequity aversion theories (Fehr & Schmidt, 1999). Here $\alpha_{ij}$ is envy towards disadvantageous inequality and $\beta_{ij}$ is guilt towards advantageous inequality.

**Example:**

$$U_i(A) = \cdots - \alpha_{ij} \times max\left(\pi_j^A - \pi_i^A, 0\right)^\gamma - \beta_{ij} \times max\left(\pi_i^A - \pi_j^A, 0\right)^\gamma$$

### *(14) Tying Self-Interest and Altruism*

This setting forces the weights on the self and other payoffs to sum to one. When it is off, self-interest $V_{ii}$ and altruism $V_{ij}$ are estimated as two free weights, so a person can value their own payoff and the other person's payoff in any combination, including caring about both a great deal or neither very much. When it is on, altruism is replaced by $1 - \mathrm{V}_{ii}$, so the two weights become a single trade-off on a fixed budget. Any payoff given more weight for the self is given that much less weight for the other, which represents the chooser as dividing a constant amount of concern between themselves and their counterpart. This constant-sum weighting is the structure used by the Constant Elasticity of Substitution form of Andreoni and Miller (2002).

**Untied example:**

$$U_i(A) = V_{ii}\left(\pi_i^A\right) + V_{ij}\left(\pi_j^A\right)$$

**Tied example:**

$$U_i(A) = V_{ii}\left(\pi_i^A\right) + (1 - V_{ii})\left(\pi_j^A\right)$$

### *(15) Variable versus Fixed Self-Interest Parameter*

We assessed whether self-interest varies across individuals or is universally fixed at 1, simplifying individual heterogeneity assumptions and following common practice (Andreoni & Miller, 2002).

**Variable example:**

$$U_i(A) = V_{ii}\left(\pi_i^A - \pi_i^B\right)^\gamma + \cdots$$

**Fixed example:**

$$U_i(A) = \left(\pi_i^A - \pi_i^B\right)^\gamma + \cdots$$

***(16 & 17) Inclusion of Altruism and Social Comparison Terms***

Finally, we explicitly tested the necessity of including altruism and social comparison terms, frequently identified as critical to social behavior (Fehr & Schmidt, 1999; Bolton & Ockenfels, 2000).

These systematic explorations resulted in $526$ distinct utility forms. We searched for the best fitting utility form, defined as minimizing the information criterion score, to ensure a robust foundation for our Bayesian cognitive modeling.

## 7.2 Methodological Approach

We select the utility function best suited to the UBM's likelihood term, we tested how all functional forms fit the human-on-human data from Experiment 2 using information criteria (IC), called AIC and BIC, that balance fit-to-data against model complexity, meaning that models are penalized for additional parameters (Akaike, 1973; Schwarz, 1978; Anderson & Burnham, 2002). For each candidate model, we fit one parameter vector per participant, summing negative log-likelihood (NLL) across all participants' choices and predictions.[9]

AIC and BIC were computed as:

$$AIC = 2k + 2 \times NLL$$

$$BIC = k \times ln(n) + 2 \times NLL$$

where $k$ is the number of free parameters and $n$ is the total number of data points.

## 7.3 Model Recovery Simulation

Yet, before interpreting the IC analysis on human data, we verified that this IC pipeline works in the first place. Earlier in this paper, we discussed how we generated synthetic participants with known parameters to test if the optimizer could recover those parameters in a simulation. Analogous to that, we ran the entire IC pipeline on simulated data where each synthetic participant generated responses via a utility function $x$, where the IC pipeline had no direct assess or foreknowledge of the identity of utility function $x$, but instead had to infer it from the data. The synthetic data was stored just like the human raw data so that every line of code involved in the IC analysis would be necessary. Matching the experiment's $N$, we simulated 73 synthetic agents with parameters sampled from the empirical distribution fitted to utility function $x$, including their softmax temperatures. We varied $x$ but focused on the utility function that actually ranked first in the IC model comparison because it the most important model to recover. Each synthetic agent made choices and predictions over up to 150 randomly sampled binary dictator games. We then ran the IC pipeline on subsets of $\{4, 28, 52, 76, 100, 124\}$ games per agent on all 526 functional forms.

[9] Utility functions were fit separately to chooser and predictor decisions. Chooser parameters directly express pure social preferences without incorporating belief updates and thus are inherently non-Bayesian. Predictor parameters, by contrast, reflect beliefs about the chooser's preferences and ideally would employ dynamic Bayesian updating. Here, for computational tractability, we evaluated both chooser and predictor parameters using a static model.

Rather than using only binary pass-fail recovery metric, we developed two continuous distance metrics, that quantify how behaviorally and structurally similar the BIC winner is to the generating model. Average Model Policy Distance (AMPD) measures the average difference in behavior between any two models given identical inputs, while conditional Hamming distance measures the structural differences between two models via the Boolean utility settings used to establish their functional forms.[10] Figure 9 reports the distance between the generating model and the top ranked BIC model and the how the BIC rankings correlate with these distance metrics. However, despite these nuanced metrics, the generating model consistently achieved the top BIC rank in any simulation with more than 76 games per synthetic participant. This demonstrates that the IC analysis can recover the true generating utility function under realistic data conditions and the experiment collected enough data per participant for reliable recovery.

[10] AMPD is a functional distance between two models measured at the level of their outputs: it averages the difference in response probabilities between two models across the complete space of binary payoff structures and a distribution of parameter draws, using the normalized Jensen-Shannon divergence as the per-game distance (bounded within $[0, 1]$, where 0 means identical policies). These are the same choice probabilities that models submit when predicting participant responses and that enter into the NLL objective, so AMPD directly measures how interchangeable two models are from the IC's perspective. AMPD is computed as follows. For each of 250 Monte Carlo iterations, a single full nine-parameter reference vector is drawn uniformly from the parameter bounds. Each model receives only the parameters it uses (its projection of the shared vector), so that $AMPD(M, M) = 0$ by construction. Given this shared parameter draw, each model computes a probability of choosing option A under the softmax decision rule across the complete set of 625 binary payoff structures. The per-game distance is the normalized Jensen-Shannon divergence between the two resulting Bernoulli choice distributions, $JSD(Bern(p) \parallel Bern(q)) \, / \, ln\, 2 \, \in \, [0, 1]$. AMPD is the mean of this quantity over all payoff structures and parameter draws. Conversely, conditional Hamming distance measures structural similarity: it counts the number of Boolean utility flags that differ between two models, restricted to flags that are freely toggled in both, factoring out flags that are forced inactive by parent-child dependencies in the utility parameterization.

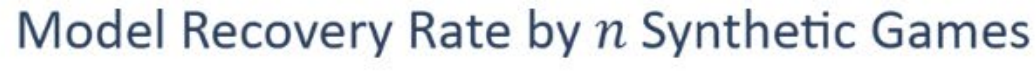

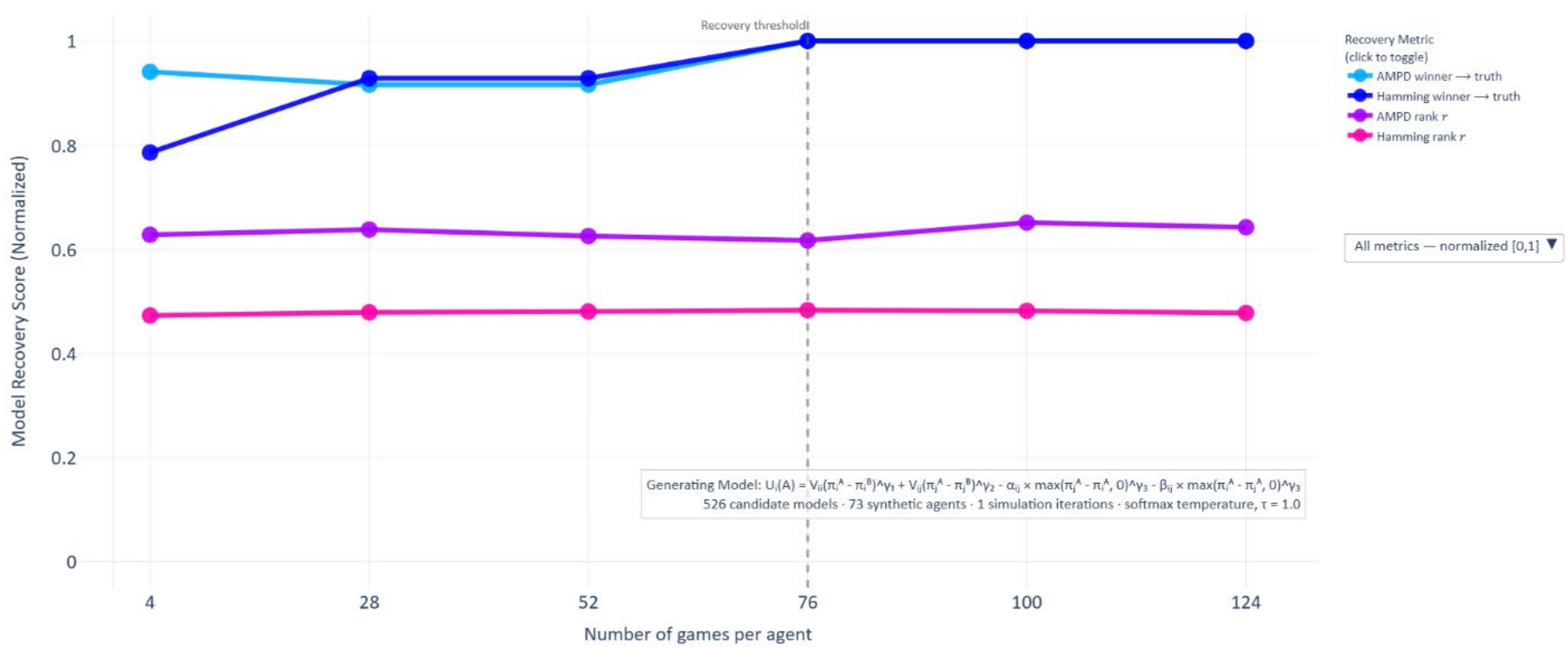

Figure 9: Model recovery as a function of games per agent.

Each synthetic agent generated choices and predictions from the IC's top-ranked utility function using parameters fitted to a real participant. The IC pipeline was run on sub-samples of games across all 526 candidate utility forms. Four recovery metrics are shown as a function of games per agent, where scores are normalized between 0 and 1, where higher scores are better. AMPD winner → truth: the Average Model Policy Distance between the BIC winner and the generating model. Conditional Hamming winner → truth: the number of live utility flags differing between the BIC winner and the generating model. This metric exhibits a ceiling effect: it is one at all data levels, indicating that even in the most data-poor condition the IC selected a model from the same structural class as the generating model. AMPD Spearman $r$ and conditional Hamming Spearman $r$: Spearman rank correlations between BIC scores and AMPD or conditional Hamming distances to the generating model across all 526 candidates, measuring how well the IC globally ordered the candidate set by proximity to the truth. Shown by the dashed vertical line, at 16 games per participant, the simulation consistently ranks the generating model as the BIC winner. The real experiment collected 120 games per participant, well above the recovery threshold.

## 7.4 Pairwise Comparison Analysis

Beyond identifying the single best utility form, it is worth learning general principles about which design features consistently help models fit the data because these principles have implications for moral psychology and can guide how new models are created, such as expanding the UBM to richer settings—e.g., games with more players, strategic reasoning, or belief-based preferences. For instance, our results suggest that models should include independent altruism and social comparison terms to reflect the finding that altruism and social comparison are distinct drivers. Because our set of 526 functional forms were created by flipping 17 Boolean settings, we summarize the average contribution flipping each setting had to BIC when all else was equal.[11]

[11] This pairwise comparison analysis is preferable to running a regression, which would have been misleading due to utility settings that are logically interdependent (creating multicollinearity).

Just understand that these recommendations are generalities that ignore exceptions and interactions between settings. For example, “apply exponents to payoffs” worsened fit on average ($\Delta$BIC $\approx +565$), but when models using payoff ratios, min-max, and conditional-welfare forms are excluded, the same setting *improves* fit (mean $\Delta$BIC $\approx$ -225).

Table 5 summarizes the results of this pairwise comparison analysis:

Table 5: Pairwise BIC contrasts for each Boolean utility setting: $\Delta BIC = BIC(setting\ off) - BIC(setting\ on)$. Negative values indicate that turning the setting on improves fit on average. “Relation” denotes whether the compared pairs are nested (parent/child), siblings (same $k$), or neither. Because some settings are only defined with subsets of the model space, the number of eligible pairs differs by row.

| Utility Setting | Relation | N Pairs | Mean ΔBIC | Med ΔBIC | Effect on Fit |
|---|---|---|---|---|---|
| **Use Exponential Parameters** | Nested | 125 | -1303.21 | -1350.56 | Strongly improves |
| **Envy/Guilt in Social Comparison** | Nested | 92 | -473.49 | -263.41 | Strongly improves |
| **Include Altruism Term** | Nested | 178 | -448.17 | -304.68 | Strongly improves |
| **Include Social Comparison** | Nested | 110 | -202.69 | -204.07 | Moderately improves |
| **General Negativity Parameters** | Nested | 99 | -104.22 | 37.32 | Moderately improves |
| **Single Exponential Parameter** | Nested | 163 | 226.83 | 113.94 | Moderately worsens |
| **Tied Self-Interest and Altruism** | Nested | 96 | 355.15 | 240.93 | Moderately worsens |
| **Fix Self-Interest Parameter** | Nested | 111 | 483.67 | -61.75 | Moderately worsens |
| **Single Payoffs (not differences)** | Siblings | 74 | -515.17 | -183.64 | Strongly improves |
| **Reference-Dependent Altruism** | Siblings | 32 | 15.58 | 7.28 | Neutral effect |
| **Reference-Dependent Utility** | Siblings | 170 | 441.4 | 479.63 | Moderately worsens |
| **Apply Exponents to Payoffs** | Siblings | 153 | 730.77 | 1038.13 | Strongly worsens |
| **Payoff Ratios (not differences)** | Siblings | 132 | 893.56 | 827.97 | Strongly worsens |
| **Relative Income Penalty** | None | 24 | -537.75 | -284.68 | Strongly improves |
| **Conditional Welfare Form** | None | 32 | -519.07 | -335.54 | Strongly improves |
| **Welfare Efficiency Term** | None | 5 | -377.77 | -329.63 | Moderately improves |
| **Min-Max Rawlsian/Leontief** | None | 72 | 630.03 | 611.37 | Strongly worsens |

**Provisional Principles and Recommendations:**

**Strongly recommended:**

- Include exponential parameters and an altruism term.
- Exclude exponents directly on payoffs, payoff ratios, and min-max forms.

**Recommended:**

- Include a social comparison term, ideally split into sides for disadvantageous (envy) and advantageous (guilt) inequality.

- Use single payoffs instead of payoff differences. Avoid reference points based on the midpoint payoff ($R = 3$).
- **Caution advised:** The neutral or weak effects of models that include negativity parameters or use $R = 3$ as a reference point likely reflect our payoff space lacking true losses. These forms might behave differently when outcomes can fall below zero.

## 7.5 Optimal Utility Functions and Nesting

Our top-performing utility functions were determined by ranking BIC scores, as BIC penalizes model complexity more strongly than AIC. Figure 10 shows the ΔBIC scores for all 526 models and Table 6 shows the top-performing models by $k$ parameters.

The optimal utility function is:

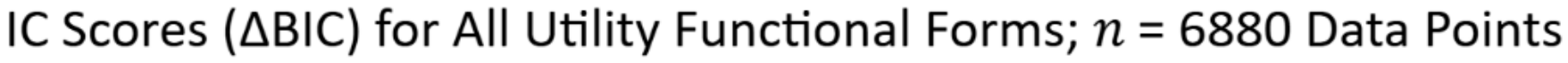

$$U_i(A) = V_{ii}\left(\pi_i^A - \pi_i^B\right)^{\gamma_1} + V_{ij}\left(\pi_j^A - \pi_j^B\right)^{\gamma_2} - \alpha_{ij} \times max\left(\pi_j^A - \pi_i^A, 0\right)^{\gamma_3} - \beta_{ij} \times max\left(\pi_i^A - \pi_j^A, 0\right)^{\gamma_3}$$

IC Scores (ΔBIC) for All Utility Functional Forms; $n$ = 6880 Data Points

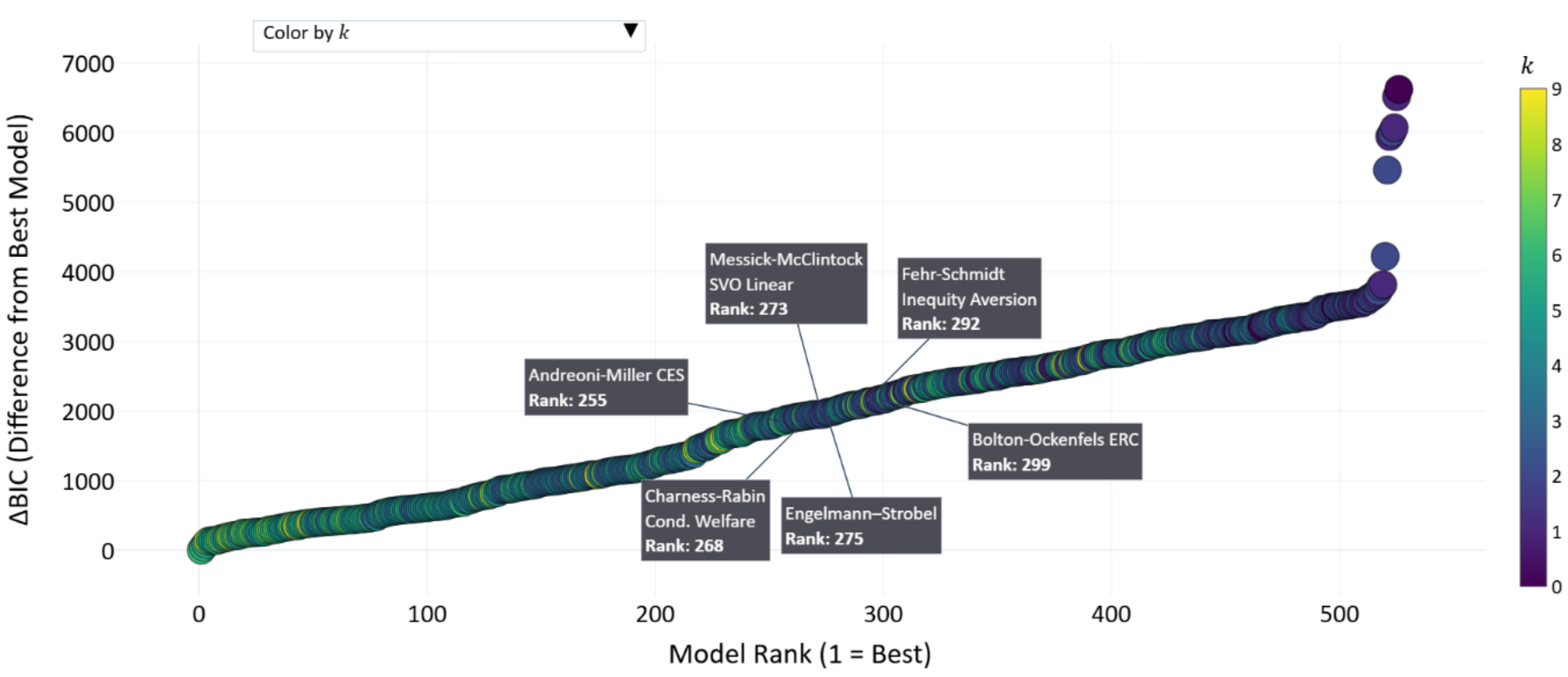

Figure 10: All 526 ΔBIC Scores Per Functional Form from Least to Greatest:
This plot shows ΔBIC for all 526 models from least to greatest, where ΔBIC is the difference between the minimum BIC score and the model's BIC score. Each point is one model. Models are colored by number of parameters ($k$). Annotations indicate the BIC scores and ranks of the canonical outcome-based utility functions that have historically dominated the literature: Messick-McClintock (1968), Fehr-Schmidt (1999), Bolton-Ockenfels (2000), Andreoni-Miller (2002), Charness-Rabin (2002), and Engelmann-Strobel (2004).

Table 6: Information Criterion Scores and Functional Forms for Top-Performing Utility Functions by Parameter Count. This table presents the nine best-fitting utility functions from our set of $526$ candidate models, organized by the number of free parameters ($k$). Utility functions are ranked within each $k$ on the basis of their Bayesian Information Criterion (BIC), a metric that balances model fit (measured by summed negative log-likelihood) against complexity. The ΔBIC column reports the BIC difference from the best-fitting model across all $526$ (the $k = 5$ model with $\Delta\mathrm{BIC} = 0$). ΔBIC values greater than $10$ indicate significantly worse model performance.

Binary indicators ($0/1$) in columns S1–S17 denote the inclusion of 17 theoretically motivated utility settings, such as reference dependence, exponent structure, and social comparison terms (see legend below). While the top-performing model includes seven parameters—self-interest, altruism, envy, guilt, and exponents for each term—the broader comparison highlights which features tend to improve fit as model complexity increases.

The equations in the lower panel correspond to the models listed in the upper panel, matched by their $k$ values. Due to formatting constraints, the top and bottom halves are separated but aligned row-wise by $k$.

While we recommend using the top performing model ($k = 7$), this table allows modelers to select the best utility functions based on the number of parameters they prefer to use, and it reveals general insights into the types of models that tend to fit the data best.

| $k$ | Loss | AIC | BIC | ΔBIC | $S_1$ | $S_2$ | $S_3$ | $S_4$ | $S_5$ | $S_6$ | $S_7$ | $S_8$ | $S_9$ | $S_{10}$ | $S_{11}$ | $S_{12}$ | $S_{13}$ | $S_{14}$ | $S_{15}$ | $S_{16}$ | $S_{17}$ |
|---|---|---|---|---|---|---|---|---|---|---|---|---|---|---|---|---|---|---|---|---|---|
| 1 | 3274.05 | 6550.10 | 6556.94 | 1443.94 | 0 | 0 | 0 | 0 | 0 | 1 | 1 | 1 | 0 | 0 | 0 | 0 | 0 | 0 | 1 | 0 | 0 |
| 2 | 2932.19 | 5868.39 | 5882.06 | 769.06 | 0 | 0 | 0 | 0 | 0 | 1 | 1 | 1 | 0 | 0 | 0 | 0 | 0 | 0 | 1 | 0 | 1 |
| 3 | 2764.46 | 5534.92 | 5555.43 | 442.43 | 0 | 0 | 0 | 0 | 0 | 1 | 1 | 1 | 0 | 0 | 1 | 0 | 1 | 0 | 1 | 1 | 0 |
| 4 | 2657.34 | 5322.68 | 5350.02 | 237.02 | 0 | 0 | 0 | 0 | 0 | 1 | 1 | 1 | 0 | 0 | 0 | 0 | 0 | 0 | 0 | 1 | 1 |
| 5 | 2628.47 | 5266.94 | 5301.12 | 188.13 | 0 | 0 | 0 | 0 | 0 | 1 | 1 | 0 | 1 | 0 | 0 | 0 | 1 | 0 | 0 | 1 | 1 |
| 6 | 2597.37 | 5206.74 | 5247.75 | 134.76 | 0 | 0 | 0 | 0 | 0 | 1 | 0 | 0 | 0 | 0 | 0 | 0 | 1 | 0 | 1 | 1 | 1 |
| 7 | 2525.57 | 5065.14 | 5113.00 | 0.00 | 0 | 0 | 0 | 0 | 0 | 1 | 0 | 0 | 0 | 0 | 0 | 0 | 1 | 0 | 0 | 1 | 1 |
| 8 | 2534.47 | 5084.93 | 5139.62 | 26.63 | 0 | 0 | 0 | 0 | 0 | 1 | 0 | 0 | 0 | 0 | 0 | 1 | 1 | 0 | 1 | 1 | 1 |
| 9 | 2525.24 | 5068.48 | 5130.01 | 17.01 | 0 | 0 | 0 | 0 | 0 | 1 | 0 | 0 | 0 | 0 | 0 | 1 | 1 | 0 | 0 | 1 | 1 |

Column Names:

$k$: Number of parameters.

loss: Summed Negative Log-Likelihood across all participant responses.

AIC/BIC: Information Criteria balancing fit and complexity (lower values indicate superior models).

ΔBIC: Difference in BIC relative to the best (lowest) scoring model ($\Delta\mathrm{BIC} > 10$ considered decisive).

S1 through S17: Utility Settings ($1 = \mathrm{true}$, $0 = \mathrm{false}$) corresponding to the enumerated settings:

1. Conditional welfare mode
2. Reference-dependent altruism
3. Rawlsian and Leontief min-max
4. Total welfare efficiency term
5. Relative income penalty term
6. Use of exponential parameters
7. Single versus individual exponents
8. Apply exponents directly to payoffs
9. Single payoffs (instead of differences)
10. Payoff ratios (instead of differences)
11. Reference-dependent utility ($R$ = 3)
12. Negativity parameters (loss aversion)
13. Negativity within social comparison
14. Tie self-interest and altruism terms
15. Variable versus fixed self-interest
16. Inclusion of social comparison
17. Inclusion of altruism

| $k$ | Equation |
|---|---|
| 1 | $U_i(A) = (\pi_i^A)^\gamma - (\pi_i^B)^\gamma$ |
| 2 | $U_i(A) = (\pi_i^A)^\gamma - (\pi_i^B)^\gamma + V_{ij}((\pi_j^A)^\gamma - (\pi_j^B)^\gamma)$ |
| 3 | $U_i(A) = (\pi_i^A)^\gamma - (R)^\gamma - \alpha_{ij} \times max((\pi_j^A)^\gamma - (\pi_i^A)^\gamma, 0) - \beta_{ij} \times max((\pi_i^A)^\gamma - (\pi_j^A)^\gamma, 0)$ |
| 4 | $U_i(A) = V_{ii}((\pi_i^A)^\gamma - (\pi_i^B)^\gamma) + V_{ij}((\pi_j^A)^\gamma - (\pi_j^B)^\gamma) - \alpha_{ij}[max((\pi_j^A)^\gamma - (\pi_i^A)^\gamma, 0) + max((\pi_i^A)^\gamma - (\pi_j^A)^\gamma, 0)]$ |
| 5 | $U_i(A) = V_{ii}(\pi_i^A)^\gamma + V_{ij}(\pi_j^A)^\gamma - \alpha_{ij} \times max(\pi_j^A - \pi_i^A, 0)^\gamma - \beta_{ij} \times max(\pi_i^A - \pi_j^A, 0)^\gamma$ |
| 6 | $U_i(A) = (\pi_i^A - \pi_i^B)^{\gamma_1} + V_{ij}(\pi_j^A - \pi_j^B)^{\gamma_2} - \alpha_{ij} \times max(\pi_j^A - \pi_i^A, 0)^{\gamma_3} - \beta_{ij} \times max(\pi_i^A - \pi_j^A, 0)^{\gamma_3}$ |
| 7 | $U_i(A) = V_{ii}(\pi_i^A - \pi_i^B)^{\gamma_1} + V_{ij}(\pi_j^A - \pi_j^B)^{\gamma_2} - \alpha_{ij} \times max(\pi_j^A - \pi_i^A, 0)^{\gamma_3} - \beta_{ij} \times max(\pi_i^A - \pi_j^A, 0)^{\gamma_3}$ |
| 8 | $U_i(A) = max(\pi_i^A - \pi_i^B, 0)^{\gamma_1} + \lambda_{ii} \times max(\pi_i^B - \pi_i^A, 0)^{\gamma_1} + V_{ij} \times max(\pi_j^A - \pi_j^B, 0)^{\gamma_2} + \lambda_{ij} \times max(\pi_j^B - \pi_j^A, 0)^{\gamma_2} - \alpha_{ij} \times max(\pi_j^A - \pi_i^A, 0)^{\gamma_3} - \beta_{ij} \times max(\pi_i^A - \pi_j^A, 0)^{\gamma_3}$ |
| 9 | $U_i(A) = V_{ii} \times max(\pi_i^A - \pi_i^B, 0)^{\gamma_1} + \Lambda_{ii} \times max(\pi_i^B - \pi_i^A, 0)^{\gamma_1} + V_{ij} \times max(\pi_j^A - \pi_j^B, 0)^{\gamma_2} + \lambda_{ij} \times max(\pi_j^B - \pi_j^A, 0)^{\gamma_2} - \alpha_{ij} \times max(\pi_j^A - \pi_i^A, 0)^{\gamma_3} - \beta_{ij} \times max(\pi_i^A - \pi_j^A, 0)^{\gamma_3}$ |

Where:

- $U_i(A)$ is the utility for player $i$ in option $A$.
- $\pi_j^B$ is the payoff for player $j$ in option $B$, where $\pi \in [1, 5]$.
- $R$ is a reference point of 3—the middle payoff in $\{1, 2, 3, 4, 5\}$.
- $V_{ii}$ is the self-interest parameter—the amount that player $i$ values player $i$.
- $V_{ij}$ is the altruism parameter—the amount that player $i$ values player $j$.
- $\alpha_{ij}$ is the social comparison parameter activated by disadvantageous inequality.
- $\beta_{ij}$ is the social comparison parameter activated by advantageous inequality.
- $\lambda_{ii}$ and $\lambda_{ij}$ are the negativity versions of the self-interest and altruism parameters.
- $\gamma$ is the exponent parameter, which may or may not be unique to each term.

## 7.6 Robustness Analysis of Model Selection

To ensure that our information criterion (IC) results were robust to the inherently stochastic nature of the optimization process, we conducted a robustness analysis based on repeated iterations of the fitting process. Each iteration attempted to discover better (lower) negative log-likelihood (NLL) losses for each candidate utility function, starting from different randomized initial parameter estimates.

We tracked two complementary metrics:

1. **Sum of Δ Minimum Loss**: This metric quantifies the incremental improvement in model fits across iterations. For each model, we recorded the minimum loss discovered up to each iteration. After every iteration, we calculated how much lower the new minimum loss (if found) was compared to the best minimum loss identified in all previous iterations. Summing these differences across all models yielded a total improvement measure for that iteration. We chose this as our primary robustness metric because diminishing returns in this metric reflect when further iterations no longer substantively improve the stability of our results.

2. **Ordinal Rank Changes**: As a complementary metric, we examined how models' ranks by BIC changed between iterations. Trust in the results of our IC analysis depends on the stability of BIC rank ordering of models across repeated analyses. We measured this by the sum of the absolute differences in rank for all models between consecutive iterations. The results from this metric corroborate those observed with the sum of Δ minimum loss.

We terminated the analysis once the sum of Δ minimum loss improvements fell below a pre-specified threshold (e.g., 10) for two consecutive rounds because this indicates convergence. The sum of Δ minimum loss decreased substantially within the first few iterations and remained below our threshold by iteration 11, triggering the stopping criterion. This convergence provides confidence that our IC-based selection of the optimal utility function is stable and reproducible. Figure 11 illustrates these robustness metrics across iterations.

Figure 11: Robustness Analysis of Information Criterion (IC) Results:
The left panel illustrates the sum of Δ minimum loss (negative log-likelihood) discovered across iterations of model fitting, indicating incremental improvements in model fit. Iterations were stopped once the sum of incremental improvements approached minimal practical significance (indicating stability). The right panel shows the sum of ordinal rank changes (by Bayesian Information Criterion; BIC) between consecutive iterations, corroborating the stability suggested by the diminishing returns observed in the sum of Δ minimum loss metric.

## 7.7 Nested Models

Many of our utility functions are nested: a child model is a special case of a more flexible parent model, obtained by fixing one or more special parameters, such as setting added exponents to one. For example, Table 7 below presents a small family of nested models. Under identical parameter bounds, a correctly optimized parent must achieve a negative log-likelihood (NLL) that is less than or equal to its child. Any case where a fitted parent's best NLL is worse than its child's is a nesting violation. Our goal is to prevent such violations that arise from optimization artifacts.

Table 7: Example nested family:
Each utility function is a parent of the model in the row below it. $k$ = number of free parameters. The special parameters column shows the values these parameters must be set to for the parent to function as the child model in the row below.

| $k$ | Utility Function | Special Parameters |
|---|---|---|
| **4** | $U_i(A) = V_{ii}\left(\pi_i^A\right)^{\gamma_1} + V_{ij}\left(\pi_j^A\right)^{\gamma_2}$ | $\gamma_1 = \gamma_2$ |
| **3** | $U_i(A) = V_{ii}\left(\pi_i^A\right)^{\gamma} + V_{ij}\left(\pi_j^A\right)^{\gamma}$ | $\gamma = 1$ |
| **2** | $U_i(A) = V_{ii}\left(\pi_i^A\right) + V_{ij}\left(\pi_j^A\right)$ | $V_{ij} = 0$ |
| **1** | $U_i(A) = V_{ii}\left(\pi_i^A\right)$ | $V_{ii} = 1$ |
| **0** | $U_i(A) = \pi_i^A$ | |

### 7.7.1 Nesting Network and Verification

We formalized a nesting network over all 526 models:

**Parent-child relationship:** A model is a parent of a child if (1) the parent contains all the parameter of the child, plus at least one additional parameter, (2) exactly one Boolean utility setting differs, and (3) at the child-anchor (the special parameter values), the parent and child produce identical utilities on all possible 625 ($5^4$) payoff structures in our dataset. We verified these criteria for all child-parent pairs.

**Sibling relationship:** Two models are siblings if they have the same number of free parameters and differ in exactly one Boolean utility setting (no strict nesting).

Table 8 below, shows the type of relationship produced between a pair of models when only one boolean utility setting differs between them.

Table 8: Boolean Utility Setting Flips and the Relationships They Produce.

| Utility Setting Flip | Relationship |
|---|---|
| Place exponents on each term | child-parent |
| Use the same exponent for all terms | child-parent |
| Include negativity (loss aversion) parameters | child-parent |
| Include guilt in the social comparison term | child-parent |
| Fix self-interest parameter to one | child-parent |
| Include an altruism term | child-parent |
| Include a social comparison term | child-parent |
| Use single payoffs, not payoff differences | sibling-sibling |
| Compute payoff ratios, not payoff differences | sibling-sibling |
| Apply exponents directly to payoffs | sibling-sibling |
| Reference-dependent utility ($R$=3) | sibling-sibling |
| Reference-dependent altruism | sibling-sibling |
| Conditional welfare form | None |
| Min-max Rawlsian-Leontief | None |
| Welfare efficiency term | None |
| Relative income penalty | None |

### 7.7.2 Parent-Fair L2 Regularization

Many of our models are quasi–scale-invariant in their weights (e.g., $V_{ii}$, $V_{ij}$,…): multiplying all social preference weights by a constant can leave choice predictions nearly unchanged for a fixed softmax temperature. This creates flat directions in the loss surface that can cause gradient-based local optimizers (like L-BFGS-B) to stop early or wander.

A small L2 penalty adds gentle curvature in these flat directions, stabilizing the search and making local refinement reliable. We keep the weight small. It is used only to help the optimizer, not to change which model wins.

A naïve L2 penalty can bias against parents (they have more free parameters), which could itself produce nesting violations. We therefore design the penalty to be parent-fair: When a parent is set to the child-anchor (i.e., the special parameters that collapse it to the child) and the shared parameters match, the parent and child incur the same penalty. So parents are penalized more only when they deviate away from the anchor that makes them behave like their children.

$$Loss = \sum NLL + \sum \rho \begin{cases} (w_+)^2 \\ \left(\frac{|w_+| + |w_-|}{2}\right)^2 \\ \left(\frac{\gamma_1 + \gamma_2 + \cdots + \gamma_n}{n} - 1\right)^2 \end{cases}$$

where:

- ρ is the penalty weight of $0.005$,
- $w_+$ are social preference weights: $\mu(V_{ij})$ or $\mu(\alpha_{ij})$,
- $w_+$ are positivity weights, while $w_-$ are negativity weights: $\mu(\lambda_{ii})$, $\mu(\lambda_{ij})$, or $\mu(\beta_{ij})$,
- γ are exponent parameters on one of the three social preference terms,
- Note: self-interest is penalized as $(\mu(V_{ii}) - 1)^2$.

Equation 1: Penalty Function:

We add a L2 penalty term to stabilize optimization. To not contribute to nesting violations, the penalty is identical for child and parent models if both models use identical parameters and the additional parent parameter(s) are set to the special points that make the parent behave like its child. The social preference weights for altruism and envy are only penalized when they are non-zero, while the self-interest weight is only penalized when it deviates from one.[12] When positivity parameters are paired with negativity parameters: $\mu(V_{ii})$ with $\mu(\lambda_{ii})$, $\mu(V_{ij})$ with $\mu(\lambda_{ij})$, and $\mu(\alpha_{ij})$ with $\mu(\beta_{ij})$, we penalize the mean of the absolute values of each pair because if a parent model differs from its child only by the exclusion versus inclusion of these negativity parameters, the child and parent are penalized the same when the positivity and negativity parameters are the same. Exponents are penalized as the mean of all exponents minus one because exponents set to one make the parent equivalent to a child without exponents. Only parameter means are penalized, never standard deviation parameters.

### 7.7.3 Optimization Schedule

To avoid nesting violations, we first fit child models and pass their best fitting parameters to their parents as an initial point for optimization, called a warm start. This helps move the optimizer out of local minima in high-dimensional parent parameter spaces. However, warm starts can create path dependence, where a child stuck in a local minima could propagate this suboptimal initialization up the model hierarchy. To mitigate this problem, we fit the models iteratively, using the same iterations within the robustness analysis, where the system progresses from exploration to exploitation. The first four iterations, all models are optimized with simulated annealing, followed by L-BGFS-B, without the help of warm starts. In the next four iterations, children provide their best-fitting parameters to their parents as warm starts but the parents select the child to take the parameters from at random, meaning that parents do not always select their best-fitting child. In the final iterations, parents always exploit the best-fitting parameters of their best-fitting child model.[13]

[12] Self-interest is penalized as $(\mu(V_{ii}) - 1)^2$ because the relevant Boolean utility setting determines if the utility function fixes the self-interest parameter to one, meaning that a child model could look like $U_i(A) = \pi_i^A + \cdots$ while the parent could look like $U_i(A) = V_{ii}(\pi_i^A) + \cdots$. By contrast, the other relevant utility settings determine if altruism terms or social comparison terms should be included at all, which is why the penalty for altruism and social comparison should bias those parameters towards zero, not one.

All model ranking (AIC/BIC) is computed from the best raw NLL attained by each model across all passes. The schedule above eliminated all but trivial (floating-point) nesting violations in practice.

Overall, these analyses justify our selection of the optimal seven-parameter utility form as a model for observers' outcome-based social preferences.

## 7.8 Does the population need more than one utility architecture?

The IC analysis identifies the single best-fitting utility architecture for the population, but population-level model selection leaves one further question open: are all participants well described by that single architecture, or do meaningful subgroups need different ones? This matters because the parameter distributions reported in the next section presume that individuals differ from one another primarily in parameter values within a common utility space. To test that presumption, we conducted an architecture-set analysis. For each $M$, we selected the set of $M$ architectures that minimized total BIC when each participant was assigned to the architecture in the set that fit them best. We then computed $A(M)$, the fraction of the fully individualized advantage captured by that $M$-architecture set, where $M = 1$ corresponds to the single IC-winning architecture and the upper bound corresponds to giving every participant their own best-fitting architecture.

The single-architecture solution already captures the overwhelming majority of explainable structure. On an absolute scale, the IC-winning architecture alone closes $95.4\%$ of the gap between chance prediction and the fully individualized ceiling. Shown in Table 9, adding a second architecture raises this to $98.1\%$, and $M = 3$, $4$, and $5$ raise it to $99.0\%$, $99.3\%$, and $99.5\%$ respectively. Most of the explainable structure is therefore carried by the single winning architecture, though a small set of architectures can improve fit for participants whose choices are better described by alternative forms.

On the relative $A(M)$ scale, the first additional architecture produced the largest gain: $M = 2$ captured $58.3\%$ of the remaining individualization advantage. The $M = 2$ set splits participants almost evenly between the IC winner and a second, structurally distinct architecture. The IC winner is a difference-based form: utility is computed from option-to-option differences in self-payoff, other-payoff, and inequality terms, each with its own exponent. The second architecture is reference-dependent: utility is computed from gains and losses around the midpoint payoff of $3$, with a shared exponent and additional loss-aversion parameters distinguishing payoffs below the reference from payoffs above it. The two forms encode different psychological accounts of how the same choices get evaluated, and $37$ of $73$ participants are better described by the second form than by the population winner.

[13] We use only local refinement in the warm start steps because standard simulated/dual annealing APIs do not accept an explicit initial guess.

These two architectures are also behaviorally distinct in our payoff space, with an Average Model Policy Distance (AMPD) of $0.128$, approximately the $90^{th}$ percentile of pairwise distances across the full candidate set. The split therefore reflects more than a symbolic difference between near-duplicate equations.

Shown in Figure 12, the compression curve does not yield a single unambiguous stopping point. Maximum curvature and meta-BIC select $M = 3$, the cumulative-gain criterion selects $M = 4$, the Kneedle elbow (Satopaa, Albrecht, Irwin, & Raghavan, 2011) selects $M = 5$, and the $1\%$ marginal-gain rule selects $M = 8$. We do not interpret the disagreement as identifying a uniquely correct number of utility architectures. It quantifies a tradeoff. For population-level description, the seven-parameter IC winner is a compact and defensible default. For applications requiring more individualized prediction, $M = 2$ or $M = 3$ captures most of the additional benefit, while $M = 5$ provides higher resolution at the cost of parsimony.

We therefore use the seven-parameter IC winner as the primary coordinate system for the population analyses that follow, while treating the architecture-set analysis as a calibration of that choice. The single architecture is not perfect for every participant, but it captures the overwhelming majority of explainable structure. The multi-architecture sets in Table 9 remain available as principled extensions for researchers who wish to trade parsimony for individual-level precision.

Table 9: Utility architectures selected at each $M$ from 1 to 5.

Each row of the upper panel is one architecture in the $M$-set. $H(M)$ is the fraction of the explainable improvement from chance to the fully individualized ceiling captured by the $M$-architecture set as a whole: $H(M) = (BIC_{chance} - BIC_{model})/(BIC_{chance} - BIC_{individualized})$. $A(M)$ is the fraction of the fully individualized advantage captured ($A(1) = 0$ by definition). $k$ gives the number of free parameters; $n$ gives the number of participants whose choices are best fit by that architecture under the $M$-set assignment; mean BIC gives the average individual BIC across those participants. The lower panel gives the equations corresponding to each architecture label. Researchers wishing to trade parsimony for individual-level precision can read $M$-set membership directly off this table.

| $M$ | $H(M)$ | $A(M)$ | Mean BIC | $k$ | $N$ | Label |
|---|---|---|---|---|---|---|
| 1 | 0.954 | 0.000 | 131.7 | 7 | 73 | A |
| 2 | 0.981 | 0.583 | 142.0 | 7 | 36 | A |
| 2 | 0.981 | 0.583 | 118.3 | 7 | 37 | B |
| 3 | 0.990 | 0.774 | 131.1 | 7 | 31 | A |
| 3 | 0.990 | 0.774 | 86.9 | 7 | 19 | C |
| 3 | 0.990 | 0.774 | 180.9 | 9 | 23 | D |
| 4 | 0.993 | 0.843 | 133.7 | 7 | 27 | A |
| 4 | 0.993 | 0.843 | 84.1 | 7 | 11 | C |
| 4 | 0.993 | 0.843 | 191.0 | 9 | 19 | D |
| 4 | 0.993 | 0.843 | 117.7 | 9 | 16 | E |
| 5 | 0.995 | 0.898 | 134.3 | 7 | 26 | A |
| 5 | 0.995 | 0.898 | 86.0 | 7 | 9 | C |
| 5 | 0.995 | 0.898 | 191.0 | 9 | 19 | D |
| 5 | 0.995 | 0.898 | 115.6 | 9 | 15 | E |
| 5 | 0.995 | 0.898 | 79.7 | 5 | 4 | F |

| Label | Equation |
|---|---|
| A | $U_i(A) = V_{ii}(\pi_i^A - \pi_i^B)^{\gamma_1} + V_{ij}(\pi_j^A - \pi_j^B)^{\gamma_2} - \alpha_{ij} \times max(\pi_j^A - \pi_i^A, 0)^{\gamma_3} - \beta_{ij} \times max(\pi_i^A - \pi_j^A, 0)^{\gamma_3}$ |
| B | $U_i(A) = V_{ii} \times max(\pi_i^{A^{\gamma_1}} - R^{\gamma_1}, 0) - \lambda_{ii} \times max(R^{\gamma_1} - \pi_i^{A^{\gamma_1}}, 0)$<br>$+ V_{ij} \times max(\pi_j^{A^{\gamma_1}} - R^{\gamma_1}, 0) - \lambda_{ij} \times max(R^{\gamma_1} - \pi_j^{A^{\gamma_1}}, 0)$<br>$- \alpha_{ij} \times max(\pi_j^{A^{\gamma_1}} - \pi_i^{A^{\gamma_1}}, 0) - \beta_{ij} \times max(\pi_i^{A^{\gamma_1}} - \pi_j^{A^{\gamma_1}}, 0)$ |
| C | $U_i(A) = V_{ii} \times max(\pi_i^{A^{\gamma_1}} - \pi_i^{B^{\gamma_1}}, 0) + V_{ij} \times max(\pi_j^{A^{\gamma_2}} - \pi_j^{B^{\gamma_2}}, 0)$<br>$- \alpha_{ij} \times max(\pi_j^{A^{\gamma_3}} - \pi_i^{A^{\gamma_3}}, 0) - \beta_{ij} \times max(\pi_i^{A^{\gamma_3}} - \pi_j^{A^{\gamma_3}}, 0)$ |
| D | $U_i(A) = V_{ii} \times max(\pi_i^A - R, 0)^{\gamma_1} - \lambda_{ii} \times max(R - \pi_i^A, 0)^{\gamma_1} + V_{ij} \times max(\pi_j^A - R, 0)^{\gamma_2}$<br>$- \lambda_{ij} \times max(R - \pi_j^A, 0)^{\gamma_2} - \alpha_{ij} \times max(\pi_j^A - \pi_i^A, 0)^{\gamma_3} - \beta_{ij} \times max(\pi_i^A - \pi_j^A, 0)^{\gamma_3}$ |
| E | $U_i(A) = V_{ii} \times max(\pi_i^{A^{\gamma_1}} - R^{\gamma_1}, 0) - \lambda_{ii} \times max(R^{\gamma_1} - \pi_i^{A^{\gamma_1}}, 0)$<br>$+ V_{ij} \times max(\pi_j^{A^{\gamma_2}} - R^{\gamma_2}, 0) - \lambda_{ij} \times max(R^{\gamma_2} - \pi_j^{A^{\gamma_2}}, 0)$<br>$- \alpha_{ij} \times max(\pi_j^{A^{\gamma_3}} - \pi_i^{A^{\gamma_3}}, 0) - \beta_{ij} \times max(\pi_i^{A^{\gamma_3}} - \pi_j^{A^{\gamma_3}}, 0)$ |
| F | $U_i(A) = V_{ii}(\pi_i^{A^{\gamma_1}} - \pi_i^{B^{\gamma_1}}) + V_{ij}(\pi_j^{A^{\gamma_1}} - \pi_j^{B^{\gamma_1}}) \ if \ \pi_i^{A^{\gamma_1}} \geq R \ else \ \lambda_{ii}(\pi_i^{A^{\gamma_1}} - \pi_i^{B^{\gamma_1}}) + \lambda_{ij}(\pi_j^{A^{\gamma_1}} - \pi_j^{B^{\gamma_1}})$ |

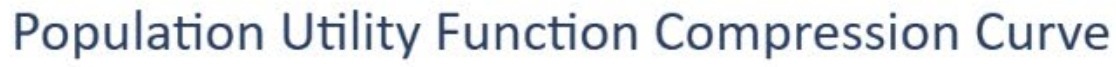

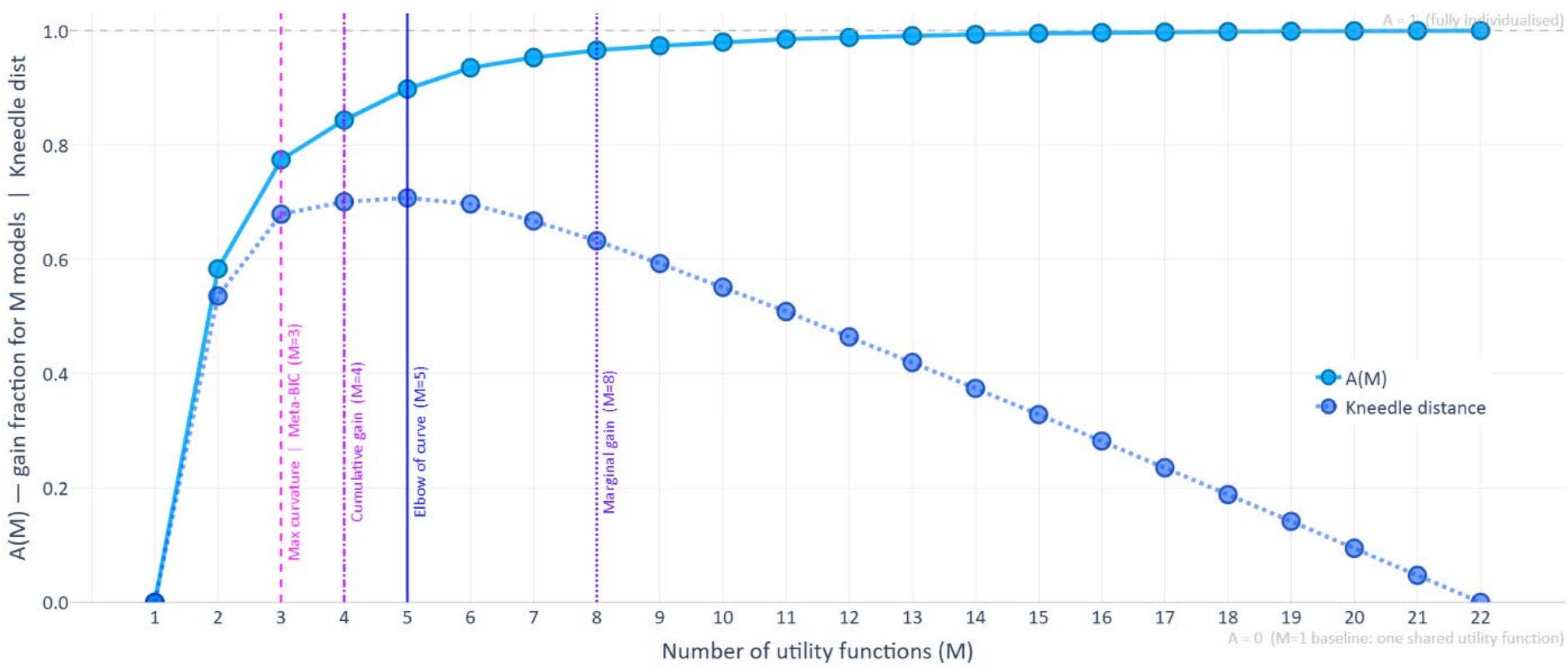

Figure 12: Population Utility Function Compression Curve.

For each value of $M$, the analysis selected the best set of $M$ utility architectures and assigned each participant to the architecture in that set that minimized their BIC. The solid blue curve shows $A(M)$, the fraction of the fully individualized advantage recovered by an $M$-architecture set. $A(M) = 0$ is the baseline in which all participants use the single population-winning architecture; $A(M) = 1$ is the fully individualized ceiling, in which every participant uses their own best-fitting architecture. The curve rises sharply from $M = 1$ to $M = 2$ and then gradually plateaus. The dotted dark-blue curve plots Kneedle distance (Satopaa, Albrecht, Irwin, & Raghavan, 2011), the vertical distance from each point on the compression curve to the straight line connecting its endpoints, after both axes have been normalized. The largest Kneedle distance, marking the curve's elbow, occurs at $M = 5$. Vertical lines indicate four stopping criteria that disagree among themselves: maximum curvature and meta-BIC select $M = 3$, cumulative gain selects $M = 4$, Kneedle selects $M = 5$, and the $1\%$ marginal-gain rule selects $M = 8$. The figure therefore quantifies a fit-parsimony tradeoff rather than identifying a uniquely correct number of architectures. Labels on the first five markers show $H(M)$, the fraction of explainable improvement from chance to the fully individualized ceiling captured by each set: $H(1) = 0.954$, $H(2) = 0.981$, $H(3) = 0.990$, $H(4) = 0.993$, $H(5) = 0.995$.

## 7.9 Relative performance of canonical models

Our model space contained every major outcome-based social-preference utility function the field has relied on. Table 10 reports the rank and ΔBIC of each canonical form against the 526-model space. While the canonical forms varied in how well they fit the data, no canonical form, however theoretically well-motivated, ranks among the top tier of models.

Table 10: Canonical models evaluated against all 526 models.

The models are ranked by BIC score. ΔBIC is the difference between the model and the BIC-minimizing winner. Each model is represented using notation standard across this paper: $\pi_i^A$ and $\pi_j^A$ for self- and other-payoffs in option A; $V_{ii}$ for self-interest; $V_{ij}$ for altruism; $\alpha_{ij}$ and $\beta_{ij}$ for envy and guilt; $\gamma$ for payoff exponents; $\lambda_{ii}$ is repurposed within the conditional welfare family to mean the value one places on oneself versus another when $\pi_i^A < \pi_j^A$. Parameters were free to vary within the same bounds placed on all other models, except $V_{ii}$ and $\lambda_{ii}$, normally within [-1, 1], were transformed the CES and conditional welfare forms via $x = (x+1)/2$ to stay within $[0, 1]$.[14]

| Rank | H(M) | ΔBIC | Model | Citation |
|---|---|---|---|---|
| 255 | 0.412 | 1847 | Constant Elasticity of Substitution | (Andreoni & Miller, 2002) |
| $U_i(A) = V_{ii}(\pi_i^A)^\gamma + (1 - V_{ii})\left(\pi_j^A\right)^\gamma$ | | | | |
| 268 | 0.384 | 1934 | Conditional Welfare | (Charness & Rabin, 2002) |
| $U_i(A) = \begin{cases} V_{ii}(\pi_i^A) + (1 - V_{ii})\pi_j^A, \ if\ \pi_i^A \geq \pi_j^A \\ \lambda_{ii}(\pi_i^A) + (1 - \lambda_{ii})\pi_j^A, \ if\ \pi_i^A < \pi_j^A \end{cases}, where\ (0 \leq V_{ii}\ and\ \lambda_{ii} \leq 1)$ | | | | |
| 273 | 0.378 | 1951 | Social Value Orientation | (Messick & McClintock, 1968; 1976) |
| $U_i(A) = V_{ii}(\pi_i^A) + V_{ij}\left(\pi_j^A\right)$ | | | | |
| 292 | 0.325 | 2118 | Inequality Aversion | (Fehr & Schmidt, 1999) |
| $U_i(A) = \pi_i^A - \alpha_{ij} \times max\left(\pi_j^A - \pi_i^A, 0\right) - \beta_{ij} \times max\left(\pi_i^A - \pi_j^A, 0\right)$ | | | | |
| 338 | 0.222 | 2441 | Maximin-Efficiency | (Engelmann & Strobel, 2004) |
| $U_i(A) = \left(1 - V_{ii} - V_{ij}\right)\pi_i^A + V_{ii}\left(\frac{\pi_i^A + \pi_j^A}{2}\right) + V_{ij} \times min\left(\pi_i^A, \pi_j^A\right)$ | | | | |
| 454 | 0.001 | 3136 | Equity, Reciprocity, Competition | (Bolton & Ockenfels, 2000) |
| $U_i(A) = V_{ii}(\pi_i^A) - \alpha_{ij}\left(\frac{\pi_i^A}{\pi_i^A + \pi_j^A} - \frac{1}{2}\right)^2$ | | | | |

[14] Some canonical models are not specified with fully explicit functional forms in their original sources. For Messick and McClintock (1968), our implementation follows the typical implementation within the field of Social Value Orientation (SVO), and the self-interest parameter is allowed to vary freely, including taking negative values, to accommodate MacCrimmon and Messick's discussion of self-sacrifice (1976). Andreoni and Miller (2002) fit a Constant Elasticity of Substitution (CES) utility function. A CES utility function describes how a person trades off two goods, here their own payoff and the other person's payoff, using a single curvature parameter that fixes how easily one is substituted for the other across the whole range of allocations. The original form ties the weights on the two payoffs together so that they sum to one, applies the same exponent to both payoffs, and then raises the whole expression to one over that exponent. Our implementation keeps the tied weights and the shared exponent but omits the final outer exponent. That outer exponent only rescales utility, so the softmax temperature parameter absorbs it and choice predictions are unaffected. The result is the behavioral core of the Andreoni and Miller form. ERC (2000) penalizes how far a person's share of the total payoff departs from an equal split. To include it in our space, we squared the relative income penalty term and used no other free exponents in the equation.

## 7.10 Parameter Recovery by $k$ Parameters

The parameter recovery simulation in section 3.2 only used a utility function with two parameters. We ran a parameter recovery simulation on models with $k$ parameters to see how well this parameter recovery result generalizes to larger models, which present more difficult optimization problems.

We replicated the conditions of Experiment 2 exactly with 73 synthetic choosers, each playing 60 randomly sampled binary dictator games, fitted using the standard simulated annealing and L-BFGS-B pipeline. For each value of $k \in \{1, 2, 3, \ldots, 9\}$, we generated synthetic choosers whose parameters were drawn uniformly from the estimation bounds, then computed Pearson $r$ between true and recovered values per parameter and in aggregate. Each value of $k$ corresponded to the BIC-winning utility function of that complexity, so the recovery test reflects not an arbitrary model but the best candidate at each tier of the IC ranking. Figure 13 shows the parameter recovery correlations by $k$.

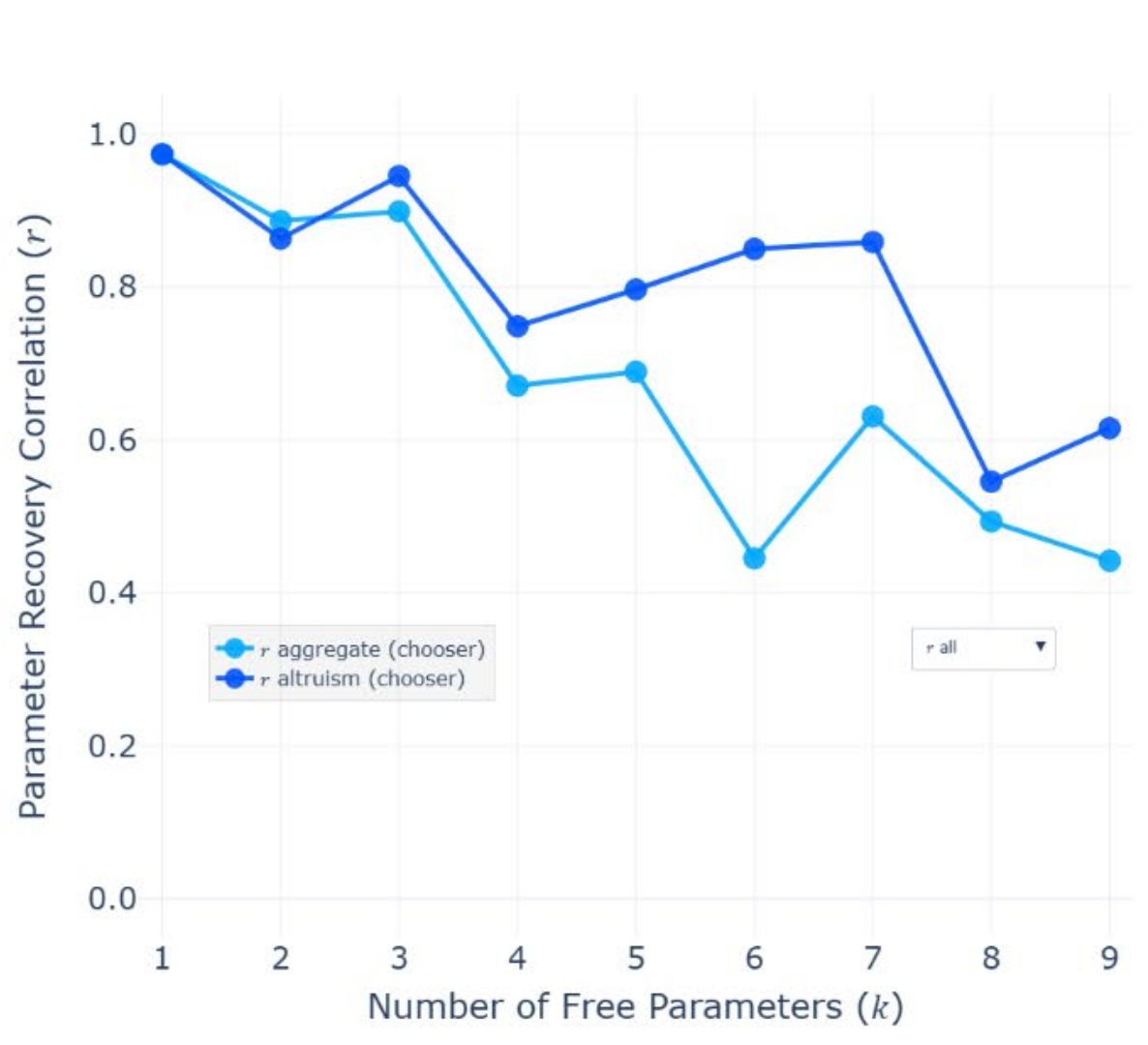

Figure 13: Parameter recovery correlations by model complexity:
Each panel plots per-parameter Pearson recovery correlations $r$ for the BIC-winning utility function at each value of $k$ from 1 to 9. Parameters $V_{ii}$ and $V_{ij}$ maintain the highest recovery across all model sizes, while the three exponent parameters ($\gamma_1$, $\gamma_2$, $\gamma_3$) span the widest range and generally recover least reliably. Within $k = 7$, $\gamma_2$ exceeds $\gamma_1$ and $\gamma_3$, reflecting that altruism curvature produces more discriminable choice patterns in this payoff design than self-interest or social-comparison curvature.

Aggregate recovery declined with model complexity, from $r = 0.97$ at $k = 1$ to $r = 0.44$ at $k = 9$. The BIC-winning seven-parameter model achieved an aggregate of $r = 0.63$. Recovery was not uniform across parameters. Within the $k = 7$ model, self-interest ($r = 0.83$) and altruism ($r = 0.86$) recovered well. Envy ($r = 0.71$) reached an acceptable level. Guilt ($r = 0.60$) and the altruism exponent $\gamma_2$ ($r = 0.67$) were moderate. The self-interest exponent $\gamma_1$ ($r = 0.39$) and the social comparison exponent $\gamma_3$ ($r = 0.37$) recovered most poorly. Per-parameter and aggregate correlations for all nine models appear in Table 11.

Table 11: Per-parameter recovery correlations for each BIC-ranked model from $k = 1$ through $k = 9$.
Each row corresponds to the BIC-winning utility function at that level of complexity. The aggregate column is the mean $r$ across all parameters present in each model. The $k = 7$ model is the BIC winner selected for all parameter estimation results. Each synthetic chooser played 60 randomly sampled binary dictator games under the same conditions as Experiment 2. All fits used the standard simulated annealing + L-BFGS-B pipeline.

| $k$ | $V_{ii}$ | $V_{ij}$ | $\alpha_{ij}$ | $\beta_{ij}$ | $\gamma_1$ | $\gamma_2$ | $\gamma_3$ | $\lambda_{ii}$ | $\lambda_{ij}$ | Agg. $r$ |
|---|---|---|---|---|---|---|---|---|---|---|
| 1 | | 0.97 | | | | | | | | 0.97 |
| 2 | | 0.86 | | | 0.91 | | | | | 0.89 |
| 3 | | 0.94 | 0.90 | | 0.85 | | | | | 0.90 |
| 4 | 0.79 | 0.75 | 0.59 | | 0.56 | | | | | 0.67 |
| 5 | 0.46 | 0.80 | 0.70 | 0.75 | 0.73 | | | | | 0.69 |
| 6 | | 0.85 | 0.45 | 0.19 | 0.68 | 0.46 | 0.05 | | | 0.45 |
| 7 | 0.83 | 0.86 | 0.71 | 0.60 | 0.39 | 0.67 | 0.37 | | | 0.63 |
| 8 | | 0.55 | 0.37 | 0.59 | 0.36 | 0.51 | 0.30 | 0.60 | 0.68 | 0.49 |
| 9 | 0.65 | 0.62 | 0.59 | 0.51 | 0.45 | 0.42 | 0.00 | 0.63 | 0.13 | 0.44 |

To calibrate the levels of credence we should assign to the parameter distribution results presented in the following section, we conducted a parametric bootstrap analysis to determine if low individual-level parameter recovery correlations reflected noise or bias. We treated each of the 73 participants' fitted parameter vectors from the IC analysis as the "true" parameters for a synthetic agent, generated 20 independent synthetic datasets under those true parameters using the same experimental conditions, refitted the model to each, and compared recovered population means to empirical ones. For each parameter we computed normalized bias (bias / empirical SD), regression slope, and Pearson $r$ across individuals. The results appear in Figure 14.

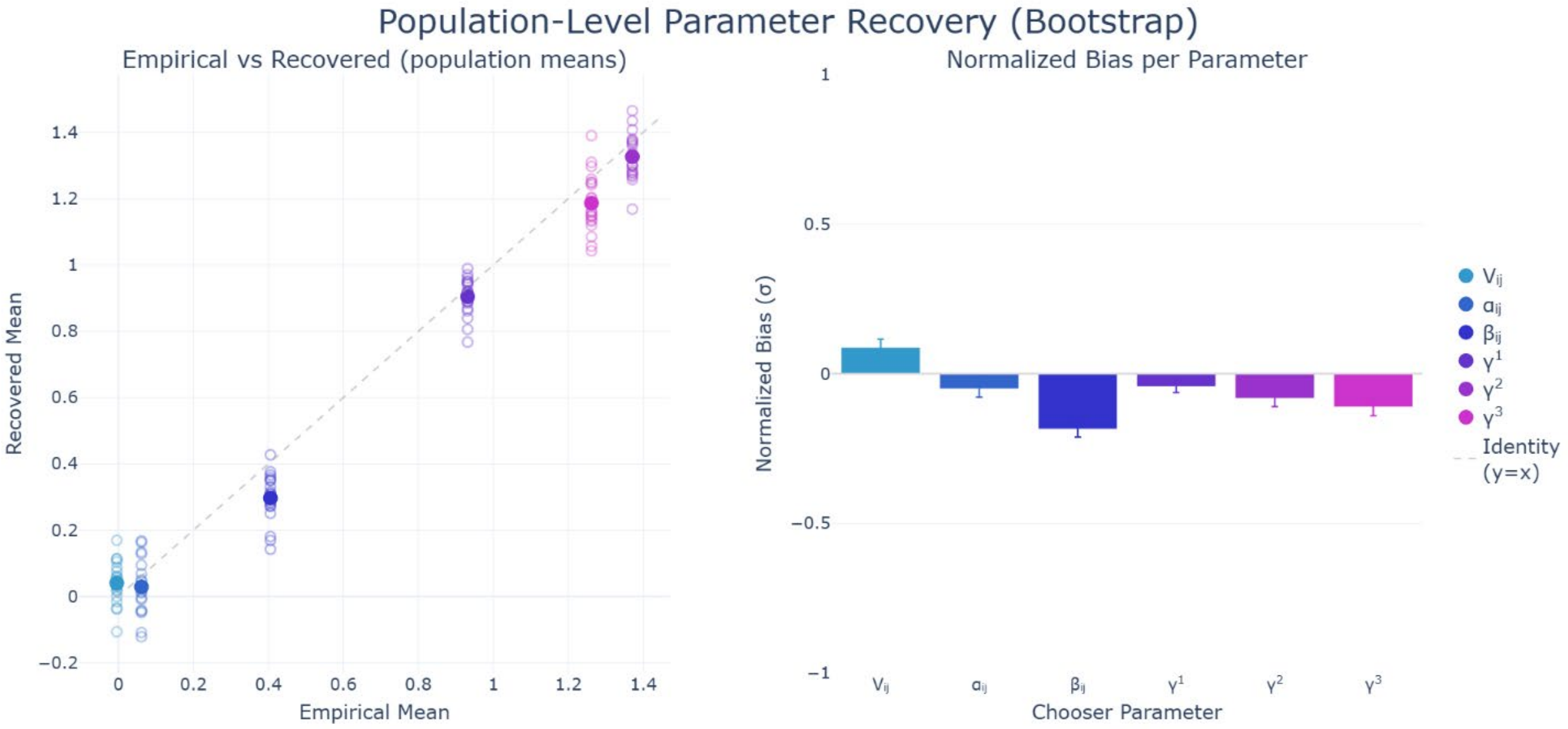

Figure 14: Parametric bootstrap population recovery for the $k = 7$ BIC-winning model:
Results of a 16-iteration parametric bootstrap treating the 73 participants' fitted parameter vectors as true parameters, generating synthetic data under those parameters, refitting, and comparing recovered population means. Left panel: Scatter of empirical population means ($x$-axis) versus recovered population means ($y$-axis), one point per parameter per iteration. Filled markers show means across iterations and hollow markers show individual iterations. The dashed identity line represents zero bias. Right panel: Normalized bias (bias / empirical SD) for each parameter, with error bars showing $\pm 1$ SE across iterations. Positive values indicate overestimation; negative values indicate underestimation. Parameters $V_{ii}$, $V_{ij}$, and $\beta_{ij}$ sit near zero or within acceptable bounds. $\gamma_1$ is positively biased (true sub-linearity is stronger than observed); $\gamma_3$ is negatively biased (true super-linearity is stronger than observed). $\gamma_2$ shows near-zero population mean bias, supporting the directional reliability of the altruism super-linearity finding.

The bootstrap analysis distinguished reliable versus unreliable parameter population means. $V_{ii}$ and $V_{ij}$ show marginal but acceptable underestimation ($\text{normalized bias} = -0.18$ for both), and $\beta_{ij}$ shows near-zero bias ($+0.03\sigma$). $\gamma_2$ has essentially zero population mean bias ($+0.001\sigma$), so the finding that altruism curvature is super-linear at the population level is well-supported despite the low individual-level correlation. $\gamma_1$ is positively biased ($+0.31\sigma$), meaning the true sub-linearity of self-interest is stronger than the fitted mean of $0.83$ suggests. $\gamma_3$ is negatively biased ($-0.37\sigma$), meaning the true super-linearity of social comparison is stronger than the fitted mean of $1.38$ reflects.

These analyses calibrate the inferences that can be drawn from the results below. The central population-level results, that guilt exceeds envy, that altruism and social comparison exponents are super-linear, and that self-interest substantially outweighs altruism, rest on means for which the bootstrap finds near-zero or directionally conservative bias. Individual-level estimates for $\gamma_1$ and $\gamma_3$ carry more noise and should not be over-interpreted at the individual level. However, the population-level directional findings are understated rather than inflated relative to the underlying true parameters.

# 8 Results: Joint Parameter Distributions in Human-on-human Experiment

Having validated the UBM in general and identified the optimal seven-parameter utility function to use within the likelihood term, we now do what this framework was built for—report the resulting social preference parameters from the human-on-human experiment that this model allowed us to discover. This experiment had participants alternate between chooser and predictor roles, providing data to distinguish between genuine social preferences (chooser parameters) and beliefs about others' social preferences (predictor parameters). Moreover, the use of randomized payoff structures facilitated robust analysis across a multidimensional preference space, enabling meaningful exploration of trade-offs between parameters like self-interest, altruism, envy, guilt, and nonlinear payoff sensitivity.

## 8.1 Population-Level Distributions of Social Preference Parameters

### 8.1.1 Self-Interest and Altruism

Participants predominantly exhibited positive self-interest and altruism parameters, with self-interest generally outweighing altruism. Population-level means and standard deviations of chooser parameters were: self-interest [$\mu(V_{ii}) = 0.67841$, $\sigma(V_{ii}) = 0.36920$] and altruism [$\mu(V_{ij}) = 0.04689$, $\sigma(V_{ij}) = 0.26501$]. Predictors' initial priors showed a similar but pattern [$\mu(V_{ii}) = 0.45530$, $\sigma(V_{ii}) = 0.45071$; $\mu(V_{ij}) = 0.14819$, $\sigma(V_{ij}) = 0.47929$], reflecting participants' accurate belief that self-interest dominates altruism, though they expected that dominance to be milder than choosers displayed.

Additionally, $33.33\%$ of choosers exhibited negative altruism parameters, indicating sadistic or spiteful preferences, and $10.96\%$ of choosers exhibited negative self-interest parameters, indicating masochistic or self-sacrificing behaviors.

After excluding these sadistic and masochistic participants, we computed the ratios between their self-interest and altruism parameters ($V_{ii}$ and $V_{ij}$). The average ratio between self-interest and altruism was $8.974$ [$V_{ii}$ / ($V_{ii}$ +$V_{ij}$): $\text{mean} = 0.755$, $\text{SD} = 0.154$, $n = 30$]. This result indicates that, on average, choosers weighted their own payoff roughly nine times more than their counterparts' payoffs.

## 8.1.2 Asymmetric Social Comparison: Envy Versus Guilt

At the population level, choosers' guilt parameters were larger than their envy parameters ($[\mu(\beta_{ij}) = 0.25591, \sigma(\beta_{ij}) = 0.40912]$ vs. $[\mu(\alpha_{ij}) = 0.01541, \sigma(\alpha_{ij}) = 0.35149]$), and 68.49% of choosers had $\beta_{ij} > \alpha_{ij}$. Predictors' priors were: ($[\mu(\beta_{ij}) = -0.03515, \sigma(\beta_{ij}) = 0.55385]$ vs. $[\mu(\alpha_{ij}) = 0.20265, \sigma(\alpha_{ij}) = 0.54697]$; 42.47% with $\beta_{ij} > \alpha_{ij}$).

Figure 15 shows how the distributions of these parameter ratios in the participant population.

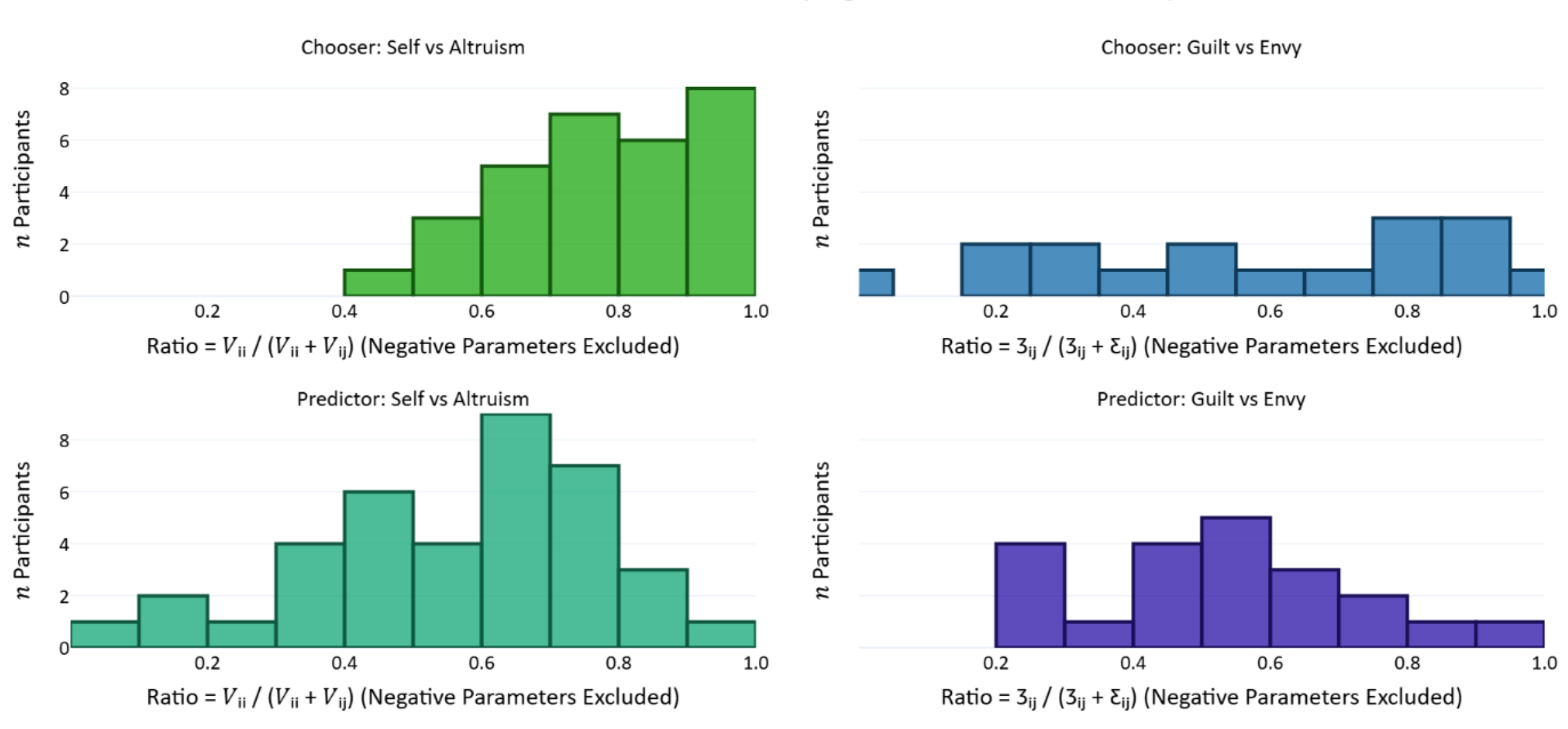

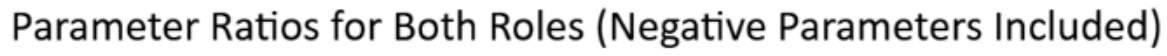

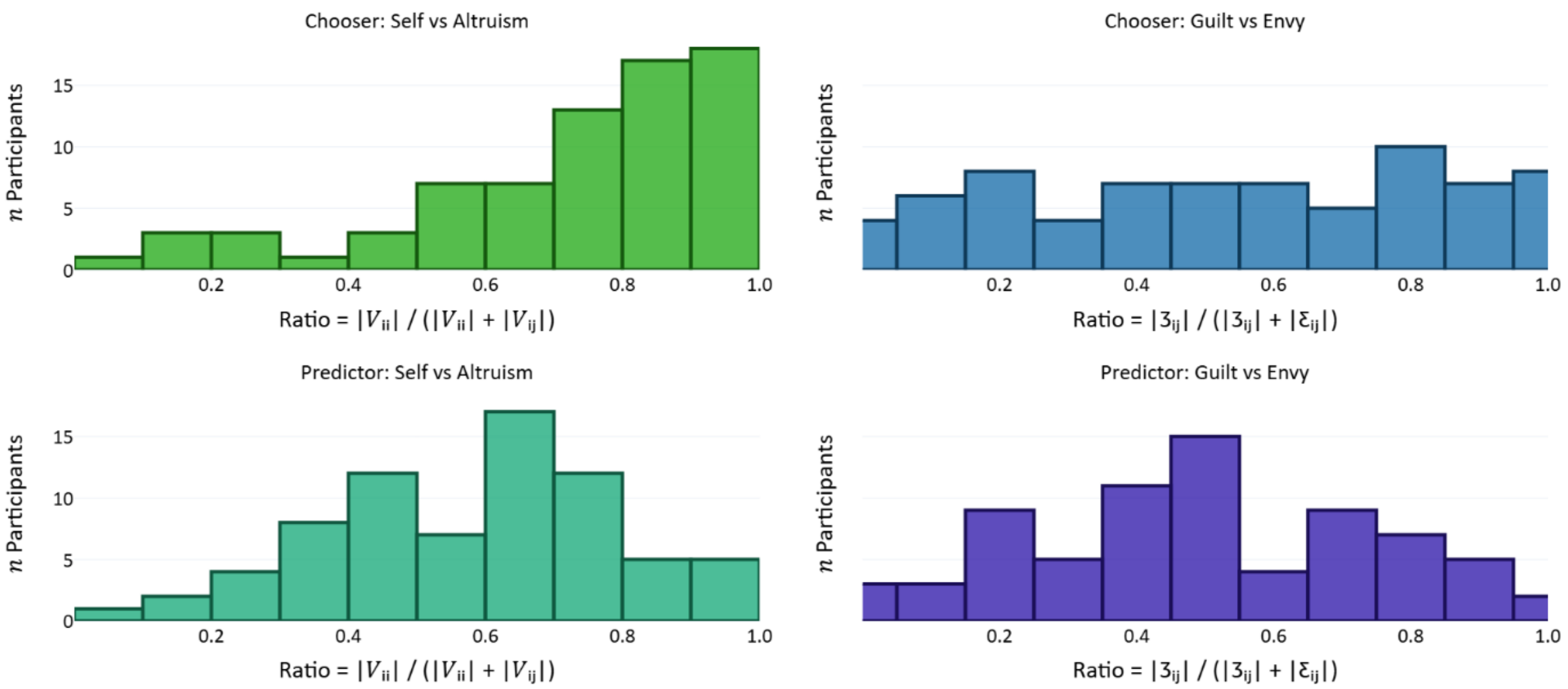

Figure 15: Histograms of Social Preference Parameter Ratios:
For each participant, we computed the ratios of some of their parameters and report the distribution of these ratios in the population. Histograms in the top panel report the distributions only of the subset of participants where both parameters in the ratio were positive ($n$ participants $\in [0, 9]$), while histograms in the bottom panel report distributions of the ratios of the absolute values of parameters ($n$ participants $\in [0, 19]$). Within each panel, the top histograms report chooser parameter ratios and the bottom histograms report predictor parameters. Also, the left histograms report ratios between self-interest and altruism, while the right histograms report ratios between guilt and envy. Panels (A) and (B) show distributions of chooser and predictor self-interest/altruism ratios, respectively; panels (C) and (D) display distributions of chooser and predictor envy/guilt ratios. Vertical dashed lines indicate mean ratios, emphasizing the asymmetrical prioritization of self-interest over altruism, and envy over guilt.

### 8.1.3 Nonlinear Payoff Sensitivity

With the seven-parameter model, exponent parameters differ by term. For choosers, $\mu(\gamma_1)$= $0.82552$ ($\sigma(\gamma_1) = 0.72610$) for self-interest, $\mu(\gamma_2)$=$1.33617$ ($\sigma(\gamma_2) = 0.67218$) for altruism, and $\mu(\gamma_3)$=$1.37846$ ($\sigma(\gamma_3) = 0.69538$). For predictor priors, $\mu(\gamma_1)$= $1.08456$ ($\sigma(\gamma_1) = 0.056325$) for self-interest, $\mu(\gamma_2)$=$0.85037$ ($\sigma(\gamma_2) = 0.48445$) for altruism, and $\mu(\gamma_3)$= $0.94475$ ($\sigma(\gamma_3) = 0.59211$).

### 8.1.4 Clarification on Parameter Reporting

Individual-level mean parameters reflect each participant's central social preference tendencies, whereas individual-level standard deviation parameters (used only in predictors' priors) reflect uncertainty about others' preferences. Population-level statistics reported here include the mean and standard deviation across these individual-level mean parameters.

Additionally, choices were likely influenced by the expectation that players would be matched in subsequent rounds, depending on the round number and matching probabilities. Knowledge of these time horizons likely altered the calculation of social preferences. For instance, many choosers may have acted more cooperatively than if they expected to never meet their counterpart again, making these repeated games less of a pure index of altruism than one-shot dictator games.

Figure 16 and Table 12 show the distribution of individual-level mean parameters.

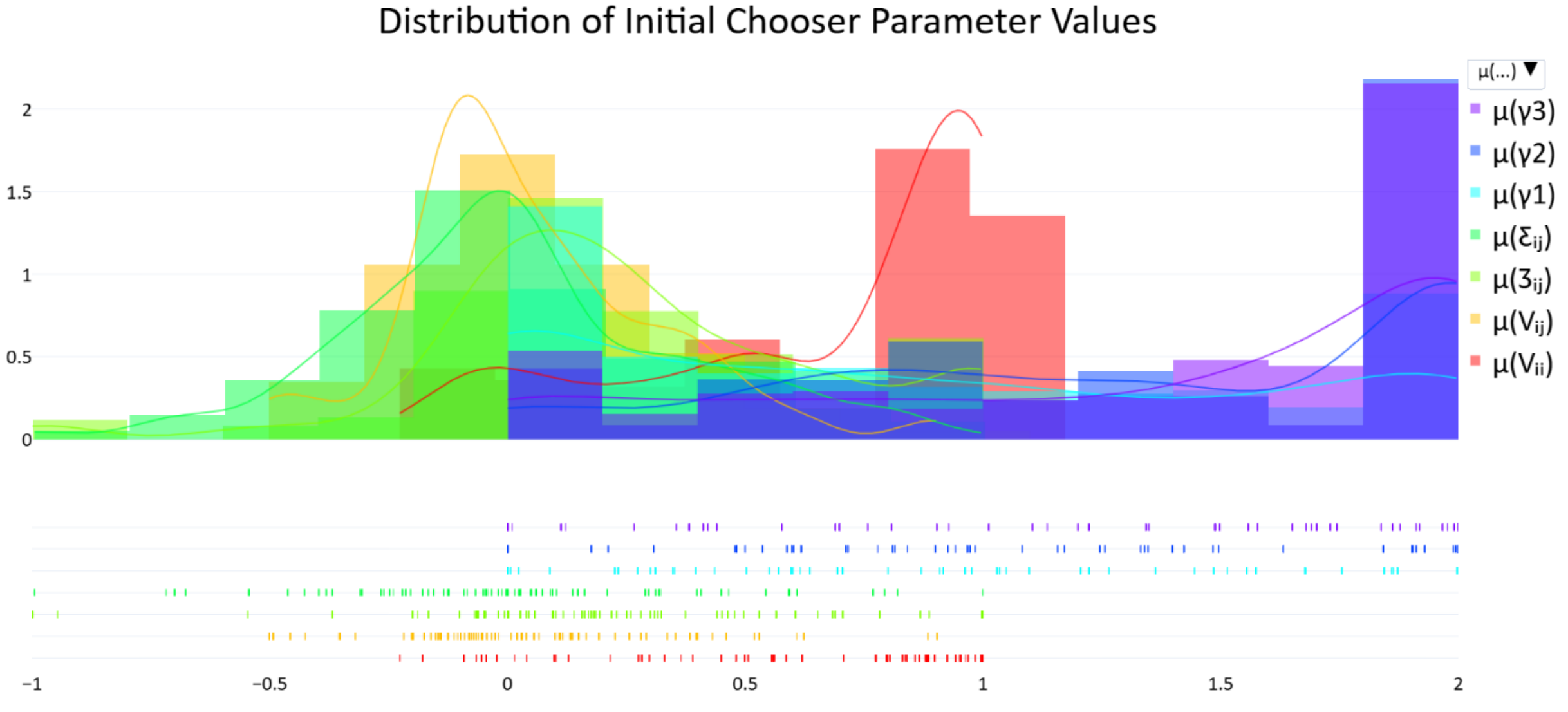

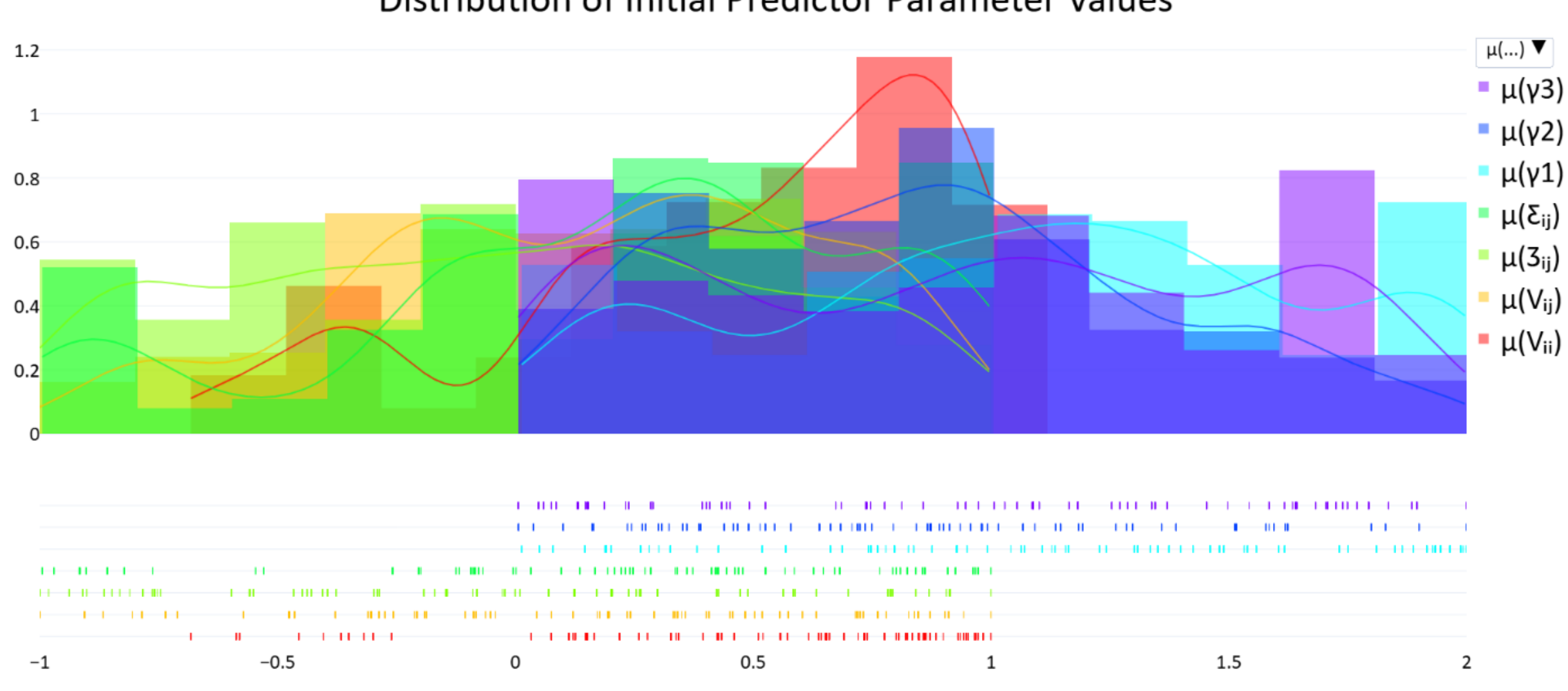

Figure 16: Population-level Distributions of Social Preference Parameters:
Panels display distributions for chooser and predictor parameters (self-interest $[V_{ii}]$, altruism $[V_{ij}]$, envy $[\alpha_{ij}]$, guilt $[\beta_{ij}]$, exponents over the self-interest, altruism, and social comparison terms $[\gamma_1, \gamma_2, \gamma_3]$). Upper histograms show chooser parameters; lower histograms show initial predictor parameters (priors).

Table 12: Population-level averages of individual mean chooser and predictor social preference parameters.

| Population Average Chooser and Predictor Parameter Values – Means and Standard Deviations | | | | |
|---|---|---|---|---|
| | **Chooser Parameters; $n = 73$** | | **Predictor Parameters; $n = 73$** | |
| **Parameter** | **Mean** | **Std Dev** | **Mean** | **Std Dev** |
| **Self-interest $V_{ii}$** | 0.67861 | 0.36920 | 0.45530 | 0.45017 |
| **Altruism $V_{ij}$** | 0.04689 | 0.26501 | 0.14819 | 0.47929 |
| **Envy $\alpha_{ij}$** | 0.01541 | 0.35149 | 0.20265 | 0.54697 |
| **Guilt $\beta_{ij}$** | 0.25591 | 0.40912 | −0.03515 | 0.55385 |
| **Exponent $\gamma_1$** | 0.82552 | 0.72610 | 1.08456 | 0.56325 |
| **Exponent $\gamma_2$** | 1.33617 | 0.67218 | 0.85037 | 0.48445 |
| **Exponent $\gamma_3$** | 1.37846 | 0.69538 | 0.94475 | 0.59211 |
| **Temperature $\tau$** | 0.97034 | 0.71975 | 1.95989 | 0.62519 |

## 8.1.5 Correlations Among Social Preference Parameters

Chooser (within-role). Several theoretically coherent relationships emerged (see Table 13). Self-interest was strongly negatively associated with $\gamma_1$ ($r = -0.445$, Holm-corrected $p_{\mathrm{h}} = .0021$), indicating that higher weighting of self-interest co-occurred with more concave scaling of one's own payoff differences. Altruism positively correlated with envy ($r = 0.457$, $p_{\mathrm{h}} = .0013$). Among the exponents, we observed negative associations between $\gamma_1$–$\gamma_3$ ($r = -0.409$, $p_{\mathrm{h}} = .0081$) and $\gamma_2$–$\gamma_3$ ($r = -0.475$, $p_{\mathrm{h}} = .0006$). The self-interest exponent $\gamma_1$ was negatively correlated with temperature $\tau$ ($r = -0.389$, $p_{\mathrm{h}} = .0160$). All other entries were non-significant after Holm correction.

Predictor (within-role). The only robust association after Holm correction was a negative correlation between the altruism exponent $\gamma_2$ and temperature $\tau$ ($r = -0.407$, $p_{\mathrm{h}} = .0099$).

Table 13: Pearson correlations coefficients among chooser and predictor parameters.
Parameters were normalized so that their absolute values summed to 1 before computing the correlations. Note that these are within-role parameter correlations.

| **Chooser Parameter Correlations** | | | | | | | | |
|---|---|---|---|---|---|---|---|---|
| | $V_{ii}$ | $V_{ij}$ | $\alpha_{ij}$ | $\beta_{ij}$ | $\gamma_1$ | $\gamma_2$ | $\gamma_3$ | $\tau$ |
| $V_{ii}$ | | | | | | | | |
| $V_{ij}$ | 0.183 | | | | | | | |
| $\alpha_{ij}$ | 0.081 | 0.457** | | | | | | |
| $\beta_{ij}$ | -0.069 | -0.183 | -0.262* | | | | | |
| $\gamma_1$ | -0.445** | 0.010 | 0.116 | 0.064 | | | | |
| $\gamma_2$ | 0.022 | -0.244* | -0.242* | 0.096 | -0.256* | | | |
| $\gamma_3$ | 0.256* | -0.050 | 0.050 | -0.246* | -0.409** | -0.479** | | |
| $\tau$ | -0.135 | -0.093 | -0.091 | -0.188 | -0.389** | -0.177 | 0.039 | --- |

| **Predictor Parameter Correlations** | | | | | | | | |
|---|---|---|---|---|---|---|---|---|
| | $\mu(V_{ii})$ | $\mu(V_{ij})$ | $\mu(\alpha_{ij})$ | $\mu(\beta_{ij})$ | $\mu(\gamma_1)$ | $\mu(\gamma_2)$ | $\mu(\gamma_3)$ | $\tau$ |
| $\mu(V_{ii})$ | | | | | | | | |
| $\mu(V_{ij})$ | -0.049 | | | | | | | |
| $\mu(\alpha_{ij})$ | 0.037 | -0.115 | | | | | | |
| $\mu(\beta_{ij})$ | 0.028 | -0.190 | -0.052 | | | | | |
| $\mu(\gamma_1)$ | 0.090 | -0.077 | 0.036 | 0.224 | | | | |
| $\mu(\gamma_2)$ | 0.011 | 0.108 | -0.110 | -0.250* | -0.066 | | | |
| $\mu(\gamma_3)$ | -0.087 | -0.006 | 0.168 | 0.153 | -0.225 | -0.333** | | |
| $\tau$ | -0.232* | -0.077 | -0.057 | 0.156 | -0.334** | -0.407** | -0.055 | --- |
| $^*p < .05$, $^{**}p < .01$ | | | | | | | | |

## 8.1.6 Cross-Role Consistency in Preferences

Because participants alternated between chooser and predictor roles, we examined cross-role consistency to assess preference projection (see Table 14). All cross-role correlations were small and non-significant after Holm correction. The results therefore provide no support for the consensus-effect hypothesis that observers project their own preferences onto others (Ross, Greene, & House, 1977; Engelmann & Strobel, 2000).

Table 14: Cross-role Parameter Correlations.
Correlation between chooser and predictor parameters, illustrating the degree of preference projection ($n = 73$). The bottom row reports $p$ values.

| **Correlation Between Chooser and Predictor Parameters** | | | | | | | | |
|---|---|---|---|---|---|---|---|---|
| | $\mu(V_{ii})$ | $\mu(V_{ij})$ | $\mu(\alpha_{ij})$ | $\mu(\beta_{ij})$ | $\mu(\gamma_1)$ | $\mu(\gamma_2)$ | $\mu(\gamma_3)$ | $\tau$ |
| $r$ | 0.062 | 0.032 | 0.081 | -0.070 | 0.204 | -0.029 | -0.043 | -0.092 |
| $p$ | 0.605 | 0.791 | 0.494 | 0.554 | 0.084 | 0.810 | 0.721 | 0.440 |

# 9 Discussion and Conclusion

To cooperate, people need to infer others' hidden motives from their actions and update those beliefs over time. This paper modeled that process with a Utility Bayesian Model: an observer maintaining continuous probabilistic beliefs over a counterpart's social preferences, reweighting the parameter space against the likelihood of each observed choice. A Bayesian learning model is only as good as the utility function inside it, and a utility function is only as good as the payoff space it was fit to. Generalizing requires testing a broad range of candidate utility forms across a broad range of payoff structures, not a handful of models across many structures and not many models across a handful of structures.

When we did that, the parameters that prior work routinely collapses turned out to be empirically separable and psychologically distinct. The seven-parameter winner beat the next-best model by 17 BIC points and beat the canonical low-dimensional alternatives by hundreds. The parameters those simpler models collapse (altruism with social comparison, envy with guilt, fixed with variable self-interest, shared with term-specific exponents) carry different psychological readings, and once each was given its own term, three findings emerged that change the picture: guilt outweighs envy, social exponents are super-linear rather than concave, and antisocial preferences are more common than previous estimates suggest. Estimating all seven parameters jointly also places every participant in the same psychological space, which makes results commensurable across studies in a way they previously were not.

## 9.1 Validation and the case for continuous representations

The Utility Bayesian Model represents an observer's beliefs about a counterpart's social preferences as a probability distribution over a continuous parameter space, reweighted after each observed choice by the likelihood of that choice under candidate parameter values. Before fitting it to human-on-human data in Experiment 2, we validated the framework in two ways. Simulations confirmed that the machinery does what it claims: the optimizer recovers known parameters, posteriors converge toward a chooser's true preferences with repeated observations, and update speed responds to prior variance and choice stochasticity in the way Bayesian theory predicts. Experiment 1, the human-on-bot experiment, then evaluated the model against a controlled ground truth by pairing participants with artificial agents whose preferences were fixed in advance. Across both checks, the continuous UBM outperformed non-Bayesian alternatives and several thousand typological Bayesian models that partitioned the preference space into discrete types.

That comparison is informative in its own right. It asks whether observers' hypothesis spaces are best represented as a fine-grained continuum or as a small discrete set of types (i.e. good versus evil). At least within the domain we tested, the answer is the continuum. Social cognition here looks less like sorting people into a few categories and more like positioning them in a graded space, updatable at fine resolution. Whether this generalizes beyond outcome-based preferences in binary dictator games is an open question, but the present results give continuous representations a stronger empirical footing than they previously had.

## 9.2 The cost of inherited simplicity

Outcome-based social preferences have largely been modeled with a handful of canonical utility forms developed between 1992 and 2004: inequity aversion (Fehr & Schmidt, 1999), ERC (Bolton & Ockenfels, 2000), Constant Elasticity of Substitution (Andreoni & Miller, 2002), social welfare (Charness & Rabin, 2002), and maximin-efficiency (Engelmann & Strobel, 2004). Subsequent innovation moved largely toward non-outcome-based motives such as reciprocity (Falk & Fischbacher, 2006) and self-image, leaving these foundational outcome-based kernels mostly untouched (Capraro, Halpern, & Perc, 2024; Fehr & Charness, 2025), and recent empirical work continues to rely on the same canonical frameworks or on even simpler two-dimensional schemes[15] (Kerschbamer, 2015; Bakker & Dijkstra, 2021; Nunnari & Pozzi, 2022; Liu & Lange, 2023; Mizuno & Shimizu, 2024; Wang, et al., 2025). At the same time, these forms have typically been fit to small, hand-selected payoff sets, frequently constant-sum, which left most of the relevant tradeoff space unsampled. Habits in model choice and habits in design choice reinforced each other: narrow payoff spaces could not discriminate among richer utility forms, so richer utility forms were not developed, so payoff spaces continued to be designed for the models already in use. The result is a literature that rests heavily on a small inheritance from the late 1990s and early 2000s, rarely tested against the wider space of candidates those original models came from.[16] Each of these canonical forms appears as a special case within the candidate set evaluated here, and none of them ranks among the top-performing models.

[15] Consider a linear utility function over two payoffs $\pi_i^A$, and $\pi_j^A$, such as $U_i(A) = V_{ii}(\pi_i^A) + V_{ij}\left(\pi_j^A\right)$. Adding a symmetric social-comparison term, $-\alpha_{ij}\left(\pi_i^A - \pi_j^A\right)$, does not add a genuinely independent parameter. Distributing the term gives $U_i(A) = \left(V_{ii} - \alpha_{ij}\right)\pi_i^A + \left(V_{ij} + \alpha_{ij}\right)\pi_j^A$, so any value of $\alpha_{ij}$ can be exactly offset by rescaling $V_{ii}$ and $V_{ij}$; the symmetric social-comparison parameter is therefore unidentified in a linear two-term utility. This redundancy is what makes the asymmetric decomposition into envy ($\alpha_{ij}$, active when$\pi_i^A < \pi_j^A$) and guilt ($\beta_{ij}$, active when $\pi_i^A > \pi_j^A$) substantive rather than cosmetic. Once the social-comparison term is split asymmetrically, no rescaling of $V_{ii}$ and $V_{ij}$ recovers it, and $\alpha_{ij}$ and $\beta_{ij}$ become independently identified.

Our IC analysis estimates the cost of this inheritance directly. Table 15 shows the BIC penalty incurred by collapsing each parameter the seven-parameter winner separates out. Removing the altruism term costs roughly $442$ BIC points and conflates altruism with social comparison. Symmetrizing social comparison costs about $233$ points and collapses envy and guilt into a single inequality term. Forcing a single exponent across terms costs about $193$ points and obscures whether nonlinearity is operating on self-payoffs, other-payoffs, or inequality. Fixing self-interest to one costs about $135$ points and precludes detection of masochism. Each of these gaps exceeds by an order of magnitude or more the conventional ΔBIC threshold of $10$ for decisive model comparison (Schwarz, 1978; Anderson & Burnham, 2002). The parameters that simpler models discard are not statistical decorations.

Table 15: Costs of simplifying the winning seven-parameter model:
The seven-parameter BIC-minimizing model has four child models. This table shows the costs of reducing the winning model to these children in added BIC and theoretically.

| **Simplification** | **+ΔBIC** | **Theoretical Cost** |
|---|---|---|
| **Remove altruism** | $\sim 442$ | Conflates altruism with social comparison |
| **Symmetrize social comparison** | $\sim 233$ | Conflates envy and guilt |
| **Force a single exponent** | $\sim 193$ | Loses specificity |
| **Fix self-interest to one** | $\sim 135$ | Precludes masochism |

The numerical penalties also correspond to specific psychological confusions. Combining altruism with social comparison forces a model to misclassify a participant who genuinely enjoys hurting others as merely 'extremely competitive,' or a participant who enjoys being superior to others as 'sadistic'[17]. Combining envy with guilt obscures the documented asymmetry of inequality aversion, which prior work has shown to be substantial in either direction (Fehr & Schmidt, 1999; Kerschbamer, 2015; Nunnari &

[16] Additions such as Kerschbamer's Equality-Equivalence Test (2015), Cappelen et al.'s**Invalid source specified.** fairness ideals, moral wiggle room, merit biases, stake effects, or reference dependence represent important elaborations but do not substantially alter this overall historical narrative

[17] The altruism-social comparison distinction is roughly analogous to the division between the morality of harm and the morality of fairness in moral foundations theory**Invalid source specified.**, although there are many forms of fairness, such as outcome equality, procedural fairness, and reciprocity.

Pozzi, 2022; Capraro, Halpern, & Perc, 2024). Fixing self-interest to one precludes detection of masochism, which our data show at non-trivial frequencies. Lower-dimensional models remain useful when the penalty for complexity is severe (Experiment 1, where higher-dimensional payoff exploration was infeasible, is one such case), but in any setting that can sustain richer estimation, the additional parameters earn their place.

## 9.3 What the broader search revealed

Four results from the population parameter distributions deserve emphasis, because in each case the broader search either sharpens a familiar finding or overturns one. Two of them bear directly on questions people have always asked about each other: how good is the typical person, and how many people are genuinely cruel.

There are more antisocial people than prior work has found. A third of choosers ($33.33\%$) placed negative weight on the other person's payoff, the signature of spite or sadism, and about one in nine ($10.96\%$) placed negative weight on their own, the signature of masochism or self-sacrifice. These rates are well above what earlier studies report. If one is willing to read a preference for making others worse off as a form of malice, then the plain implication is that malice of this kind is not rare. It sits in roughly a third of this sample, at least toward anonymous strangers. The most likely objection is that some of this is strategy rather than disposition, since punishing a stranger can pay off when you expect to meet again (Abbink & Sadrieh, 2009; Herrmann & Orzen, 2008). That may explain part of the spite and competitiveness, but it cannot explain masochism, which lowers one's own payoff for no strategic return. The simplest account of why these types show up here and not in earlier work is that earlier utility forms lacked the parameters needed to see them. What you cannot represent, you cannot detect.

Most people are genuinely kind, but self-interest outweighs that kindness by roughly nine to one. Setting aside the antisocial minority, the typical chooser valued their own payoff about nine times as heavily as a stranger's ($V_{ii}/V_{ij}$ ratio $\approx 8.97$), yet still placed reliably positive weight on that stranger. This shows that genuine concern for others is the norm but concern for oneself is much stronger. If altruism is taken as one ingredient of what people mean by goodness, then the average person here is kind, but self-regarding first. Two caveats keep this from being read too flatly, and both are developed below. The nine-to-one figure is an average over how people treat anonymous strangers; it says nothing about how they treat people they know, and ordinary life is mostly the latter. And it turns out to depend on whether the other person is ahead or behind: as the guilt finding will show, much of what looks like weak altruism is actually a willingness to help when someone has fallen behind, which behaves quite differently from a flat nine-to-one discount. The number is a real summary of one situation, not a verdict on human nature.

This nine-to-one ratio is also cleaner than the comparable numbers from standard dictator games (Engel, 2011; Falk, et al., 2018) because ordinary dictator-game giving is not a pure measure of altruism.[18] Because our design varies payoffs widely, it separates altruism from the distaste for inequality without leaning on the assumptions earlier designs require, so the altruism estimate here is no longer quietly absorbing the work of other motives. That separation is what sets up the next result. With self-interest this dominant, the motivational pull usually credited to altruism, the pull to lift the other person up, has to be coming from somewhere else, because most people are still doing the lifting. The guilt finding tells us where it comes from.

Curvature on the social terms runs the opposite way from the textbook prediction. Curvature on the chooser's own payoff was near-linear and slightly concave ($\mu(\gamma_1) \approx 0.83$), the familiar shape of diminishing returns. But the exponents on altruism and social comparison came out greater than one ($\mu(\gamma_2) \approx 1.34$; $\mu(\gamma_3) \approx 1.38$), meaning people react more sharply to large gifts and large inequalities than to small ones, not less. This runs against the standard assumption that every subjective magnitude flattens as it grows (Gossen, 1854; Marshall, 1920; Kahneman & Tversky, 1979; Andreoni & Miller, 2002). It is the kind of pattern only visible once each term carries its own exponent, which is why a search over canonical forms could not have surfaced it. What drives the super-linearity is still open: it may reflect a threshold below which small inequities are shrugged off, an escalating moral salience of large ones, or a non-preference process such as attention that the exponent is partly soaking up. Each is testable.

Guilt outweighs envy, reversing the canonical Fehr-Schmidt ordering. At the population level, $\mu(\beta_{ij}) \approx 0.26$ exceeded $\mu(\alpha_{ij}) \approx 0.015$, and about 69% of choosers had $\beta_{ij} > \alpha_{ij}$. This contrasts with most of the inequality-aversion literature (Fehr & Schmidt, 1999; Loewenstein, Thompson, & Bazerman, 1989; Nunnari & Pozzi, 2022). Part of that earlier ordering may itself be a methodological artifact. Earlier studies typically combined narrow payoff designs (constant-sum, or otherwise unable to separate altruism from social comparison) with utility forms that omitted altruism or constrained $\alpha_{ij} \geq \beta_{ij}$, both of which inflate α relative to $\beta_{ij}$. So part of the canonical ordering reflects designs that prevent guilt from emerging even when present in the data.

---

[18] A linear two-term utility function such as $U_i(A) = V_{ii}(\pi_i^A) + V_{ij}(\pi_j^A)$ always prefers an extreme split: keep everything or give everything. Standard dictator-game data show interior splits only after the utility is *modified*, either by adding a social-comparison penalty (e.g., $-\Sigma_{ij}(\pi_i^A - \pi_j^A)$) or by raising each pay-off to a concave exponent $U_i(A) = V_{ii}(\pi_i^A)^\gamma + V_{ij}(\pi_j^A)^\gamma$, and only once it is embedded in a choice rule such as a softmax. So observed giving reflects whatever extra structure the chosen model imposes, which is why inferring altruism from giving alone is unreliable. Our varied-payoff design recovers interior choices without needing those extra terms, which is what lets altruism and social comparison come apart. Note that our winning model does include exponents but the point is only that they are not required to measure altruism here.

Because the reversal is so unexpected, we verified independently that it is not an artifact of the seven-parameter model's flexibility. We constructed two minimal Fehr-Schmidt-style agents, linear utilities with no exponents and no reference dependence, identical in every respect except their inequality-aversion ratios: an envious agent with $\alpha_{ij} = 0.75$ and $\beta_{ij} = 0.25$, and a guilt-heavy agent with $\alpha_{ij} = 0.25$ and $\beta_{ij} = 0.75$. We then scored each agent against all participants' observed choices across a grid of self-interest ($V_{ii}$) and altruism ($V_{ij}$) weights (Figure 17). Stripping the utility form down to its canonical inequality-aversion ingredients puts the comparison on the same footing as the original Fehr-Schmidt framework. Which agent fits better turns out to depend on where in ($V_{ii}$, $V_{ij}$) space the scoring is done. The guilt-heavy agent wins throughout the high-$V_{ii}$, low-$V_{ij}$ region, exactly where our population is concentrated ($\mu(V_{ii}) \approx 0.68$, $\mu(V_{ij}) \approx 0.047$). The envious agent dominates only in the opposite corner, where altruism is large relative to self-interest.

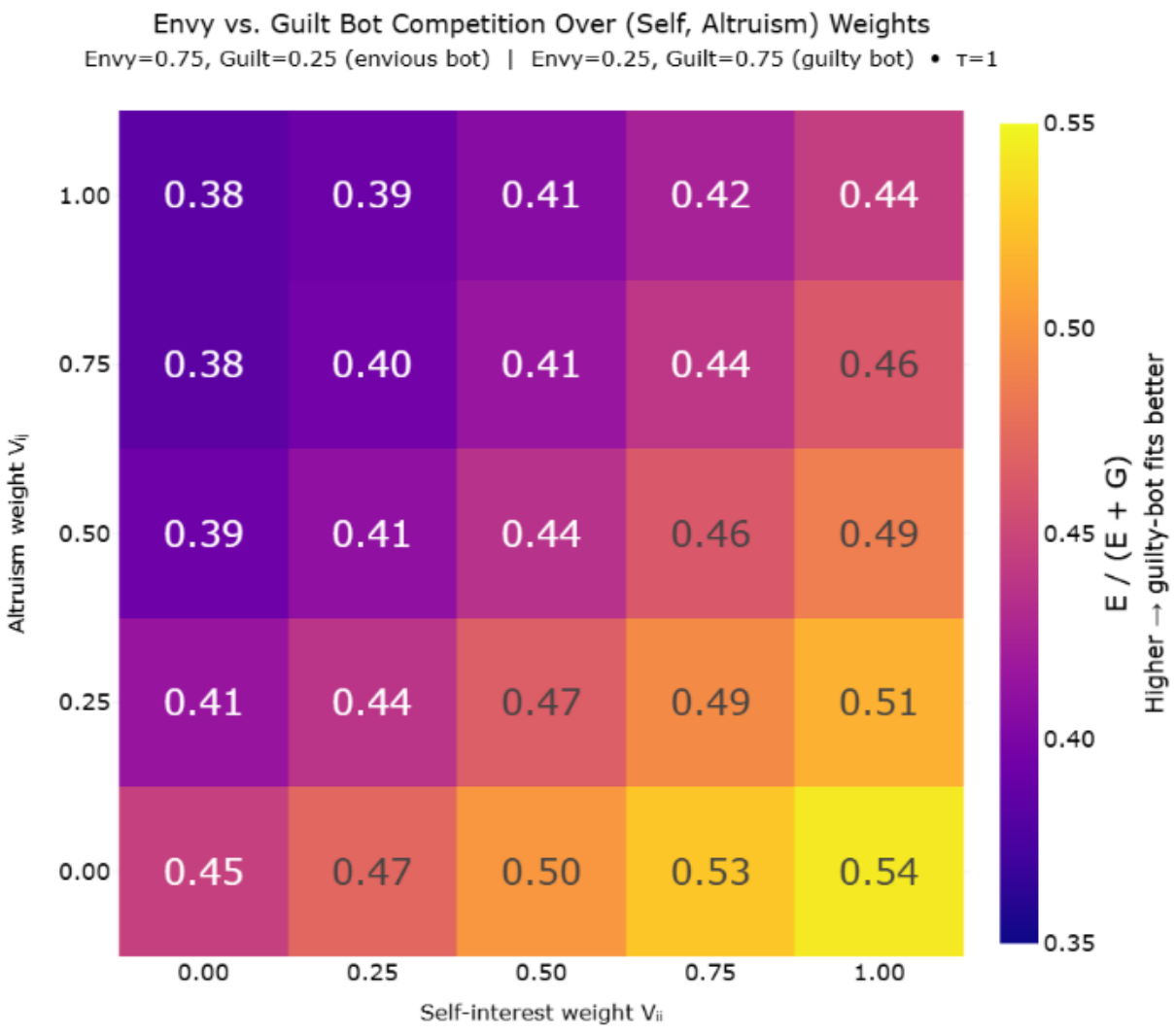

Figure 17: Model Comparison Between Envious ($\alpha_{ij} > \beta_{ij}$) and Guilt-Weighted ($\beta_{ij} > \alpha_{ij}$) Agents Across ($V_{ii}$, $V_{ij}$) grid: Cells show the proportion of NLL attributable to the guilt-heavy agent relative to the total error from both agents; values above 0.5 indicate better fit by the guilt-heavy agent. The guilt-heavy agent fits human behavior more closely when self-interest is high and altruism is low, the region where most participants' fitted parameters lie. The envious agent dominates only when self-interest is low and altruism is high.

The mechanism becomes interpretable in light of the previous finding. When self-interest is strong and altruism is weak, the motivational role usually carried by altruism, increasing the other person's payoff, has to be carried by something else, because most people are still doing it. Guilt steps in: aversion to advantageous inequality picks up the slack and produces the same kind of choices that strong altruism would produce, but conditional on the other person being worse off. The pattern can be read in the other direction too. When $V_{ii}$ is large, $\alpha_{ij}$ contributes little additional explanatory value because most of the envy-active contrasts are already explained by self-interest alone. When $V_{ij}$ is large, $\beta_{ij}$ loses its job because unconditional altruism curbs advantageous inequality. So the relative strength of $\alpha_{ij}$ and $\beta_{ij}$ depends on the balance of self- and other-directed motives, and in our population that balance favors $\beta_{ij}$.

What Fehr and Schmidt called guilt[19] may therefore be better understood as conditional altruism: a willingness to raise others up to one's own level, without going beyond it. This is, in spirit, the intuition behind social-welfare models that weight the worse-off party more heavily (Charness & Rabin, 2002), recovered here not as a built-in assumption but as an emergent property of how guilt and self-interest interact. The practical implication matters. Cooperative behavior does not require a population with strong unconditional altruism. It can be sustained when the trigger is a concrete disparity rather than an abstract concern for others' welfare. A population that looks modestly altruistic by standard dictator-game measures can still be reliably motivated to close gaps, which is the more common shape of real-world helping.

## 9.4 A shared coordinate system

Because the seven parameters are estimated jointly and refer to the same underlying utility space, every participant in this study occupies a point in the same seven-dimensional system, and that system can in principle be extended to other studies that adopt the same parameterization. This matters because prior work has reported social-preference parameters in isolated pairs or within two-dimensional SVO frameworks (Messick & McClintock, 1968; Van Lange, 1999; Murphy, Ackermann, & Handgraaf, 2011), and the resulting estimates are difficult to combine. Two studies that each report an 'altruism' parameter may not be measuring the same thing if one uses a model that includes social comparison and the other does not, because the absent term forces its psychological work onto whichever term is available to absorb it. The pieces do not fit together, not because the underlying construct is incoherent, but because the parameter spaces are incommensurable.

---

[19] Fehr and Schmidt's (1999) label "guilt" is a misnomer in two ways. First, in ordinary usage guilt is a reactive emotion triggered by an action one has performed, not a preference over outcome states one passively occupies (Tangney, 1991; Tracy & Robins, 2004). Second, the β parameter requires no belief about having wronged anyone. It is activated simply by finding oneself ahead, regardless of how the disparity arose. This is distinct from "guilt aversion" in the psychological-game literature, where guilt refers to disappointing another's expectations (Battigalli & Dufwenberg, Guilt in games, 2007). "Advantageous inequality aversion" captures the behavioral signature without the causal connotation.

A common parameter space changes this. Cross-study comparisons become meaningful, meta-analyses become possible, and models can use accumulated empirical distributions to set informed priors instead of starting from uniform every time. Each subsequent study that fits the same parameterization adds to a growing public stock of priors that the next study can draw on.

## 9.5 Future directions

Future studies should expand this research to paid participants across many cultures (Henrich, Heine, & Norenzayan, 2010), and should include negative payoffs. The utility settings that make payoffs reference-point dependent and that include negativity parameters similar to loss aversion underperformed in our IC analysis, but those are precisely the terms most likely to matter once payoffs can fall below the reference point. Our results should not be read as strong evidence against either kind of structure.

Beyond such expansions in sample and design, the framework rests on two complementary commitments: invertibility and generalizability. The first means that any forward model expressing choice probabilities as a function of candidate parameter values can be turned into an inference engine for those parameters. While the UBM here inverted a simple utility function that represented outcome-based social preferences within a choice probability function, much more sophisticated forward models can be inverted also. Once inverted, a model of reciprocity becomes a model of inferring others' propensities to reciprocate. Once inverted, a model of limited strategic reasoning depth becomes a model of inferring the reasoning depth of others. Once inverted, a model of attributions of moral responsibility becomes a model of inferring how others attribute moral responsibility. Such complexities, once inverted, allow the framework to generalize to the full space of games and to capture a greater richness of moral psychology, including how we come to learn the moral psychology of others.

As one example among many, the UBM can be combined with a model of higher-order Theory of Mind reasoning so that the resulting model handles multi-step games with multiple choosers. The winning seven-parameter utility function would be retained, with priors informed by distributions accumulated from work such as the present study, but in place of literal payoffs it would consume expected payoffs forecasted via recursive simulation of the future choices of self and others, that is, by recursive Theory of Mind. That recursion is bounded because thinking about what others are thinking is cognitively costly and degrades with depth (Hedden & Zhang, 2002; Stanley, 2026; Kleiman-Weiner, Ho, Austerweil, Littman, & Tenenbaum, 2016; Wu, et al., 2021), so the forward model would introduce a reasoning-depth parameter governing how many recursive levels the chooser simulates before falling back on a non-mentalistic default at further-out nodes, such as treating downstream choices as if they were governed by baseline probabilities rather than by explicit forecasts of others' decisions.

Inverting this more sophisticated forward model would do something the original UBM cannot. Even reasoning-depth parameters can be inferred from observed choices when those choices depend on recursive mentalizing, recovered alongside the social-preference parameters. Two observers can hold identical social preferences but reason to different depths, a possibility long recognized in level-$k$ frameworks of strategic reasoning (Stahl & Wilson, 1995; Nagel, 1995) but not previously integrable with continuous Bayesian models of preference learning, and their choices in sequential games differ in characteristic ways; observed choices therefore identify both. Combining the UBM with Theory of Mind in this way also lets observers forecast their own future belief updates. Consider a multi-step trust game in which player A currently values player B highly. Player B may betray; player A may later choose to punish or reward. Player A, currently inclined to reward, can nevertheless anticipate that following a betrayal their valuation of player B will be revised downward and they will choose to punish. Inverting a model of recursive belief updating produces this form of forecasting one's own learning and is essential for explaining cooperation in settings where today's choices depend on anticipating reactions to tomorrow's outcomes.

The same principle generalizes. A forward model that conditions utility on what each agent believes the others believe yields, on inversion, a model of inferring belief-based preferences such as guilt aversion and the desire to meet others' expectations (Charness & Dufwenberg, 2006; Battigalli & Dufwenberg, 2007; 2009; 2022). A forward model that incorporates causal contribution and foreseeability into the evaluation of agents yields, on inversion, a model of inferring how each observer attributes moral responsibility (Cushman, Young, & Hauser, 2006; Young & Saxe, 2009; Halpern & Kleiman-Weiner, 2018). A forward model that lets the chooser's utility depend on what they believe others see them choosing yields, on inversion, a model of inferring sensitivity to self- and social-image concerns (Dana, Weber, & Kuang, 2007; Andreoni & Bernheim, 2009). A forward model that conditions current utility on the history of interaction within a dyad yields, on inversion, a model of inferring individual differences in reciprocity and the rate at which preferences themselves drift over relationships.

Generalizability is the framework's second commitment, and randomizing payoffs across a fully crossed space, as we did here, was only its first step. The argument that motivated it, that models calibrated to a narrow slice of situational variation can only speak to that slice, extends naturally beyond payoffs to the structure of games themselves. Future work should fit forward models like the UBM to broader spaces: games that vary not only payoffs but branching factor, depth, node types, the presence of chance, the distribution of choosers across the tree, knowledge asymmetries between agents, and the cross-game histories that link encounters. Expanding the model and expanding the games it handles are two sides of the same project. Game trees can encode promises, threats, asymmetric knowledge, choices under uncertainty, and the temporal structure of relationships, but a model captures these only insofar as its forward representations are sophisticated enough to reason over them. Once inverted, a forward model rich enough to navigate the full space of such games becomes an inference engine for the social and moral parameters that drive human behavior across the situations game trees encode.

## 9.6 From inferring values to judging character

The inversions described above all recover what an agent wants: its preferences, its propensity to reciprocate, its sensitivity to others' beliefs. But inferring what an agent wants is only a short distance from a judgment of the agent's character. What separates them is a function that reads the parameters through a lens of moral salience and projects them across the situations one expects to encounter. By this view, character is not a parameter but a projected policy: what this agent would do across a space of morally relevant situations, weighted by how much those situations matter and how likely they are to arise. The machinery this paper validates, inferring dispositions from ambiguous choices, is the first component such a judgment requires.

This matters because character judgment is a distinct mode of moral judgment. Judging an outcome asks what happened, judging an action asks whether it was permitted, and judging character asks what kind of agent produced it. These can come apart. An agent of good character can cause harm by accident and an action can be forbidden even when its outcome is harmless. A model that collapses character into outcome utility cannot represent the difference, the same way a utility function without a separate altruism term cannot tell genuine spite from mere competition, as we found here. The parameters a model omits do not vanish. Their work is silently reassigned to whatever terms remain.

The reason to expect that any sufficiently capable agent, human or artificial, will need to judge character rather than compute consequences alone is a matter of tractability. Evaluating every decision by its full expected utility is intractable for bounded agents, which is why humans lean on heuristics: rules about which actions are forbidden, and judgments about which agents can be trusted (Sunstein, 2005; Maier, Cheung, & Lieder, 2026). Character judgment is one such heuristic, a compression of an agent's likely behavior into a portable summary that can be carried into situations not yet encountered. An artificial agent operating under the same bounds will plausibly need the same compression. And the most important target of that judgment may be the agent itself. An AI that can ask what kind of agent it is across the space of moral situations, measured against an ideal standard rather than only against its instructions, is doing something neither reward modeling nor rule-following quite captures. The inversion demonstrated here, recovering dispositions from choices, is a step toward the machinery such self-evaluation would require.

## 9.7 The Reach of Social Preferences

Every result in this paper concerns the outcome-based social preferences that binary dictator games activate. It is fair to ask how far that subject extends. We believe the reach of the preferences themselves is an open question, but the reach of the four dimensions that activate them is not. These dimensions are mathematical, and mathematics is not limited to one laboratory or one species.

A binary dictator game presents two allocations, each with a payoff for the chooser and a payoff for the other player, represented by four numbers $\{\pi_i^A, \pi_i^B, \pi_j^A, \pi_j^B\}$. The same structure can be written as four dimensions, $[\pi_i^A - \pi_i^B, \pi_j^A - \pi_j^B, \pi_i^A - \pi_j^A, \pi_i^A]$, capturing what the chooser stands to gain, what the other stands to gain, the gap between them, and the absolute stakes. An agent could represent the same information as ratios, as deviations from a reference point, with or without nonlinear scaling, so the exact form of these dimensions is not universal, but their meaning is. Four questions are being asked: how much of what I want do I get, how much of what you want do you get, how much more do I get than you, and how much is there to get at all.

These properties are inevitable features of social choices, which are decisions involving more than one option and more than one agent. A situation with a single option is not a choice, and a situation with a single agent is not social. As argued before, adding complexity, whether more options, more agents, or more game levels, only stacks further factors on top of these four properties and never removes them. Even when all the payoffs are equal or zero, the dimensions still exist at their zero coordinates. The choice is simply trivial, and an agent whose every situation was flat in this way would have no occasion to want anything or to think at all. Whenever an agent wants things and lives in an environment with nonuniform rewards alongside other agents who want things, these four dimensions must exist.

The ubiquity of these properties does not imply that all agents come to care about them, but there are strong reasons to expect that they would. An agent that wants things must attend to how much of what it wants it gets. An agent that knows it lives among others who want things, and whose actions may affect it, is nearly compelled to track how much of what they want they get, a problem studied at length in evolutionary biology and evolutionary game theory (Hamilton, 1964; Trivers, 1971; Smith & Price, 1973; Axelrod & Hamilton, 1981). The work in that tradition suggests that a positive weight on another's payoff, genuine altruism, evolves under recurring conditions such as kinship, repeated interaction, and reputation. This also explains why the stranger dimensions one could write down in principle, such as one pairing the chooser's payoff in option A against the other's payoff in option B, would never acquire psychological weight. They answer no question an agent has reason to ask. How regular the conditions are that make it adaptive to be sensitive to social comparison or nonlinear marginal value are open questions.

The four dimensions belong to the fitness landscape of any world that contains distinct, persistent, mutually identifiable individuals who want things, whose wants sometimes conflict, and who cannot read one another's minds. That is not a claim about biology or about Earth. It is a claim about what social choice is. It is often said that mathematics and physics would be the common ground between us and any technological civilization elsewhere. The dimensions of social choice have the same character, because they are mathematics, and a civilization built from individuals meeting those few conditions would face the same four dimensions in its choices that we face in ours. Whether it would come to weigh them as we do, with altruism, comparison, and bending utility, is the open question above. But it would be asking the same four questions. Its members might then be located, as our participants are located here, as points in a shared coordinate system, differing in their weights along axes that any of them would recognize.

## 9.8 Closing

Social preferences have been studied for decades, but mostly through narrow keyholes: a handful of utility forms, in a handful of games, over a handful of payoff structures. What those keyholes showed was real, but partial. When we opened the door, the view cohered: more parameters were worth their complexity costs, the distinctions that simpler models erase turned out to matter, and the findings that emerged were both more surprising and more unified than the inherited picture suggested. The framework introduced here opens a larger space in which questions about human social motivation can be asked more precisely, answered more reliably, and compared more meaningfully across studies.

What we have shown is the simplest version of a more general idea: that the values behind a choice can be recovered from the choice itself, at fine resolution and with calibrated uncertainty. The same move, applied to richer models of how agents reason and choose, reaches further. Inferring what others value, and judging what kind of agent holds those values, is a capacity we will want not only in ourselves but in the artificial agents we build to act among us.